\documentclass[fleqn,usenatbib]{mnras}

\usepackage{newtxtext,newtxmath}

\usepackage[T1]{fontenc}

\DeclareRobustCommand{\VAN}[3]{#2}
\let\VANthebibliography\thebibliography
\def\thebibliography{\DeclareRobustCommand{\VAN}[3]{##3}\VANthebibliography}

\usepackage{graphicx}
\usepackage{amsmath}
\usepackage{caption}

\newcommand{\lya}{\mbox{Ly\,{\sc $\alpha$}}}
\newcommand{\hi}{\mbox{H\,{\sc i}}}
\newcommand{\hii}{\mbox{H\,{\sc ii}}}

\newcommand{\heii}{\mbox{He\,{\sc ii}}}

\title[Emergence of dust-free galaxies at $z>10$]{\vspace{-0.2cm} The ultraviolet continuum slopes of high-redshift galaxies: evidence for the emergence of dust-free stellar populations at $\mathbf{z > 10}$ \vspace{-0.2cm}}

\author[F. Cullen et al.]{F. Cullen $^{1}$\thanks{E-mail: fergus.cullen@ed.ac.uk},
D.\,J. McLeod$^{1}$, 
R. J. McLure$^{1}$, 
J. S. Dunlop$^{1}$, 
C. T. Donnan$^{1}$, 
A. C. Carnall$^{1}$, 
L. C. Keating$^{1}$, \and  
D. Magee$^{2}$,
K. Z. Arellano-Cordova$^{1}$, 
R.\,A.\,A. Bowler$^{3}$,
R. Begley$^{1}$, 
S. R. Flury$^{4}$, 
M. L. Hamadouche$^{1}$, \and
T. M. Stanton$^{1}$
\footnotesize\\\\
$^{1}$Institute for Astronomy, University of Edinburgh, Royal Observatory, Edinburgh, EH9 3HJ, UK\\
$^{2}$ Department of Astronomy and Astrophysics, UCO/Lick Observatory, University of California, Santa Cruz, CA 95064, USA\\
$^{3}$Jodrell Bank Centre for Astrophysics, University of Manchester, Oxford Road, Manchester, M13 9PL, UK\\
$^{4}$Department of Astronomy, University of Massachusetts Amherst, Amherst, MA 01002, United States\vspace{-0.3cm}\\
}

\date{Accepted XXX. Received YYY; in original form ZZZ\vspace{-0.2cm}}

\pubyear{2022}

\begin{document}
\label{firstpage}
\pagerange{\pageref{firstpage}--\pageref{lastpage}}
\maketitle

\begin{abstract}
We present an analysis of the ultraviolet (UV) continuum slopes ($\beta$) for a sample of $172$ galaxy candidates at $8 < z_{\mathrm{phot}} < 16$ selected from a combination of \emph{JWST} NIRCam imaging and COSMOS/UltraVISTA ground-based near-infrared imaging.
Focusing primarily on a new sample of $121$ galaxies at $\langle z \rangle \simeq 11$ selected from $\simeq 320$ arcmin$^2$ of public \emph{JWST} imaging data across $15$ independent data sets, we investigate the evolution of $\beta$ in the galaxy population at $z \geq 9$.
We find a significant trend between $\beta$ and redshift, with the inverse-variance weighted mean UV slope evolving from $\langle \beta \rangle = -2.17 \pm 0.06$ at $z = 9.5$ to $\langle \beta \rangle = -2.59 \pm 0.06$ at $z = 11.5$.
Based on a comparison with stellar population models including nebular continuum emission, we find that at $z>10.5$ the average UV continuum slope is consistent with the intrinsic blue limit of dust-free stellar populations $(\beta_{\mathrm{int}} \simeq -2.6)$.
These results suggest that the moderately dust-reddened galaxy population at $z < 10$ was essentially unattenuated at $z \simeq 11$.
The extremely blue galaxies being uncovered at $z>10$ place important constraints on dust attenuation in galaxies in the early Universe, and imply that the already observed galaxy population is likely supplying an ionising photon budget capable of maintaining ionised IGM fractions of $\gtrsim 5$ per cent at $z\simeq11$.
\end{abstract}

\begin{keywords}
galaxies: evolution - galaxies: formation - galaxies: high-redshift - galaxies: starburst -  dark ages, reionisation, first stars
\vspace{-0.5cm}
\end{keywords}



\section{Introduction}

The capability of \emph{JWST} to provide very deep, near-/mid-infrared imaging and spectroscopy at ${\lambda > 2\mu \mathrm{m}}$ is revolutionising our understanding of the earliest galaxies.
Before the launch of \emph{JWST}, a small number of $z>10$ galaxy candidates had been discovered using the \emph{Hubble Space Telescope} (\emph{HST}) \citep[e.g.][]{ellis2013, oesch2016}.
Now, deep NIRCam imaging surveys are revealing hundreds of these systems \citep[e.g.][]{finkelstein2022, adams2023_arxiv, adams2023, austin2023, bouwens2023, casey2023, donnan2023a, donnan2023b, finkelstein2023_apj, harikane2023a, hainline2023, leung2023, mcleod2024, robertson2023, donnan2024}.
Moreover, increasing numbers of these galaxies have been confirmed spectroscopically \citep[e.g.][]{arrabalharo2023a, arrabalharo2023b, bunker2023, curtis-lake2023, fujimoto2023, harikane2023b, hsiao2023, roberts-borsani2023, tang2023}.
The discovery of a large abundance of galaxies at $z>10$ has been one of the key early \emph{JWST} results.

The fact that we are finding large numbers of $z>10$ galaxies, including a number of relatively ultraviolet (UV) bright objects \citep[$M_{\mathrm{UV}} \leq -20$; e.g.][]{naidu2022, castellano2023, casey2023}, represents a challenge for models of early galaxy formation.
At $z \lesssim 8$ the evolution of the UV luminosity function (LF) can be readily explained assuming no redshift evolution in the star formation efficiency \cite[e.g.][]{bouwens2015, mason2015, mashian2016, tacchella2018, harikane2022}.
If this model is extrapolated to $z>8$, a continued evolution of the UV LF is predicted, but this is in tension with the \emph{JWST} observations \citep[e.g.][]{finkelstein2023_arxiv, harikane2023a}.
In fact, current constraints suggest that the UV LF does not evolve at all between $z=9$ and $z=11$ \citep[at least at $M_{\mathrm{UV}} \lesssim -20$;][]{mcleod2024}.

A number of explanations for the non-evolving UV LF have been proposed including (i) an evolving star-formation efficiency, for example via a reduction in the stellar feedback efficiency at ${z>10}$, including the potential for `feedback-free' star formation \citep{dekel2023, yung2023}; (ii) a bias towards young, highly star-forming galaxies up-scattered with respect to the median UV magnitude versus halo mass relation \citep{mason2023, shen2023}; (iii) inherent uncertainties related to the spectral energy distributions (SEDs) of low-metallicity stellar populations \citep[][]{inayoshi2022}; and (iv) a transition to essentially dust-free star formation at $z>10$ \citep{ferrara2023, ferrara2023b, ziparo2023}.
In this paper, we focus on the last of these scenarios, presenting an analysis of the UV continuum SEDs of galaxy candidates in the redshift range $8<z_{\mathrm{phot}}<16$ to investigate whether a transition to zero/negligible dust attenuation at $z>10$ is consistent with the current data.

It is well established that the stellar UV continuum, which can be described by a power-law ($f_{\lambda} \propto \lambda^{\beta}$ for $\lambda_{\mathrm{rest}}\simeq0.12-0.3 \, \mu \mathrm{m}$), is an excellent probe of dust obscuration in star-forming galaxies \citep[e.g.][]{calzetti1994, meurer1999}.
In general, the UV continuum slope, $\beta$, is sensitive to the light-weighted age, metallicity and dust attenuation of the population of massive stars in a galaxy (i.e. O- and B-type stars, with ages $\lesssim 100$ Myr).
However, at the highest redshifts - and more broadly for all young star-forming galaxies - age and metallicity effects become subdominant, and $\beta$ is especially sensitive to dust \citep[e.g.][]{tacchella2022}.
Indeed, this connection between $\beta$ and dust attenuation has been demonstrated directly in studies sensitive to infrared dust emission up to $z\simeq8$ \citep[see e.g.][]{bowler2023}.
By providing deep infrared imaging out to $\lambda = 5 \, \mu \mathrm{m}$, \emph{JWST}/NIRCam now enables robust estimates of $\beta$ for galaxies at $z>10$ \citep[e.g.][]{cullen2023, topping2023}, probing dust in galaxies at the earliest cosmic epochs.

In an earlier work, we presented an examination of the UV continuum slopes of galaxies at $z > 8$ selected from a combination of early \emph{JWST}/NIRCam imaging and ground-based near-infrared imaging from COSMOS/UltraVISTA \citep{cullen2023}.
This initial sample comprised $61$ galaxies at a mean redshift of $\langle z \rangle=10$ spanning a factor of $\simeq 80$ in UV luminosity ($-22.6 < M_{\mathrm{UV}} < -17.9$).
We found $\beta$ slopes that were, on average, bluer than their lower-redshift counterparts at fixed $M_{\mathrm{UV}}$.
These results had been tentatively anticipated with \emph{HST} \citep[e.g.][]{dunlop2013, wilkins2016, bahtawdeker2021, tacchella2022} and were in agreement with a number of theoretical model predictions \citep[e.g.][]{yung2019a, vijayan2021, kannan2022}.
Independent \emph{JWST} analyses painted a similar picture \citep[e.g.][]{topping2022, nanayakkara2023}.
These early studies of $z \simeq 10$ objects suggested stellar populations consistent with the young, low-metallicity, and moderately dust-reddened stellar populations anticipated by theoretical models.

Here, we expand upon our initial analysis using a new sample of galaxy candidates at $z>9$ selected from a number of public \emph{JWST} imaging surveys.
Our new set of candidates builds on the sample presented in \citet{mcleod2024}, which was used to robustly estimate the ${z\simeq11}$ luminosity function.
These new $z>9$ galaxy candidates are selected across $\simeq 320$ arcmin$^2$ of deep \emph{JWST}/NIRCam imaging, which represents a $\simeq 7 \times$ increase in area compared to \citet{cullen2023}. 
As a result, we can now significantly improve upon our previous analysis in terms of the total number of $z>9$ candidates.
Crucially, our new sample now contains a significant number of galaxy candidates at $z>10$, with an average redshift of $\langle z \rangle = 10.7$, and enables us to place robust constraints on the evolution of $\beta$ up to $z\simeq12$.

The new analysis presented here complements the recent study of \citet{topping2023}, who present an investigation of the UV continuum slopes of galaxies up to $z\simeq12$ in the JADES survey \citep{eisenstein2023}, finding that by $z=12$ the average value of $\beta$ is extremely blue ($\beta \simeq -2.5$).
In this paper, we explore evidence for a similar trend in an independent sample of galaxies selected across a wide range of current \emph{JWST} imaging data sets.
Our sample is selected across an area $\simeq 2 \times$ larger than the full JADES area ($320$ versus $175$ arcmin$^2$) and, crucially, across $11$ independent, noncontiguous fields, thereby mitigating the effect of cosmic variance.
Although the JADES imaging used in \citet{topping2023} is deeper, the increase in area means that our sample is on average brighter, and the resulting samples sizes at $z>9$ are comparable.

The paper is structured as follows.
In Section \ref{sec:sample_and_properties} we describe the data and sample selection, and provide details of the sample properties and our method for determining $\beta$.
In Section \ref{sec:results} we present the results of our analysis, focusing on the evolution of $\beta$ with $z$ and absolute UV magnitude ($M_{\mathrm{UV}}$).
Our primary new finding is that, by $z\simeq11$, the typical UV continuum slope of the galaxy population is extremely blue ($\beta \simeq -2.6$).
In Section \ref{sec:discussion} we discuss the implications of this result and demonstrate, based on an analysis of theoretical stellar population models, that the UV continuum slopes of galaxies at $z \gtrsim 10$ are consistent with dust-free star formation at this epoch.
Finally, we summarise our main conclusions in Section \ref{sec:conclusions}.
Throughout we use the AB magnitude system \citep{oke1974,oke1983}, and assume a standard cosmological model with $H_0=70$\,km s$^{-1}$ Mpc$^{-1}$, $\Omega_m=0.3$ and $\Omega_{\Lambda}=0.7$.

\section{Sample Selection and Properties}\label{sec:sample_and_properties}

Our primary high-redshift galaxy sample is drawn from the wide-area \emph{JWST} search for $z>8.5$ galaxies presented in \citet{mcleod2024}.
We additionally incorporated three other ultra deep datasets not included in the original \citet{mcleod2024} sample: NGDEEP \citep{Bagley2023}, the first data release of the \emph{JWST} Advanced Deep Extragalactic Survey (JADES, \citealt{eisenstein2023}), and the additional $\simeq25$ arcmin$^2$ region taken in parallel to the recent UNCOVER NIRSpec observations (hereafter ``UNCOVER-South'').
This primary sample was augmented by additional galaxies at $z > 7.5$ drawn from COSMOS/UltraVISTA and \emph{JWST} that were presented in our earlier work \citep{cullen2023}.
In the following, we discuss the sample selection and properties of the various data sets used in this work. 

\subsection{The wide-area \emph{JWST} sample}\label{subsec:widea-area-catalogue}

Our wide-area \emph{JWST} sample was drawn from $15$ independent publicly-available \emph{JWST} imaging datasets within $11$ extragalactic fields, covering an on-sky area of $\simeq 320 \, \mathrm{arcmin}^2$.
A list of these fields, including the proposal ID and the PI name of each dataset is given in Appendix \ref{appendix:datasets} (table \ref{tab: surveys}).
We also include references to the relevant survey paper, ancillary data, and lensing maps where applicable.
With the exception of NGDEEP, JADES, and UNCOVER-South, all of the datasets presented here are the same as those described in \citet{mcleod2024}; we refer interested readers to Section 2 of \citet{mcleod2024} for a detailed description of the datasets in common.

Our inclusion of three new fields yields a further $50$ sources compared to the \citet{mcleod2024} sample.
We have identified $9$ galaxy candidates at $z\geq9$ in our latest reduction of the NGDEEP observations \citep{Bagley2023} (see below for candidate selection details). 
The NGDEEP dataset covers the \textit{HST} UDF parallel 2 field \citep{oesch2007, bouwens2011, ellis2013}, with another NIRCam module covering an adjacent area contained within the GOODS-South CANDELS footprint; therefore, we supplemented the NGDEEP NIRCam imaging with the ACS F435W, F606W, F775W, F814W and F850LP imaging released as part of the Hubble Legacy Fields \citep{Illingworth2013, Whitaker2019}.
We also included data from the first data release of the \emph{JWST} Advanced Deep Extragalactic Survey (JADES, \citealt{eisenstein2023})\footnote{Accessed via https://archive.stsci.edu/hlsp/jades}.
The first data release \citep{Rieke2023} covers the `deep' portion of the imaging across an area of $\simeq 25 \ \mathrm{arcmin}^2$, from which we have selected a further $27$ $z\geq9$ galaxy candidates. 
Finally, we include $14$ $z>9$ candidates that were found in the $\simeq25$ arcmin$^2$ of UNCOVER-South.
The coordinates, $z_{\mathrm{phot}}$ and $M_{\mathrm{UV}}$ for these new candidates are given in Table \ref{tab: extrafieldobjects}.

All of the \emph{JWST}/NIRCam data in each field was reduced using the Primer Enhanced NIRCam Image Processing Library (\texttt{pencil}) software, a customised version of the \textit{JWST} pipeline version 1.6.2 (Magee et al. in prep). 
The CRDS pmap varied slightly between each dataset due to differences in the date on which the data were taken, reduced and analysed, although each pmap was sufficiently up-to-date ($\geq 0989$) to take into account the most recent zero-point calibrations.
Again, we refer interested readers to \citet{mcleod2024} for a detailed description of the various \texttt{pencil} pipeline steps and the full data reduction routine.
In Table \ref{tab: depths} we report the global 5$\sigma$ limiting magnitudes of the \textit{HST} and \textit{JWST} imaging in each field, calculated using 0.35$^{\prime\prime}-$diameter apertures and corrected to total assuming a point-source correction \citep[see][]{mcleod2024}.

    \begin{figure}
        \centerline{\includegraphics[width=\columnwidth]{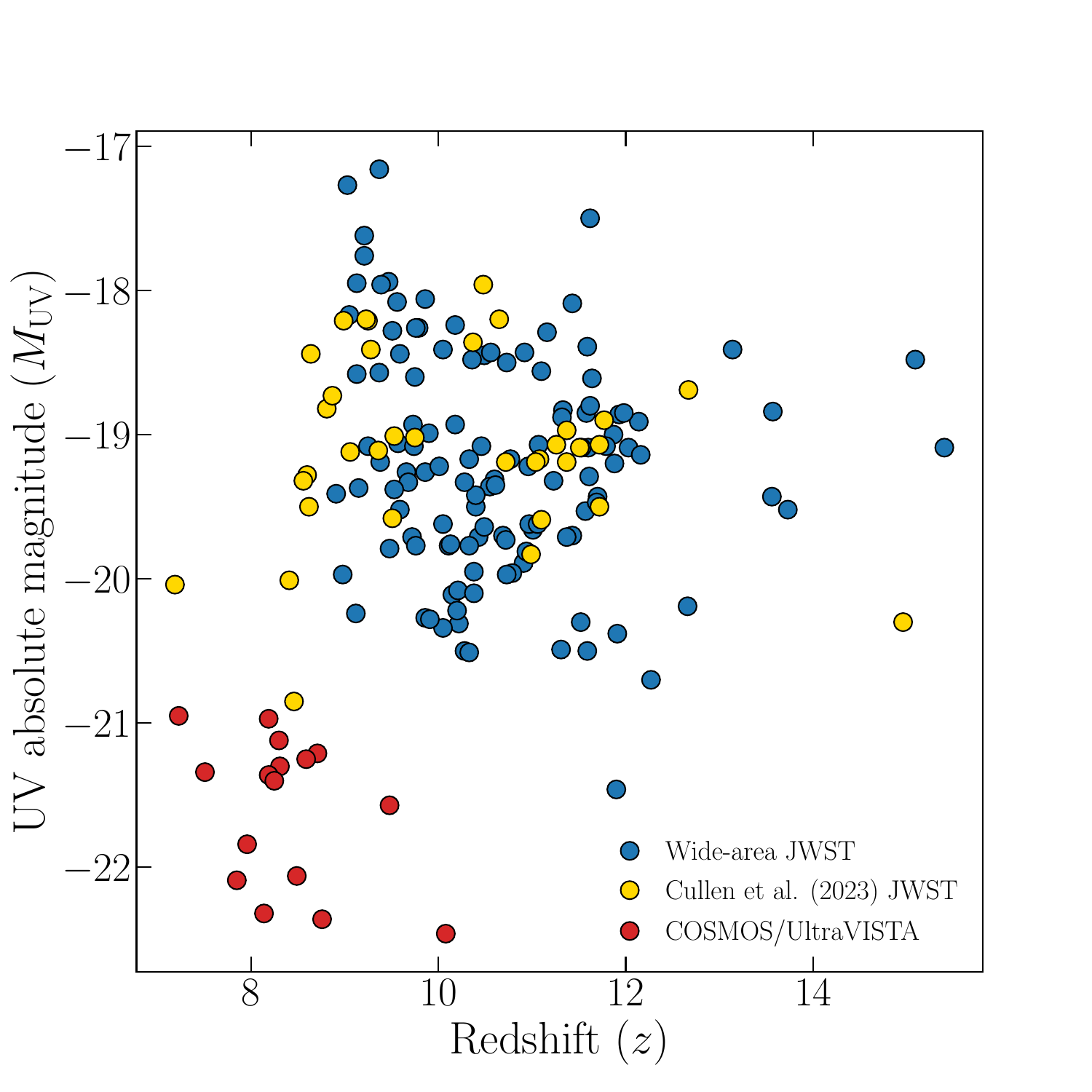}}
        \caption{The distribution of $M_{\mathrm{UV}}$ and photometric redshift for our primary wide-area \emph{JWST} sample (blue points) and the sample from \citet{cullen2023} comprised of ground-based COSMOS/UltraVISTA and \emph{JWST}-selected candidates (red and yellow points respectively).
        The combined sample covers the redshift range $7.6 < z_{\mathrm{phot}} < 15.9$ (i.e. spanning the time period $\simeq 250 - 700$ Myr after the Big Bang) and the $M_{\mathrm{UV}}$ range $-22.7 \leq M_{\mathrm{UV}} \leq -17.2$ (i.e. a factor of $\simeq 150$ in UV luminosity).}
        \label{fig:muv_z_sample}
    \end{figure}

\subsubsection{Construction of galaxy catalogues}

Prior to galaxy candidate selection, we first homogenised the point spread function (PSF) of each of our images to match the PSF full width half maximum (FWHM) of the F444W imaging. 
This enabled us to measure consistent photometry across all of the photometric filters, removing any systematics that arise as a result of differences in the PSF curve of growth.

For each data set, we constructed multiwavelength photometry catalogues running \textsc{Source Extractor} \citep{Bertin1996} in dual image mode and performing aperture photometry in 0.35$^{\prime\prime}-$diameter apertures. 
Catalogues were created primarily with the F200W imaging as the detection band, although we also created additional F277W-, F356W- and F444W-detected catalogues in order to include any sources that had been missed due to a low signal-to-noise ratio (SNR) in the F200W imaging.

To calculate photometric uncertainties, we adopted the method described in \citet{mcleod2024}.
We first generated a grid of nonoverlapping apertures of diameter 0.35$^{\prime\prime}$ spanning the full field of view and identified apertures corresponding to blank sky using a \textsc{Source Extractor} segmentation map. 
For a given object, we measured the median absolute deviation (MAD) of the nearest ${150-200}$ blank sky $0.35^{\prime\prime}-$ diameter apertures on an object-by-object basis and scaled to the standard deviation using $\sigma \simeq 1.4826 \, \mathrm{MAD}$. 
At this initial stage, we retained all sources detected at $\geq5\sigma$ in our detection catalogues. 
Finally, to reduce the possibility of artefact contamination, we required a $\geq3\sigma$ detection in any one of the other detection bands.

We performed spectral energy distribution (SED) fitting on all sources in our initial catalogues using the SED fitting code \textsc{LePhare} \citep{Arnouts2011}, adopting the \citet{bruzual2003} stellar population synthesis (SPS) library, a \citet{Chabrier2003} IMF and the \cite{calzetti2000} dust attenuation law. 
We modelled all sources assuming a declining star formation history with $\tau$ ranging from $0.1$ to $15$ Gyr and stellar metallicities of $0.2 \, \mathrm{Z}_{\odot}$ and $\mathrm{Z}_{\odot}$.
We allowed $A_{\mathrm V}$ to vary between $0$ and $6$ and accounted for intergalactic medium (IGM) absorption using the prescription of \cite{Madau1995}. 

We measured $M_{\mathrm{UV}}$ for each candidate from the best-fitting SED using a top-hat filter centred on $\lambda_{\mathrm{rest}} = 1500$\,\AA \ with $\Delta \lambda = 100$\,\AA. To correct our $M_{\mathrm{UV}}$ to total magnitudes, we scaled to the FLUX\_AUTO measurements and then applied an additional correction of $\simeq 10$ per cent to account for light outside the Kron aperture \citep{mcleod2024}.
For the cluster fields, we corrected $M_{\mathrm{UV}}$ for gravitational lensing by determining the magnification factor ($\mu$) using publicly available lensing maps (see Table \ref{tab: surveys}). 
For fields where multiple lensing maps are available, we adopted the median value of $\mu$ across all maps. 
For UNCOVER-South, where there are no public lensing maps, we adopted a flat ${\mu=1.2\pm0.2}$ magnification, motivated by the typical $\mu$ values around the southern boundary of the \cite{furtak2023} UNCOVER lensing map. 
While $\beta$ is unaffected by changes in $\mu$, we include an additional systematic uncertainty in $M_{\mathrm{UV}}$ to account for this assumed $\mu$ floor.

\subsubsection{High-redshift galaxy candidate selection}

To select our final sample, we first retained galaxies with a photometric redshift $z_{\mathrm{phot}}>8.5$ , a goodness of fit of $\chi^{2}_{\nu} \leq 10$, and at least $\Delta \chi^{2}\geq4$ between the goodness of fit of the primary redshift solution and secondary lower-redshift solution. 
To ensure a more robust final sample, we then required a detection of $\geq8\sigma$ in at least one of the F150W, F200W, or F277W filters. 
We also required a non-detection ($\leq 2\sigma$) in any of the F090W, F115W and \textit{HST} ACS imaging filters that were available for a given data set. 
In practice, the $\leq 2\sigma$ in F115W criterion restricts our sample to galaxies with $z_{\mathrm{phot}} \geq 9$.
Where there was a lack of F115W imaging, we added a further criterion, whereby we required either a ($\leq 2\sigma$) non-detection in F150W, or at least a factor of two increase in flux between F150W and F200W. 
This last criterion confined us to redshifts $z\gtrsim11.5$ for those datasets, but mitigated any potential contamination arising from F090W-F150W dropouts that have highly uncertain photometric redshifts.
For each candidate, additional checks were performed using the photometric redshift code \texttt{EAZY} \citep{brammer2008} as described in \citet{mcleod2024}.

Across the $11$ fields, we selected a total of $121$ galaxy candidates at $z_{\mathrm{phot}} \geq 9$.
Throughout this paper, we refer to these galaxies as our primary wide-area \emph{JWST} sample.
We use this sample to study the relationship between $\beta$ and redshift over the redshift range $9 \leq z \leq 12$ in Section \ref{subsec:beta_vs_redshift}.
The distribution of $M_{\mathrm{UV}}$ and $z_{\mathrm{phot}}$ is shown by the blue data points in Fig. \ref{fig:muv_z_sample}. 
The median values are $M_{\mathrm{UV}}=-19.2$ and $z_{\mathrm{phot}}=10.6$.

    \begin{figure*}
        \centerline{\includegraphics[width=0.81\textwidth]{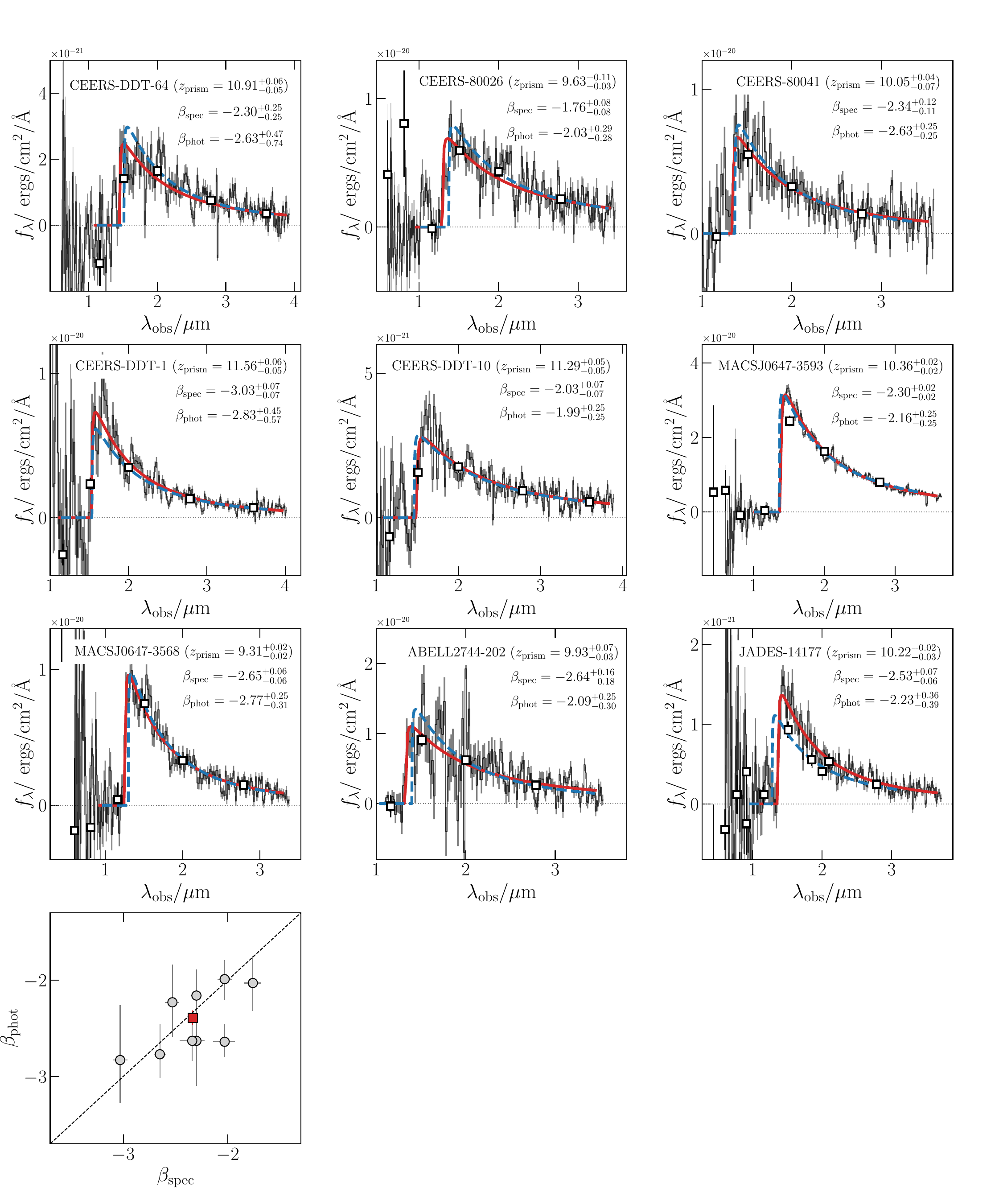}}
        \caption{A comparison of UV continuum slope fits to photometry ($\beta_{\mathrm{phot}}$) and spectroscopy ($\beta_{\mathrm{spec}}$) for the nine galaxies in our sample with \emph{JWST}/NIRCam and NIRSpec/PRISM observations.
        In the first nine panels, the black curve is the NIRSpec/PRISM spectrum, with the grey shading representing the $\pm 1 \sigma$ error spectrum.
        The square data points show the galaxy photometry, with all photometry at $\lambda \geq 0.9 \, \mu \mathrm{m}$ coming from \emph{JWST}/NIRCam.
        The solid red and dashed blue curves are the best-fitting models for the spectrum and photometry, respectively.
        In the majority of cases, the photometric and spectroscopic fits are fully consistent; however, for some galaxies, small differences in the best-fitting redshift and $\beta$ are evident.
        The tenth panel (lower left) shows the comparison between $\beta_{\mathrm{spec}}$ and $\beta_{\mathrm{phot}}$.
        The grey circular points show the nine individual galaxies and the red square point is the inverse variance-weighted mean.
        In general, we find very good agreement between the two estimates with $\langle \beta_{\mathrm{spec}} \rangle = -2.33 \pm 0.02$  and $\langle \beta_{\mathrm{phot}} \rangle = -2.39 \pm 0.07$.}
        \label{fig:uvslope_fit_examples}
    \end{figure*}
    
\subsection{Additional galaxy candidates: the combined sample}

The primary wide-area \emph{JWST} sample was augmented using the sample analysed in our previous study of the UV continuum slope \citep{cullen2023}.
This sample was initially selected and presented in the UV luminosity function study of \citet{donnan2023a} and combines a ground-based sample of bright galaxies at $z>7.5$ selected from the COSMOS/UltraVISTA field, and a \emph{JWST}-selected sample at $z > 8$ from early NIRCam imaging in SMACS J0723, GLASS, and CEERS.

The selection criteria for this sample were less stringent \citep[i.e. requiring only a $5\sigma$ detection in bands red-ward of the Lyman break; for full details see][]{donnan2023a} and therefore contained a number of candidates not selected in the new wide-area \emph{JWST} sample.
We retained the 35 galaxies from the \citet{cullen2023} \emph{JWST} sample that were not in our wide-area selection.
We included all 16 galaxies from the ground-based COSMOS/UltraVISTA sample; these galaxies, although at a lower median redshift ($z=8.3$), provide the valuable dynamic range in $M_{\mathrm{UV}}$ needed for studying the relationship between UV continuum slope and galaxy luminosity (which we explore in Section \ref{subsec:beta_muv_relations}).

In total, we included $51$ galaxy candidates from the sample presented in \citet{donnan2023a} and \citet{cullen2023}.
The distributions of $M_{\mathrm{UV}}$ and $z_{\mathrm{phot}}$ for the COSMOS/UltraVISTA and \emph{JWST} galaxies are shown by the red and yellow data points in Fig. \ref{fig:muv_z_sample}, respectively. 
The median values for the COSMOS/UltraVISTA sample are $M_{\mathrm{UV}}=-21.8$ and $z_{\mathrm{phot}}=8.3$; for the \emph{JWST} sample, the median values are $M_{\mathrm{UV}}=-19.0$ and $z_{\mathrm{phot}}=9.9$.
We note that, because of the less stringent S/N constraints, the $\beta$ and $M_{\mathrm{UV}}$ values of these galaxies have larger uncertainties compared to our primary wide-area sample.

Combined with the primary wide-area sample, the total sample contains $172$ candidates at $z>7.5$ (Fig. \ref{fig:muv_z_sample}); we refer to it as the `combined' sample throughout this paper.
This sample is used primarily in our analysis of the $\beta-M_{\mathrm{UV}}$ relation in Section \ref{subsec:beta_muv_relations}.
The median values of the absolute UV magnitude and redshift for the combined sample are $M_{\mathrm{UV}}=-19.4$ and $z_{\mathrm{phot}}=10.3$.
    
\subsection{Measuring the UV continuum slope}\label{subsec:measuring_uv_slope}

The primary goal of this paper is to study the UV continuum slopes of our high-redshift galaxy candidates.
The slope of the UV continuum in the rest frame is characterised by a power-law index, $\beta$, where ${f_{\lambda} \propto \lambda^{\beta}}$.
To determine $\beta$, we modelled the galaxy photometry covering rest-frame wavelengths $\lambda_{\rm rest} \leq 3000 \,${\AA} as a pure power law that includes both the effects of intergalactic medium (IGM) absorption and the \lya \ damping wing.
At $\lambda_{\rm rest} > 1216 \,${\AA}, we modelled the effect of the \lya \ damping wing using the prescription of \citet{miralda-escude1998} (equation 2).
At wavelengths below \lya \ ($\lambda_{\rm rest} \leq 1216 \,${\AA}), we assumed complete IGM attenuation (i.e. $f_{\lambda_{\rm rest} \leq 1216}=0$).

Our model consisted of three free parameters: (i) $\beta$, the UV continuum slope; (ii) $z$, the redshift of the galaxy; and (iii) $x_{\mathrm{HI}}$, the neutral hydrogen fraction of the surrounding IGM.
To sample the full posterior distribution, we used the nested sampling code \texttt{dynesty} \citep{speagle2020} assuming the following flat priors: ${-10 \leq \beta \leq 10}$; ${5 \leq z \leq 20}$ and $0 \leq x_{\mathrm{HI}} \leq 1.0$.
Examples of fits to nine galaxies in our sample are shown in Fig. \ref{fig:uvslope_fit_examples} (blue dashed lines).

We found that for a small minority of objects ($\simeq 10$ per cent of the full sample), our fits returned unrealistically small uncertainties ($\sigma_{\beta} < 0.1$) that significantly biased the population average estimates.
Based on our $\beta$-recovery simulations (Section \ref{subsec:recovery_sims}), we found that for the typical global imaging depths of our dataset, $\beta$ could be recovered for the brightest sources ($M_{\mathrm{UV}} \lesssim - 20$) to within ${\Delta \beta = \pm 0.25}$.
We therefore adopted this value as a minimum floor on the $\beta$ uncertainty.
Throughout this paper, unless otherwise stated, we adopt the best-fitting redshifts and corresponding uncertainties derived from these fits in our analysis.
We use these redshifts to calculate the best-fit values of $M_{\mathrm{UV}}$ and their corresponding uncertainties.
The derived values of $\beta$, $z_{\mathrm{phot}}$, and $M_{\mathrm{UV}}$ for our primary wide-area \emph{JWST} sample are given in Table \ref{tab:sample_beta_z_muv}.

Our method differs slightly from other similar studies in the literature which predominantly fix the redshift to $z_{\mathrm{phot}}$ and fit only to the photometric data redward of the Lyman break \citep[e.g.][]{topping2023, morales2023}.
However, we have verified that both methods ultimately yield comparable results, with offsets of $\Delta \beta < 0.1$.
For example, if we adopt the methodology described in \citet{topping2023} for our sample, we find a median offset of $\Delta \beta < +0.04$.
Although we advocate for our adopted method, which has the benefit of incorporating all photometric data and fully accounting for redshift uncertainties, our results remain essentially unchanged if we follow other methodologies common in the literature.

\subsubsection{The effect of \lya \ damping wings and proximate DLAs}

Evidence for \lya \ damping wings and even strong proximate damped \lya \ systems (DLAs) has been revealed from early \emph{JWST} NIRSpec/PRISM spectroscopy of galaxies at $z>8$ \citep{curtis-lake2023, heintz2023, umeda2023}. 
Both effects act to soften the \lya \ break and are a potential source of systematic bias in our derived value of $\beta$.

To account for the effect of the \lya \ damping wing, we included $x_{\mathrm{HI}}$ as a free parameter in our fits as described above.
Although it is not possible to constrain $x_{\mathrm{HI}}$ from photometric data alone, our method allowed us to marginalise over the uncertainties related to the unknown \lya \ damping wing strength.
We found that marginalising over $x_{\mathrm{HI}}$ had a minor effect on our results compared to assuming a fixed value of $x_{\mathrm{HI}}=1.0$; the median value of $\beta$ became marginally redder (${\Delta \beta = 0.02}$) and the median uncertainties were unchanged.
Object-to-object variations of up to ${\Delta \beta = \pm 0.3}$ were observed, but these were completely consistent with scatter due to the typical photometric uncertainties.
In general, we find that the average value of $\beta$ derived from broadband photometry is not strongly affected by the unknown strength of the \lya \ damping wing.

Recently, \citet{heintz2023} presented evidence for excess DLA absorption in three galaxies at $z > 8$.
\citet{heintz2023} argue that the strong absorption is likely due to large \hi \ gas reservoirs in the interstellar and circumgalactic medium along the line-of-sight; this leads to an attenuation of flux redward of \lya \ in excess of the neutral IGM damping wing (see, for example, Figure 1 of \citealt{heintz2023} and the discussion in \citealt{keating2023}).
We investigated the effect of DLA absorption by including a DLA at the source redshift with a neutral hydrogen column density of $N_{\rm HI} = 10^{22.5} \, \rm{cm}^{-2}$ in our model and re-fitting the sample.
This column density is the upper limit of the \citet{heintz2023} measurements, and also at the upper end of the $N_{\rm HI}$ distribution measured from the afterglow spectra of gamma-ray bursts (GRBs) \citep{tanvir2019}.
Again, we find that the overall effect on $\beta$ is negligible (median ${\Delta \beta = -0.01}$ with respect to the default assumptions).
The offset becomes more significant at the lowest redshifts where the difference between a DLA and the \lya \ damping wing is more pronounced.
For sources at $z<9.5$, the median offset is ${\Delta \beta = -0.05}$, however, this is still not large enough to affect the results presented here.

\subsection{Spectroscopic Sample}\label{subsec:spectroscopic_sample}

For nine galaxies in our primary sample, fully reduced \emph{JWST} NIRSpec/PRISM observations were available through the DAWN \emph{JWST} Archive (DJA)\footnote{https://dawn-cph.github.io/dja/index.html}.
We used these spectroscopic data to assess the accuracy of our photometric $\beta$ estimates.

The DJA reductions were performed using the software packages \texttt{grizli} \citep{grizli-brammer} and \texttt{msaexp} \citep{msaexp-brammer}.
The basic details of the data reduction are presented in \citet{valentino2023} and \citet{heintz2023}.
The NIRSpec spectra were affected by wavelength-dependent slit losses due to objects not being fully encompassed within the small MSA slitlets ($0.20^{\prime\prime} \times 0.46^{\prime\prime}$)  and/or slit misalignment.
To calibrate the 1D spectra, we integrated them through the available NIRCam filters and scaled the values to match the observed photometry, employing a linear interpolation between each band.
The nine flux-calibrated spectra are shown in Fig. \ref{fig:uvslope_fit_examples}.

The method used to fit $\beta$ from spectroscopic data is essentially the same as the method described above (Section \ref{subsec:measuring_uv_slope}).
The only difference is that, in the spectroscopic case, the model is smoothed to match the resolution of the NIRSpec/PRISM data rather than integrated through the relevant photometric filters.
To smooth the models we first resampled to four pixels per full width half maximum (FWHM) element and then convolved with a Gaussian with fixed pixel width of $\sigma_{\mathrm{pix}} = 1.7$.
Examples of these model fits are shown in Fig. \ref{fig:uvslope_fit_examples}.

\subsubsection{Comparing $\beta_{\mathrm{spec}}$ and $\beta_{\mathrm{phot}}$}

In Fig. \ref{fig:uvslope_fit_examples} we show the fits to the photometry (blue) and to the flux-calibrated spectroscopy (red) for the nine galaxies in our spectroscopic sample.
In general, we find excellent agreement between the two estimates. 
The lower right panel in Fig. \ref{fig:uvslope_fit_examples} shows the comparison of $\beta_{\mathrm{spec}}$ and $\beta_{\mathrm{phot}}$, where it can be seen that, for $\mathrm{N}=8/9$ of the galaxies, the two values are consistent within $1 \sigma$.
The inverse-variance weighted mean values of the two estimates are fully consistent with $\langle \beta_{\mathrm{spec}} \rangle = -2.33 \pm 0.02$ and $\langle \beta_{\mathrm{phot}} \rangle = -2.39 \pm 0.07$.

We find tentative evidence to suggest that, on average, $\beta_{\mathrm{phot}}$ is systematically lower (bluer) than $\beta_{\mathrm{spec}}$.
For the cases in which ${\beta_{\mathrm{phot}} < \beta_{\mathrm{spec}}}$, we find that the redshift estimated from the photometry is always larger than that estimated from the spectra, with ${\langle z_{\mathrm{phot}} - z_{\mathrm{spec}} \rangle \simeq 0.5}$.
Our sample size is clearly small, but other authors \citep[e.g.][]{arrabalharo2023a} have observed a tendency for $z_{\mathrm{phot}}$ to be systematically biased high at $z>8$ which, if confirmed, might suggest a general bias towards bluer $\beta_{\mathrm{phot}}$.
However, our results suggest that the average effect is likely to be small ($\Delta \beta \simeq 0.05$).
Larger samples of deep NIRSpec/PRISM spectra should soon be available to help accurately assess potential biases in $\langle \beta_{\mathrm{phot}} \rangle$ due to $z_{\mathrm{phot}}$ systematics (e.g. CAPERS; GO 6368; PI: M. Dickinson).

\begin{table}
    \centering
    \caption{The best-fitting UV continuum slopes ($\beta$) for our new wide-area JWST sample of galaxy candidates at $z>9$.
    The first column gives the source ID.
    Columns two and three give the best-fitting photometric redshift ($z_{\rm phot}$) and absolute UV magnitude ($M_{\rm UV}$) from our fitting procedure described in Section \ref{subsec:measuring_uv_slope}.
    Column four gives the derived UV continuum slope $\beta$.}
    \label{tab:sample_beta_z_muv}
    \renewcommand{\arraystretch}{1.35} 
    \begin{tabular}{lccc} 
        \hline
        ID & $z_{\rm phot}$ & $M_{\rm UV}$ & $\beta$ \\
        \hline
        RXJ-2129-3021681 & $8.98^{+1.12}_{-0.55}$ & $-19.97^{+0.10}_{-0.16}$ & $-2.56^{+0.27}_{-0.42}$\\
        RXJ-2129-16261 & $9.12^{+0.87}_{-0.40}$ & $-20.24^{+0.09}_{-0.12}$ & $-2.48^{+0.25}_{-0.32}$\\
        CEERS-570 & $10.91^{+0.43}_{-0.60}$ & $-19.89^{+0.25}_{-0.12}$ & $-3.86^{+0.90}_{-1.12}$\\
        CEERS-2543 & $11.43^{+2.07}_{-0.74}$ & $-19.70^{+0.21}_{-0.19}$ & $-2.63^{+0.47}_{-0.74}$\\
        CEERS-3530 & $10.28^{+0.28}_{-0.45}$ & $-20.50^{+0.19}_{-0.13}$ & $-2.03^{+0.29}_{-0.28}$\\
        CEERS-3961 & $11.37^{+0.24}_{-0.26}$ & $-19.71^{+0.09}_{-0.06}$ & $-3.99^{+0.45}_{-0.58}$\\
        CEERS-6303 & $9.86^{+0.55}_{-0.70}$ & $-20.27^{+0.16}_{-0.20}$ & $-2.11^{+0.34}_{-0.49}$\\
        CEERS-7927 & $10.15^{+0.10}_{-0.11}$ & $-20.11^{+0.05}_{-0.04}$ & $-1.90^{+0.25}_{-0.25}$\\
        CEERS-8536 & $11.91^{+0.63}_{-0.35}$ & $-20.38^{+0.10}_{-0.11}$ & $-2.36^{+0.26}_{-0.29}$\\
        CEERS-9617 & $10.22^{+0.26}_{-0.37}$ & $-20.31^{+0.12}_{-0.10}$ & $-2.63^{+0.25}_{-0.25}$\\
        CEERS-10389 & $10.69^{+0.83}_{-0.65}$ & $-19.70^{+0.25}_{-0.19}$ & $-2.64^{+0.57}_{-0.60}$\\
        CEERS-15760 & $11.59^{+0.66}_{-0.44}$ & $-20.50^{+0.16}_{-0.10}$ & $-2.83^{+0.45}_{-0.57}$\\
        CEERS-17070 & $11.01^{+0.36}_{-0.38}$ & $-19.66^{+0.15}_{-0.12}$ & $-1.99^{+0.25}_{-0.25}$\\
        CEERS-17402 & $9.90^{+0.96}_{-0.90}$ & $-18.99^{+0.35}_{-0.35}$ & $-1.03^{+0.40}_{-0.48}$\\
        CEERS-17416 & $9.86^{+0.64}_{-0.63}$ & $-19.26^{+0.12}_{-0.20}$ & $-2.81^{+0.39}_{-0.55}$\\
        CEERS-20465 & $11.31^{+0.20}_{-0.20}$ & $-20.49^{+0.05}_{-0.05}$ & $-2.68^{+0.25}_{-0.25}$\\
        CEERS-20687 & $10.97^{+0.21}_{-0.24}$ & $-19.62^{+0.10}_{-0.07}$ & $-3.04^{+0.35}_{-0.34}$\\
        CEERS-22229 & $10.01^{+0.65}_{-0.44}$ & $-19.22^{+0.13}_{-0.18}$ & $-3.06^{+0.68}_{-1.10}$\\
        CEERS-26592 & $11.52^{+0.75}_{-0.74}$ & $-20.30^{+0.37}_{-0.15}$ & $-3.51^{+1.02}_{-1.31}$\\
        CEERS-37967 & $13.73^{+0.81}_{-1.20}$ & $-19.52^{+0.18}_{-0.27}$ & $-2.94^{+0.45}_{-0.79}$\\
        CEERS-39491 & $10.60^{+0.50}_{-0.60}$ & $-19.31^{+0.25}_{-0.15}$ & $-2.73^{+0.59}_{-0.66}$\\
        CEERS-51314 & $10.43^{+0.11}_{-0.11}$ & $-19.71^{+0.04}_{-0.04}$ & $-0.88^{+0.25}_{-0.25}$\\
        CEERS-1032400 & $10.05^{+0.53}_{-0.54}$ & $-20.34^{+0.15}_{-0.18}$ & $-2.38^{+0.40}_{-0.52}$\\
        CEERS-1099004 & $9.72^{+0.35}_{-0.29}$ & $-19.71^{+0.12}_{-0.14}$ & $-1.21^{+0.25}_{-0.25}$\\
        CEERS-3069854 & $10.11^{+0.28}_{-0.75}$ & $-19.77^{+0.20}_{-0.11}$ & $-1.88^{+0.25}_{-0.25}$\\
        MACS-0647-20148 & $10.21^{+0.15}_{-0.18}$ & $-20.08^{+0.08}_{-0.06}$ & $-2.16^{+0.25}_{-0.25}$\\
        MACS-0647-20158 & $9.66^{+0.40}_{-0.36}$ & $-19.26^{+0.07}_{-0.08}$ & $-2.77^{+0.25}_{-0.31}$\\
        WHL-0137-10187 & $9.21^{+0.15}_{-0.11}$ & $-17.62^{+0.04}_{-0.04}$ & $-2.52^{+0.25}_{-0.25}$\\
        WHL-0137-22312 & $10.79^{+0.90}_{-0.95}$ & $-19.96^{+0.35}_{-0.22}$ & $-2.52^{+0.61}_{-0.73}$\\
        WHL-0137-21722 & $9.38^{+0.75}_{-0.52}$ & $-19.19^{+0.17}_{-0.13}$ & $-2.29^{+0.44}_{-0.51}$\\
        WHL-0137-2004877 & $9.37^{+1.06}_{-0.78}$ & $-18.57^{+0.34}_{-0.30}$ & $-1.93^{+0.54}_{-0.54}$\\
        NEPTDF-12882 & $10.94^{+0.12}_{-0.13}$ & $-19.81^{+0.04}_{-0.04}$ & $-2.49^{+0.25}_{-0.25}$\\
        NEPTDF-1042051 & $9.91^{+0.47}_{-0.43}$ & $-20.28^{+0.15}_{-0.20}$ & $-1.33^{+0.25}_{-0.30}$\\
        J1235-8807 & $11.87^{+2.33}_{-0.42}$ & $-19.00^{+0.12}_{-0.08}$ & $-3.66^{+0.80}_{-2.26}$\\
        J1235-1046989 & $11.33^{+4.14}_{-1.74}$ & $-18.83^{+1.19}_{-0.42}$ & $-2.50^{+4.77}_{-2.34}$\\
        J1235-2056758 & $10.40^{+0.09}_{-0.08}$ & $-19.50^{+0.03}_{-0.03}$ & $-2.13^{+0.25}_{-0.25}$\\
        Quintet-9091 & $11.90^{+0.30}_{-0.23}$ & $-21.46^{+0.07}_{-0.06}$ & $-2.80^{+0.25}_{-0.25}$\\
        SMACS-0723-7442 & $11.57^{+0.09}_{-0.07}$ & $-19.53^{+0.02}_{-0.02}$ & $-2.69^{+0.25}_{-0.25}$\\
        DDT-2756-1001979 & $10.33^{+0.15}_{-0.22}$ & $-20.51^{+0.10}_{-0.06}$ & $-2.60^{+0.25}_{-0.25}$\\
        DDT-2756-1004647 & $11.06^{+0.14}_{-0.14}$ & $-19.62^{+0.05}_{-0.04}$ & $-2.05^{+0.25}_{-0.25}$\\
        DDT-2756-1010177 & $9.57^{+0.35}_{-0.28}$ & $-19.06^{+0.07}_{-0.09}$ & $-1.76^{+0.25}_{-0.25}$\\
        DDT-2756-1010612 & $9.59^{+0.68}_{-0.61}$ & $-19.52^{+0.24}_{-0.21}$ & $-1.73^{+0.45}_{-0.40}$\\
        DDT-2756-1023292 & $11.58^{+0.34}_{-0.28}$ & $-18.85^{+0.07}_{-0.06}$ & $-2.73^{+0.25}_{-0.25}$\\
        \hline
    \end{tabular}
\end{table}

\begin{table}
\ContinuedFloat
    \centering
    \caption{Continued.}
    \label{tab:sample1}
    \def\arraystretch{1.35}
    \begin{tabular}{lccc} 
        \hline
        ID & $z_{\rm phot}$ & $M_{\rm UV}$ & $\beta$ \\
        \hline
        GLASS-1481 & $10.73^{+0.42}_{-0.70}$ & $-18.50^{+0.27}_{-0.14}$ & $-2.10^{+0.32}_{-0.26}$\\
        GLASS-3283 & $10.20^{+0.08}_{-0.07}$ & $-20.22^{+0.03}_{-0.03}$ & $-1.92^{+0.25}_{-0.25}$\\
        GLASS-9023 & $9.56^{+0.91}_{-0.60}$ & $-18.08^{+0.14}_{-0.26}$ & $-2.24^{+0.30}_{-0.46}$\\
        GLASS-9974 & $11.79^{+0.34}_{-0.24}$ & $-19.08^{+0.04}_{-0.05}$ & $-2.58^{+0.25}_{-0.25}$\\
        GLASS-12329 & $10.77^{+0.31}_{-0.42}$ & $-19.17^{+0.20}_{-0.11}$ & $-2.86^{+0.50}_{-0.42}$\\
        GLASS-15864 & $10.13^{+0.28}_{-0.39}$ & $-19.76^{+0.14}_{-0.13}$ & $-2.17^{+0.25}_{-0.25}$\\
        GLASS-17225 & $10.46^{+0.54}_{-0.61}$ & $-19.08^{+0.17}_{-0.13}$ & $-2.52^{+0.39}_{-0.48}$\\
        GLASS-28072 & $12.27^{+1.26}_{-0.46}$ & $-20.70^{+0.14}_{-0.15}$ & $-2.67^{+0.37}_{-0.44}$\\
        GLASS-29748 & $10.38^{+0.11}_{-0.12}$ & $-20.10^{+0.05}_{-0.04}$ & $-2.40^{+0.25}_{-0.25}$\\
        GLASS-30367 & $10.38^{+0.08}_{-0.08}$ & $-19.95^{+0.02}_{-0.02}$ & $-2.72^{+0.25}_{-0.25}$\\
        GLASS-32411 & $9.68^{+0.44}_{-0.28}$ & $-19.33^{+0.06}_{-0.11}$ & $-2.19^{+0.25}_{-0.25}$\\
        GLASS-33570 & $13.56^{+0.81}_{-0.83}$ & $-19.43^{+0.11}_{-0.27}$ & $-2.78^{+0.42}_{-0.80}$\\
        UNCOVER-3658 & $9.75^{+0.88}_{-0.68}$ & $-18.60^{+0.15}_{-0.25}$ & $-2.67^{+0.72}_{-1.32}$\\
        UNCOVER-3892 & $10.92^{+0.30}_{-0.35}$ & $-18.43^{+0.12}_{-0.07}$ & $-4.79^{+0.83}_{-0.93}$\\
        UNCOVER-6502 & $11.43^{+0.72}_{-0.61}$ & $-18.09^{+0.21}_{-0.12}$ & $-2.85^{+0.53}_{-0.61}$\\
        UNCOVER-12816 & $11.93^{+1.10}_{-0.47}$ & $-18.86^{+0.15}_{-0.12}$ & $-2.62^{+0.42}_{-0.49}$\\
        UNCOVER-15452 & $10.49^{+0.15}_{-0.15}$ & $-18.45^{+0.06}_{-0.06}$ & $-2.64^{+0.25}_{-0.25}$\\
        UNCOVER-23585 & $9.05^{+0.10}_{-0.09}$ & $-18.17^{+0.03}_{-0.03}$ & $-2.02^{+0.25}_{-0.25}$\\
        UNCOVER-30621 & $12.03^{+0.20}_{-0.16}$ & $-19.09^{+0.04}_{-0.04}$ & $-1.49^{+0.25}_{-0.25}$\\
        UNCOVER-30657 & $11.70^{+0.19}_{-0.18}$ & $-19.43^{+0.05}_{-0.05}$ & $-1.65^{+0.25}_{-0.25}$\\
        UNCOVER-32757 & $10.33^{+0.22}_{-0.29}$ & $-19.17^{+0.14}_{-0.10}$ & $-2.44^{+0.29}_{-0.26}$\\
        UNCOVER-47904 & $10.36^{+0.10}_{-0.10}$ & $-18.48^{+0.04}_{-0.04}$ & $-3.27^{+0.25}_{-0.25}$\\
        UNCOVER-48395 & $10.05^{+0.23}_{-0.33}$ & $-19.62^{+0.08}_{-0.08}$ & $-2.54^{+0.25}_{-0.25}$\\
        UNCOVER-53033 & $10.33^{+0.79}_{-1.47}$ & $-19.77^{+0.41}_{-0.29}$ & $-2.06^{+0.59}_{-0.81}$\\
        UNCOVER-53293 & $12.16^{+2.40}_{-0.64}$ & $-19.14^{+0.13}_{-0.44}$ & $-2.71^{+0.38}_{-1.22}$\\
        UNCOVER-2019859 & $13.57^{+0.80}_{-0.97}$ & $-18.84^{+0.16}_{-0.21}$ & $-3.06^{+0.43}_{-0.57}$\\
        UNCOVER-2037921 & $10.55^{+0.20}_{-0.22}$ & $-19.36^{+0.09}_{-0.07}$ & $-1.44^{+0.25}_{-0.25}$\\
        UNCOVER-3061884 & $10.40^{+0.31}_{-0.99}$ & $-19.42^{+0.31}_{-0.13}$ & $-2.29^{+0.46}_{-0.36}$\\
        JADES-3213 & $10.73^{+0.13}_{-0.16}$ & $-19.97^{+0.07}_{-0.06}$ & $-2.82^{+0.25}_{-0.25}$\\
        JADES-8695 & $10.18^{+0.42}_{-0.50}$ & $-18.93^{+0.15}_{-0.15}$ & $-2.09^{+0.29}_{-0.32}$\\
        JADES-9079 & $10.56^{+0.27}_{-0.32}$ & $-18.43^{+0.14}_{-0.10}$ & $-3.53^{+0.42}_{-0.37}$\\
        JADES-9320 & $11.64^{+1.21}_{-0.68}$ & $-18.61^{+0.22}_{-0.15}$ & $-2.88^{+0.53}_{-0.60}$\\
        JADES-9585 & $9.03^{+0.28}_{-0.23}$ & $-17.27^{+0.11}_{-0.10}$ & $-2.31^{+0.26}_{-0.25}$\\
        JADES-9996 & $12.14^{+1.01}_{-0.29}$ & $-18.91^{+0.08}_{-0.13}$ & $-2.14^{+0.25}_{-0.25}$\\
        JADES-11367 & $13.14^{+0.80}_{-0.83}$ & $-18.41^{+0.13}_{-0.18}$ & $-2.76^{+0.44}_{-0.58}$\\
        JADES-12993 & $9.79^{+0.38}_{-0.28}$ & $-18.26^{+0.07}_{-0.11}$ & $-2.59^{+0.27}_{-0.34}$\\
        JADES-14905 & $11.32^{+0.29}_{-0.34}$ & $-18.88^{+0.12}_{-0.09}$ & $-3.00^{+0.33}_{-0.36}$\\
        JADES-16929 & $10.18^{+0.40}_{-0.54}$ & $-18.24^{+0.18}_{-0.18}$ & $-2.34^{+0.31}_{-0.39}$\\
        JADES-19283 & $9.59^{+0.81}_{-0.61}$ & $-18.44^{+0.20}_{-0.25}$ & $-1.93^{+0.40}_{-0.47}$\\
        JADES-20624 & $9.53^{+0.57}_{-0.22}$ & $-19.38^{+0.06}_{-0.12}$ & $-2.44^{+0.25}_{-0.25}$\\
        JADES-33477 & $11.62^{+0.25}_{-0.23}$ & $-18.80^{+0.08}_{-0.07}$ & $-2.60^{+0.25}_{-0.25}$\\
        JADES-34414 & $11.16^{+0.22}_{-0.25}$ & $-18.29^{+0.09}_{-0.07}$ & $-3.14^{+0.27}_{-0.26}$\\
        JADES-45865 & $9.47^{+0.87}_{-0.50}$ & $-17.94^{+0.16}_{-0.25}$ & $-1.83^{+0.28}_{-0.39}$\\
        JADES-48153 & $9.39^{+0.86}_{-0.41}$ & $-17.96^{+0.14}_{-0.20}$ & $-2.49^{+0.29}_{-0.36}$\\
        JADES-50455 & $9.51^{+0.48}_{-0.38}$ & $-18.28^{+0.12}_{-0.11}$ & $-2.23^{+0.36}_{-0.39}$\\
        JADES-68549 & $11.88^{+0.14}_{-0.14}$ & $-19.20^{+0.04}_{-0.04}$ & $-3.14^{+0.25}_{-0.25}$\\
        JADES-69507 & $10.05^{+0.41}_{-0.51}$ & $-18.41^{+0.11}_{-0.17}$ & $-2.66^{+0.29}_{-0.39}$\\
        \hline
    \end{tabular}
\end{table}

\begin{table}
\ContinuedFloat
    \centering
    \caption{Continued.}
    \label{tab:sample1}
    \def\arraystretch{1.35}
    \begin{tabular}{lccc} 
        \hline
        ID & $z_{\rm phot}$ & $M_{\rm UV}$ & $\beta$ \\
        \hline
        JADES-69979 & $9.13^{+1.22}_{-0.42}$ & $-18.58^{+0.12}_{-0.21}$ & $-2.97^{+0.51}_{-0.85}$\\
        JADES-1015339 & $9.86^{+0.61}_{-0.57}$ & $-18.06^{+0.16}_{-0.18}$ & $-2.24^{+0.38}_{-0.43}$\\
        JADES-1047091 & $11.62^{+1.02}_{-0.83}$ & $-17.50^{+0.32}_{-0.20}$ & $-2.31^{+0.52}_{-0.50}$\\
        JADES-1058823 & $15.09^{+0.81}_{-0.75}$ & $-18.48^{+0.21}_{-0.16}$ & $-1.35^{+0.27}_{-0.28}$\\
        JADES-1125442 & $9.76^{+0.67}_{-0.52}$ & $-18.26^{+0.13}_{-0.23}$ & $-2.56^{+0.37}_{-0.52}$\\
        JADES-2016436 & $9.37^{+0.99}_{-0.92}$ & $-17.16^{+0.41}_{-0.38}$ & $-1.14^{+0.50}_{-0.48}$\\
        JADES-2084090 & $11.59^{+0.31}_{-0.31}$ & $-18.39^{+0.12}_{-0.09}$ & $-2.99^{+0.34}_{-0.36}$\\
        JADES-2103879 & $11.98^{+1.08}_{-0.41}$ & $-18.85^{+0.12}_{-0.12}$ & $-2.72^{+0.33}_{-0.39}$\\
        NGDEEP-17469 & $11.10^{+0.28}_{-0.32}$ & $-18.56^{+0.14}_{-0.08}$ & $-3.28^{+0.40}_{-0.41}$\\
        NGDEEP-23088 & $11.07^{+0.27}_{-0.27}$ & $-19.07^{+0.11}_{-0.08}$ & $-2.75^{+0.27}_{-0.26}$\\
        NGDEEP-26794 & $10.72^{+0.18}_{-0.23}$ & $-19.73^{+0.10}_{-0.07}$ & $-3.26^{+0.29}_{-0.26}$\\
        NGDEEP-51475 & $9.21^{+1.17}_{-0.52}$ & $-17.76^{+0.34}_{-0.30}$ & $-1.58^{+0.67}_{-0.40}$\\
        NGDEEP-51925 & $11.23^{+0.35}_{-0.39}$ & $-19.32^{+0.19}_{-0.13}$ & $-2.55^{+0.38}_{-0.32}$\\
        NGDEEP-54829 & $10.49^{+0.10}_{-0.11}$ & $-19.64^{+0.05}_{-0.04}$ & $-2.38^{+0.25}_{-0.25}$\\
        NGDEEP-1003576 & $15.40^{+0.29}_{-0.29}$ & $-19.09^{+0.08}_{-0.07}$ & $-2.56^{+0.25}_{-0.25}$\\
        NGDEEP-1003750 & $11.69^{+1.96}_{-0.49}$ & $-19.47^{+0.16}_{-0.26}$ & $-2.25^{+0.33}_{-0.51}$\\
        NGDEEP-1026946 & $9.13^{+0.08}_{-0.07}$ & $-17.95^{+0.02}_{-0.02}$ & $-2.57^{+0.25}_{-0.25}$\\
        UNCOVER-SOUTH-7302 & $9.48^{+0.97}_{-0.47}$ & $-19.79^{+0.12}_{-0.26}$ & $-2.47^{+0.25}_{-0.43}$\\
        UNCOVER-SOUTH-9195 & $9.15^{+1.23}_{-0.43}$ & $-19.37^{+0.14}_{-0.26}$ & $-2.49^{+0.27}_{-0.40}$\\
        UNCOVER-SOUTH-22977 & $9.25^{+0.08}_{-0.07}$ & $-19.08^{+0.02}_{-0.02}$ & $-2.27^{+0.25}_{-0.25}$\\
        UNCOVER-SOUTH-26383 & $9.73^{+0.20}_{-0.17}$ & $-18.93^{+0.02}_{-0.03}$ & $-2.55^{+0.25}_{-0.25}$\\
        UNCOVER-SOUTH-31496 & $11.60^{+0.56}_{-0.42}$ & $-19.09^{+0.13}_{-0.12}$ & $-2.41^{+0.28}_{-0.29}$\\
        UNCOVER-SOUTH-33059 & $9.76^{+0.21}_{-0.22}$ & $-19.77^{+0.04}_{-0.05}$ & $-2.04^{+0.25}_{-0.25}$\\
        UNCOVER-SOUTH-43514 & $8.91^{+0.78}_{-0.37}$ & $-19.41^{+0.13}_{-0.11}$ & $-2.62^{+0.31}_{-0.36}$\\
        UNCOVER-SOUTH-45015 & $9.74^{+0.38}_{-0.32}$ & $-19.08^{+0.08}_{-0.11}$ & $-2.20^{+0.25}_{-0.32}$\\
        UNCOVER-SOUTH-51653 & $11.53^{+0.31}_{-0.28}$ & $-19.09^{+0.09}_{-0.07}$ & $-2.60^{+0.25}_{-0.25}$\\
        UNCOVER-SOUTH-57007 & $10.61^{+0.15}_{-0.16}$ & $-19.35^{+0.07}_{-0.06}$ & $-2.79^{+0.25}_{-0.25}$\\
        UNCOVER-SOUTH-59562 & $11.61^{+0.09}_{-0.09}$ & $-19.29^{+0.02}_{-0.02}$ & $-2.22^{+0.25}_{-0.25}$\\
        UNCOVER-SOUTH-1017193 & $12.66^{+1.76}_{-0.77}$ & $-20.19^{+0.17}_{-0.27}$ & $-3.11^{+0.56}_{-1.16}$\\
        UNCOVER-SOUTH-1089623 & $10.96^{+0.34}_{-0.41}$ & $-19.22^{+0.17}_{-0.12}$ & $-2.04^{+0.25}_{-0.25}$\\
        UNCOVER-SOUTH-2017743 & $10.28^{+0.40}_{-0.65}$ & $-19.33^{+0.24}_{-0.17}$ & $-1.91^{+0.33}_{-0.31}$\\
        \hline
    \end{tabular}
\end{table}

\section{Results}\label{sec:results}

\begin{figure}
        \centerline{\includegraphics[width=\columnwidth]{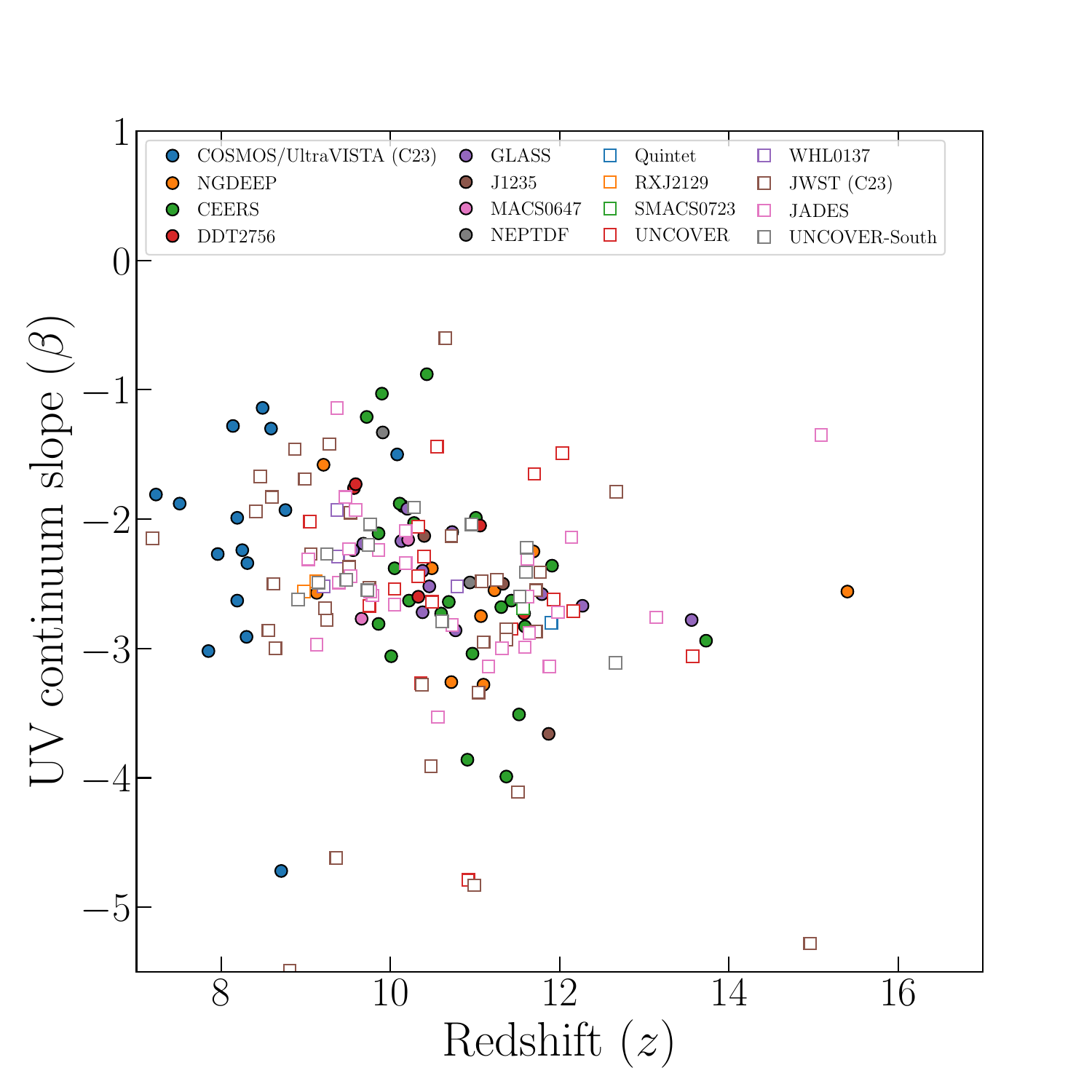}}
        \vspace{-0.25cm}
        \centerline{\includegraphics[width=\columnwidth]{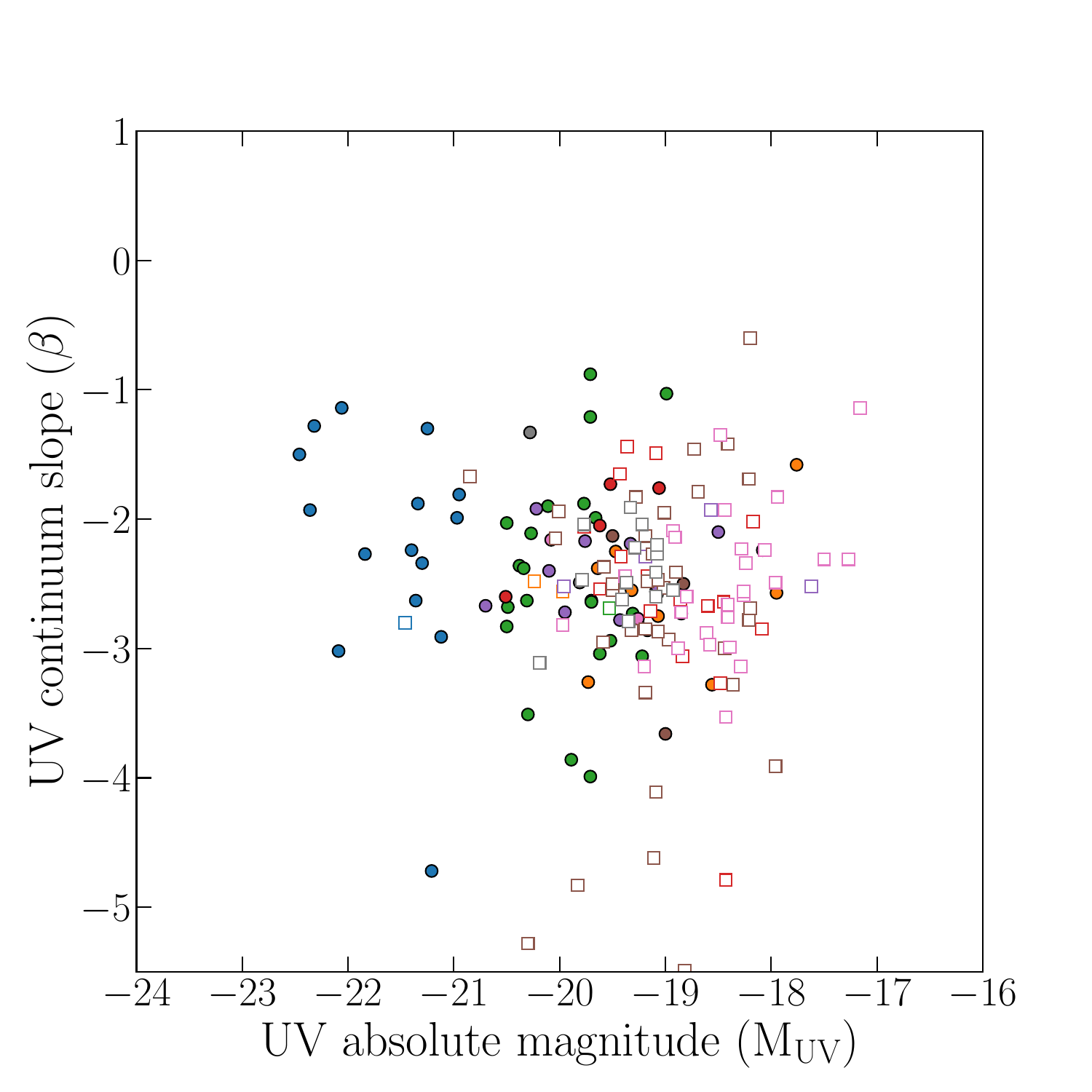}}
        \caption{Plots of UV continuum slope $\beta$ versus redshift (top) and versus absolute UV magnitude $M_{\rm UV}$ (bottom) for the galaxies in our combined sample sample. 
        In each panel, the points are colour- and symbol-coded by dataset (detailed in the legend in the top panel).
        The blue circular and brown square data points are taken from our previous sample presented in \citet{cullen2023}.
        All other data points come from our new wide-area \emph{JWST} sample (see Section \ref{sec:sample_and_properties} for full sample details and Table \ref{tab: surveys} for an overview of the wide-area sample).}
        \label{fig:sample_properties_beta}
    \end{figure}

In Fig. \ref{fig:sample_properties_beta} we plot the $\beta$ values for our combined sample as a function of redshift and $M_{\mathrm{UV}}$ and highlight the diverse selection of datasets used in this study.
The overall range of $\beta$ values in our new wide-area \emph{JWST} sample is consistent with that presented in \citet{cullen2023}, with measured values between $\beta \simeq -5$ and $\beta \simeq -1$, and we observe a number of similar features in the data.

First we again see a large scatter in $\beta$, with individual estimates as blue as $\beta \simeq -5$.
These extremely blue values ($\beta < -3$) are driven, at least in part, by the well-known blue bias in the estimates of $\beta$ at faint magnitudes \citep[e.g.][]{dunlop2012, rogers2014, cullen2023} and by the significant uncertainties for individual objects (the median uncertainty for the combined sample is ${\langle \sigma_{\beta} \rangle = 0.5}$). 
Unsurprisingly, the scatter is noticeably reduced in our higher S/N wide-area sample, which contains fewer extremely blue outliers; for example, $22$ per cent of the \citet{cullen2023} sample have $\beta < -3$ compared to $10$ per cent of the wide-area sample.
Nevertheless, it is important to note that the existence of objects with $\beta \lesssim -4$ $-$ even in robust, high S/N, samples $-$ emphasises the necessity of adopting empirical approaches to measuring $\beta$ that will not artificially truncate the full observed distribution \citep[as can happen when measuring $\beta$ via fitting stellar population models; see e.g.][]{rogers2013}.

    \begin{table}
        \centering
        \caption{Average $\beta$ values and standard errors derived for our wide-area \emph{JWST} and combined samples.
        The first column gives the sample name (as referred to in the text).
        In the second column we report the inverse-variance weighted mean and standard error of the individual $\beta$ values. 
        In the third column we report the median and $\sigma_{\rm MAD}$ of the individual $M_{\rm UV}$ values, where ${\sigma_{\rm MAD} = 1.483 \times \rm{MAD}}$ and MAD refers to the median absolute deviation.}
        \renewcommand{\arraystretch}{1.35}
        \begin{tabular}{l|c|c}
        \hline
        Sample & $\langle \beta \rangle$ & $\langle M_{\rm UV} \rangle$ \\
        \hline
         Wide-area (primary sample) & $-2.37 \pm 0.03$  &  $-19.2\pm0.8$ \\
         Combined \citep[including][]{cullen2023} & $-2.32 \pm 0.03$  &  $-19.3\pm0.8$ \\
        \hline
        \end{tabular}
        \label{tab:beta}
    \end{table}

In Table \ref{tab:beta} we report the inverse-variance weighted mean $\beta$ values for the primary sample and for the combined sample.
For our primary sample, we find ${\langle \beta \rangle = -2.37 \pm 0.03}$.
As in \citet{cullen2023}, we prefer the weighted mean over a simple median to mitigate against the blue bias in the $\beta$ scatter at faint magnitudes (this blue scatter is clearly visible in the bottom panel of Fig. \ref{fig:uvslope_fit_examples}).
For the combined sample (i.e. including the \citealp{cullen2023} sources) we find ${\langle \beta \rangle = -2.32 \pm 0.03}$.
This marginally redder value is primarily due to the inclusion of bright objects at $z<10$ from the ground-based COSMOS/UltraVISTA survey in the combined sample, as well as the slightly lower median redshift (see Fig. \ref{fig:muv_z_sample}, and the results presented in Sections \ref{subsec:beta_vs_redshift} and \ref{subsec:beta_muv_relations} below).

    \begin{figure*}
        \centerline{\includegraphics[width=16.5cm]{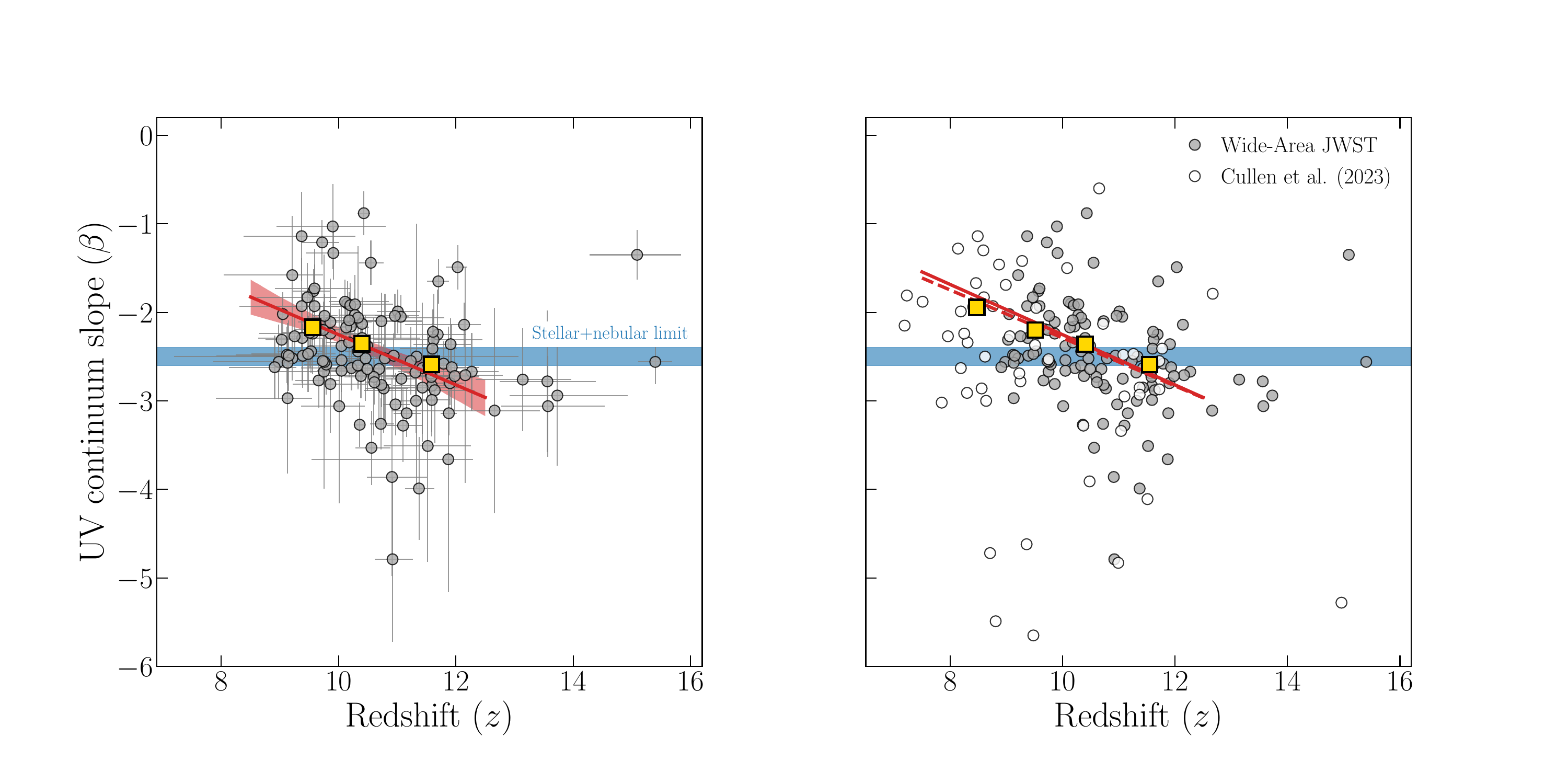}}
        \caption{Plots of UV continuum slope ($\beta$) versus redshift ($z$) at $z>8$.
        The left-hand panel shows the individual measurements for our primary wide-area \emph{JWST} sample (grey circular data points) and the inverse-variance weighted mean values in three redshift bins (yellow square data points).
        The solid red line shows the best-fitting $\beta-z$ relation fitted to the individual measurements, which has a slope of $\mathrm{d}\beta/\mathrm{d}z=-0.28\pm0.05$.
        The light-red shaded region shows the $95$ per cent confidence interval.
        The horizontal blue shading represents $\beta$ in the range $-2.6$ to $-2.4$, approximately the minimum value expected for a standard stellar population assuming no dust and a maximum contribution of the nebular continuum \citep[see Section \ref{subsec:dust_free_discussion} and][]{cullen2017, reddy2018}.
        In the highest redshift bin ($11 < z <12$), the population mean value approaches this limit, indicating galaxies unattenuated by dust.
        In the right-hand panel, we show the same $\beta-z$ relation for our combined sample \citep[i.e., including the galaxy candidates from][as indicated in the legend]{cullen2023}.
        For clarity, we omit error bars in this panel.
        At $z>9$, the \emph{JWST}-selected galaxies of \citet{cullen2023} are consistent with the same trend.
        Our data show evidence for a steep decline in $\beta$ with redshift at $z>9$, indicating that galaxies are growing progressively more dust and metal poor at early cosmic times.
        By $z\simeq 11$, the population-average $\beta$ is consistent with negligible, even zero, dust attenuation.}
        \label{fig:beta-versus-redshift}
    \end{figure*}

The $\langle \beta \rangle$ values reported in Table \ref{tab:beta} are somewhat bluer than the value of ${\langle \beta \rangle = -2.10 \pm 0.05}$  we reported in \citet{cullen2023}.
This is in part due to the fact that the new sample is marginally fainter, but the main reason is the higher proportion of candidates at $z>10$.
In our new wide-area \emph{JWST} sample $\simeq 70$ per cent ($84/121$) of the candidates have $z_{\mathrm{phot}} > 10$, compared to 43 per cent ($26/61$) in \citet{cullen2023}.
The implication here is that galaxies at $z>10$ are significantly bluer, and this can be seen in the top panel of Fig. \ref{fig:sample_properties_beta}.
This clear evolution to bluer $\beta$ is the main result of this study (we present a detailed investigation of the $\beta-z$ trend in Section \ref{subsec:beta_vs_redshift} below).

It is worth noting that $-$ as in \citet{cullen2023} $-$ the full sample average values reported in Table \ref{tab:beta} are not more extreme than the bluest galaxies observed in the local Universe \citep[e.g. NGC 1705; $\beta=-2.46$, $M_{\mathrm{UV}}=-18$;][]{calzetti1994, vazquez2004}.
As will be discussed in the following, we still find that up to $z\simeq10.5$ the typical value of $\langle \beta \rangle$ is consistent with local blue starburst galaxies.
However, we now find that at ${z>10.5}$ a significant fraction of galaxies have extremely blue UV continuum slopes with ${\langle \beta \rangle \simeq -2.6}$.
While ${\beta=-2.6}$ represents the extreme blue-end of objects observed in the local Universe \citep[e.g.][]{chisholm2022}, we find that this is the \emph{typical} value of $\beta$ for galaxies at $z\simeq 11$ with absolute UV magnitudes in the range $-20 \lesssim M_\mathrm{UV} \lesssim -17$.

In the following, we present a detailed analysis of the UV continuum slopes of our galaxy candidates and their dependence on redshift and UV luminosity ($M_{\mathrm{UV}}$). 
We begin in Section \ref{subsec:beta_vs_redshift} by investigating the evolution of $\langle \beta \rangle$ with redshift and present the main result of this work.
Then, in Section \ref{subsec:beta_muv_relations}, we derive the $\beta-M_{\mathrm{UV}}$ relation as a function of redshift for our sample, providing an update on the relation derived in \citet{cullen2023}.

\subsection{The evolution of $\beta$ with redshift}\label{subsec:beta_vs_redshift}

An evolution to bluer UV colours at higher redshifts is expected as galaxies become progressively metal and dust-poor at earlier cosmic times.
Before \emph{JWST}, this trend had been observed up to $z\simeq8$ \citep[e.g.][]{finkelstein2012, bouwens2012, dunlop2013, bouwens2014, wilkins2016}.
Recently, \citet{topping2023} have presented evidence for continued evolution beyond $z=8$, finding extremely blue average UV slopes of ${\langle \beta \rangle \simeq -2.5}$ at $z=12$ (we discuss the comparison between our results and \citealp{topping2023} below).

In Fig. \ref{fig:beta-versus-redshift} we show the $\beta-z$ relation for our primary sample (left-hand panel) and combined sample (right-hand panel).
Focusing first on the primary sample, it can be seen that we observe a clear $\beta-z$ trend in our data, with $\beta$ becoming progressively bluer at higher redshifts.
Fitting a linear relation to the individual data points (and accounting for the uncertainties in both $\beta$ and $z$) yields
\begin{equation}\label{eq:beta-redshift}
    \beta = -0.28^{\,+0.05}_{-0.05} \, z + 0.59^{\,+0.50}_{-0.49},
\end{equation}
where we have restricted the fit to $z \leq 12$ due to the lack of candidates at higher redshifts.
In Fig. \ref{fig:beta-versus-redshift} we also show the inverse-variance weighted mean $\beta$ values in bins of redshift. 
These binned values (reported in Table \ref{tab:beta-redshift}) are fully consistent with the fit to the individual objects.
We see that the typical $\beta$ changes from $\langle \beta \rangle = -2.59$ at $z \simeq 11.5$ to $\langle \beta \rangle = -2.17$ at $z \simeq 9.5$, a relatively rapid evolution of $\Delta \langle \beta \rangle \simeq 0.4$ over $\simeq 100$ Myr of cosmic time.

    \begin{table}
        \centering
        \caption{Average $\beta$ values and standard errors derived for our wide-area \emph{JWST} sample and combined sample as a function of redshift.
        The first column defines the redshift range.
        In the second column, we report the inverse-variance weighted mean and standard error of the individual $\beta$ values. 
        In the third column, we report the median redshift ($z$) and $\sigma_{\rm MAD}$.
        In the fourth column, we report the median $M_{\rm UV}$ and $\sigma_{\rm MAD}$.
        The values in columns two and three correspond to the yellow square data points shown in Fig. \ref{fig:beta-versus-redshift}.
        Our sample shows a clear trend in $\langle \beta \rangle$ with redshift.
        Crucially, the $\langle \beta \rangle$ values are not expected to be strongly biased due to the $\beta-M_{\mathrm{UV}}$ relation, as the $M_{\mathrm{UV}}$ distributions are similar in each redshift bin (with the exception of the $8 < z < 9$ bin for the combined sample; see text for discussion).
        In the highest redshift bin, the extremely blue value of $\langle \beta \rangle \simeq -2.6$ is consistent with dust-free stellar populations (see Section \ref{sec:discussion}).}
        \renewcommand{\arraystretch}{1.35}
        \begin{tabular}{r|c|r|r}
        \hline
        \multicolumn{1}{c}{Redshift range}&\multicolumn{1}{c}{$\langle \beta \rangle$}&\multicolumn{1}{c}{$\langle z \rangle$}&\multicolumn{1}{c}{$\langle M_{\mathrm{UV}} \rangle$} \\
        \hline
        \multicolumn{4}{|c|}{Wide area sample} \\
        \hline
         $9 < z < 10$  & $-2.17 \pm 0.06$  &  $9.6 \pm 0.3$  & $-18.9 \pm 1.0$ \\
         $10 < z < 11$ & $-2.36 \pm 0.05$  &  $10.4 \pm 0.3$ & $-19.6 \pm 0.6$ \\
         $11 < z < 12$ & $-2.59 \pm 0.06$  &  $11.6 \pm 0.3$ & $-19.1 \pm 0.6$ \\
        \hline
        \multicolumn{4}{|c|}{Combined sample} \\
        \hline
         $7.5 < z < 9$ & $-1.94 \pm 0.07$  &  $8.5 \pm 0.4$. & $-21.0 \pm 1.4$ \\
         $9 < z < 10$  & $-2.20 \pm 0.05$  &  $9.5 \pm 0.3$  &  $-19.0 \pm 1.0$ \\
         $10 < z < 11$ & $-2.36 \pm 0.05$  &  $10.4 \pm 0.3$ &  $-19.6 \pm 0.7$ \\
         $11 < z < 12$ & $-2.59 \pm 0.06$  &  $11.6 \pm 0.3$ &  $-19.1 \pm 0.4$ \\
        \hline
        \end{tabular}
        \label{tab:beta-redshift}
    \end{table}
    
We find that the redshift evolution observed in Fig. \ref{fig:beta-versus-redshift} is unlikely to be strongly affected by differences in the typical $M_{\mathrm{UV}}$ as a function of redshift.
Significant differences in $M_{\mathrm{UV}}$ could result in biases due to the known relation between $\beta$ and $M_{\mathrm{UV}}$ that has been demonstrated in numerous studies up to $z\simeq10$ (e.g.  \citealp{rogers2014, bouwens2014, cullen2023}; the $\beta-M_{\mathrm{UV}}$ relation for our new sample will be derived in Section \ref{subsec:beta_muv_relations}).
However, from Fig. \ref{fig:muv_z_sample} it can be seen that the $M_{\mathrm{UV}}$ distribution of the primary sample does not evolve strongly between $z=9$ and $z=12$.
From Table \ref{tab:beta-redshift}, the difference in $\langle M_{\mathrm{UV}} \rangle$ between adjacent redshift bins is approximately $|\Delta \langle M_{\mathrm{UV}} \rangle| \simeq 0.5$.
If we assume $\mathrm{d}\beta/\mathrm{d}M_{\mathrm{UV}}=-0.15$ (see Section \ref{subsec:beta_muv_relations}), this difference translates to $| \Delta \langle \beta \rangle | \simeq 0.08$, which cannot account for the observed changes in $\langle \beta \rangle$.
In Section \ref{subsec:recovery_sims} and Appendix \ref{app:beta_recovery_tests}, we provide further detailed discussions of possible selection effects and measurement biases present in our analysis.
In a variety of tests, we conclude that the $\beta-z$ trend is a real feature of our data.

Including the additional candidates from the combined sample, we find that they are fully consistent with the $\beta-z$ relation derived from the primary wide-area \emph{JWST} sample (right-hand panel of Fig. \ref{fig:beta-versus-redshift}).
At $z>9$, the \emph{JWST}-selected galaxies from \citet{cullen2023} follow the same trend.
In particular, it can be seen that these candidates also show extremely blue UV slopes at $z\simeq11$.
At $z<9$, the addition of ground-based COSMOS/UltraVISTA candidates from the \citet{cullen2023} sample provides an additional redshift bin at $\langle z \rangle = 8.5$.
The value of $\langle \beta \rangle = -1.94 \pm 0.07$ in this bin is consistent with an extrapolation of Equation \ref{eq:beta-redshift}.
However, we note that the galaxies in the ${7.5 < z < 9.0}$ redshift bin are significantly brighter by $\Delta M_{\mathrm{UV}} \simeq -2$, meaning that the average is likely biased with respect to the higher redshift bins (by up to $\Delta \beta \simeq 0.25$).
Taking into account this bias would suggest that the evolution between $z=8$ and $z=9$ is likely to be shallower if compared to a sample at $z=8$ with the same magnitude.
The focus of this paper is on the evolution in $\beta$ at $z>9$, but future \emph{JWST} studies will enable a robust determination of the redshift-evolution of $\beta$ for samples matched in magnitude across a wider redshift range.

\subsubsection{The transition to extremely dust-poor stellar populations}

    \begin{figure}
        \centerline{\includegraphics[width=\columnwidth]{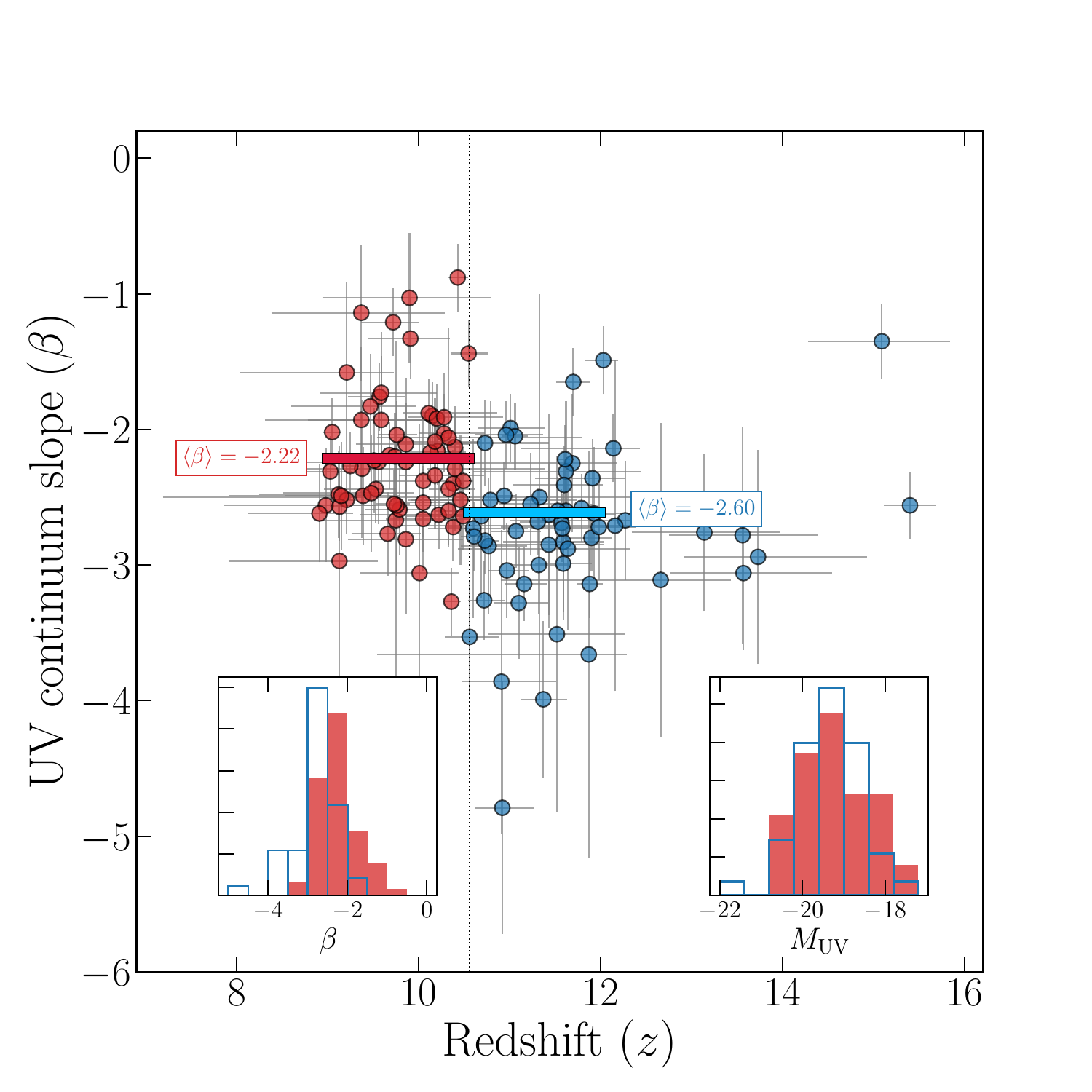}}
        \caption{Plot of $\beta$ versus redshift for our primary wide-area \emph{JWST} sample at $z \geq 9$  (coloured data points) highlighting the transition to dust-free stellar populations at $z \simeq 10.5$.
        Below $z\simeq10.5$ (black dotted vertical line), the typical value of the UV slope is $\langle \beta \rangle =-2.22$ (red solid line and red data points); this value is blue, but not inconsistent with moderate amounts of dust attenuation ($A_{\mathrm{UV}}\simeq0.5-1$, assuming $\beta_{\mathrm{int}}=-2.6$).
        Above this redshift, however, we find a typical value of $\langle \beta \rangle =-2.60$ (blue solid line and blue data points), which is at the extreme blue end of any galaxy observed locally \citep[e.g.][]{chisholm2022}.
        Based on stellar population modelling of young, metal-poor galaxies, $\beta \simeq -2.6$ is the intrinsic UV slope expected assuming a maximum nebular continuum contribution (i.e. $f_{\mathrm{esc}}=0$) but an absence of dust (see Section \ref{subsec:dust_free_discussion} and discussions in \citealp{robertson2010}; \citealp{cullen2017} and \citealp{reddy2018}).
        In the inset panels, we show the distribution of $\beta$ (left) and $M_{\mathrm{UV}}$ (right) above (red) and below (blue) $z\simeq10.5$.
        Due to the similar distributions in $M_{\mathrm{UV}}$ (both have $\langle M_{\mathrm{UV}} \rangle = -19.2$), we do not expect the difference in the $\beta$ distribution to be strongly biased by differences in galaxy luminosity.
        Our results suggest uniformly extremely blue $\beta$ values at $z>10.5$.}
        \label{fig:beta_z_redshift_transition}
    \end{figure}

The key result of this paper, highlighted in Fig. \ref{fig:beta-versus-redshift}, is that by ${z\simeq11}$ the average UV continuum slope of the galaxy population with ${-21.5 < M_{\mathrm{UV}} < -17.5}$ is consistent with the dust-free limit expected from standard stellar population models and nebular physics.
As we discuss in more detail in Section \ref{subsec:dust_free_discussion}, this limit assumes a young, dust-free, stellar population surrounded by ionised gas such that the ionising continuum escape fraction is 0 per cent and the strength of the nebular continuum emission is maximised.
Under these assumptions, the bluest UV continuum slopes expected are $\beta \simeq -2.4$ to $-2.6$ \citep[e.g.][]{robertson2010, cullen2017, reddy2018}.
This theoretical limit appears to be validated by observations of dust-poor local star-forming galaxies \citep[e.g.][]{chisholm2022}.
Any value bluer than this limit implies a non-zero escape fraction of ionising photons. 
Crucially, since dust acts to redden $\beta$\footnote{$\Delta \beta = f \, A_{1600}$ where $0.5 < f < 1.0$ for standard dust attenuation curve assumptions \citep[e.g. Calzetti to SMC;][]{mclure2018}.}, this limit also implies that it is extremely difficult to obtain UV slopes of $\beta \lesssim -2.6$ in the presence of dust attenuation.
As can be seen in Fig. \ref{fig:beta-versus-redshift}, our results suggest that this limit is reached by $z\simeq11$, implying that galaxies at this redshift are uniformly extremely dust-poor and essentially unattenuated.

The rapid transition to extremely blue $\beta$ values at the highest redshifts in our sample is highlighted further in Fig. \ref{fig:beta_z_redshift_transition}.
Considering the steep nature of the $\beta-z$ relation (Fig. \ref{fig:beta-versus-redshift}), we fitted a toy model to our wide-area \emph{JWST} sample with the following form:
\begin{equation}\label{eq:step_function_model}
    \begin{split}
    \langle \beta \rangle = \beta_0 \ (\mathrm{for} \ z < z_{\mathrm{t}}); \\
    \langle \beta \rangle = \beta_1 \ (\mathrm{for} \ z \geq z_{\mathrm{t}}),
    \end{split}
\end{equation}
where $z_{\mathrm{t}}$ is the redshift at which the population-average value of $\beta$ transitions from $\beta_0$ to $\beta_1$.
Although this model is not physically motivated, it gives a useful indication of the approximate redshift above which galaxies are becoming uniformly extremely blue.
    
Fitting the model in Equation \ref{eq:step_function_model} to our primary wide-area sample yields:
\begin{equation}\label{eq:step_model_results}
    \begin{split}
    & \beta_0=-2.22\pm0.05; \\
    & z_{\mathrm{t}}=10.56^{\,+0.03}_{-0.01}; \\
    & \beta_1=-2.60\pm0.05.
    \end{split}
\end{equation}
The fit is shown in Fig. \ref{fig:beta_z_redshift_transition}.
Again, we can be sure that this difference in $\langle \beta \rangle$ is not strongly biased due to differences in $M_{\mathrm{UV}}$ by comparing the $M_{\mathrm{UV}}$ distribution above and below $z_{\mathrm{t}}$.
The inset panel of Fig. \ref{fig:beta_z_redshift_transition} (lower right) shows that the $M_{\mathrm{UV}}$ distributions of the upper and lower redshift samples are fully consistent; both have ${\langle M_{\mathrm{UV}} \rangle = -19.2}$ and a standard KS test returns a significance ($p-$value) of $p=0.34$, consistent with the null hypothesis that both samples are drawn from the same $M_{\mathrm{UV}}$ distribution.

Interestingly, we find that the step function model (Fig. \ref{fig:beta_z_redshift_transition}) provides a better fit to our data than the linear model (Fig. \ref{fig:beta-versus-redshift}).
However, based on the reduced chi-squared values ($\chi^2_{\nu}$), neither model provides a formally statistically acceptable fit.
For the linear model, we find ${\chi^2_{\nu}=3.16}$ compared to ${\chi^2_{\nu}=2.01}$ for the step function model.
These large $\chi^2_{\nu}$ values are most likely due to a large intrinsic scatter in $\beta$ at a fixed redshift.
A given redshift bin will span a range of UV magnitude with a width of $\Delta M_{\mathrm{UV}} \simeq 3$ (Fig. \ref{fig:muv_z_sample}), corresponding to $\Delta \beta \simeq 0.45$ assuming $\mathrm{d}\beta/\mathrm{d}M_{\mathrm{UV}}=-0.15$ (see Section \ref{subsec:beta_muv_relations}).
Taking this into account, the fact that neither fit is formally acceptable is not particularly concerning.
Although neither fit can be preferred, it is intriguing that the current data prefer a sharp transition above $z \simeq 10.5$.
Larger samples, ideally with a larger fraction of robust spectroscopic redshifts, are needed to clarify the precise nature of the rate of evolution of $\langle \beta \rangle$ at $z>9$.

    \begin{figure}
        \centerline{\includegraphics[width=\columnwidth]{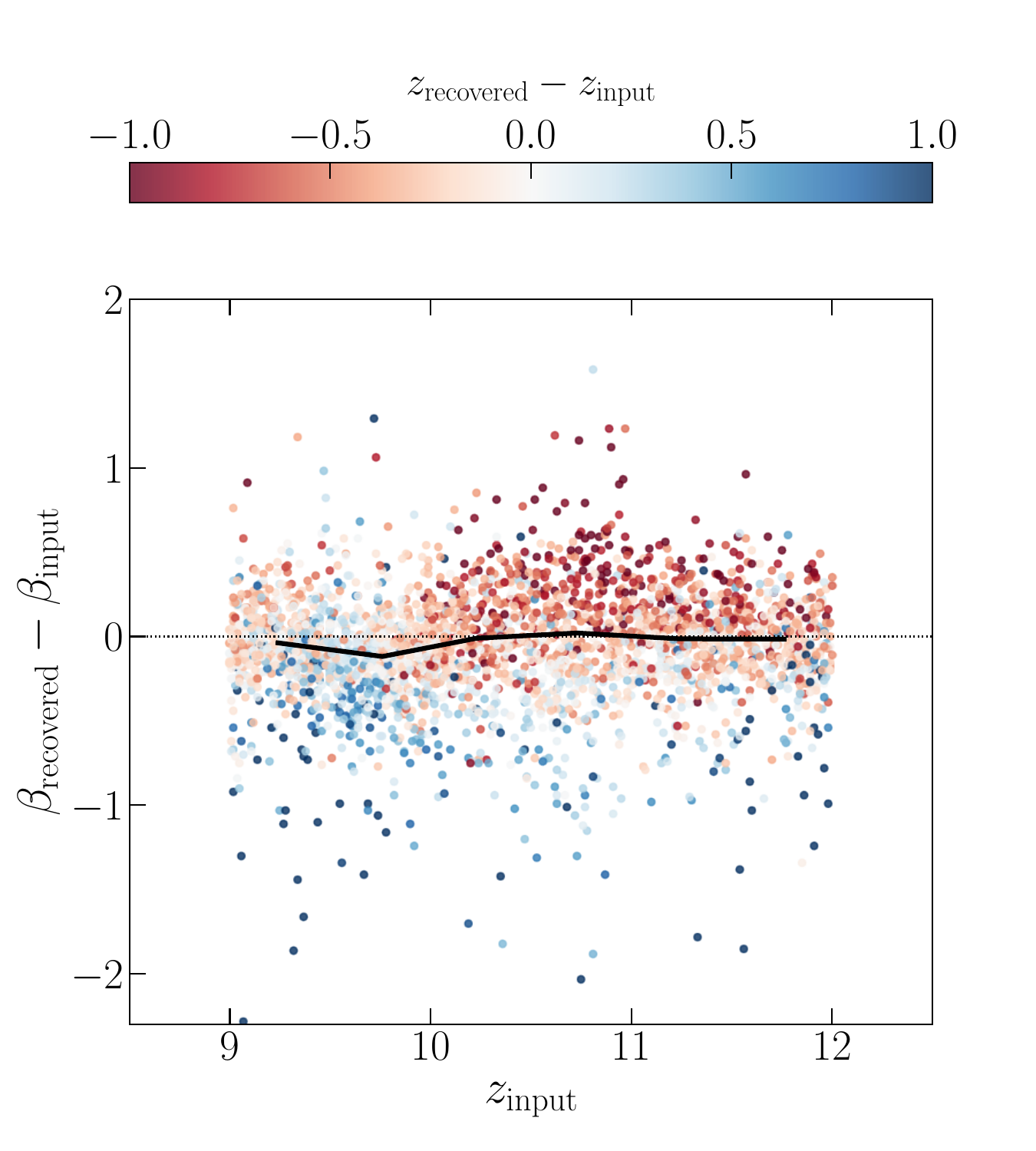}}
        \caption{Plot of the UV continuum slope bias ($\Delta \beta = \beta_{\rm recovered} - \beta_{\rm input}$) as a function of $z$ for of 20,000 simulated galaxies at $9 < z < 12$ with $\beta_{\rm int}=-2.2$ (see Section \ref{subsec:recovery_sims} for a description of the simulations).
        Each data point represents an individual galaxy from the simulation and is colour-coded according to the difference between the recovered and input redshifts ($z_{\rm recovered} - z_{\rm input}$).
        The solid black line shows the running inverse-variance weighted mean value of $\Delta \beta$ as a function redshift.
        In general, the bias is negligible at all redshifts ($\Delta \beta \simeq -0.01$).
        }
        \label{fig:uvslope_recovery}
    \end{figure}
    
Taken together, the model fits in Figs. \ref{fig:beta-versus-redshift} and \ref{fig:beta_z_redshift_transition} suggest that at $z=11$ the typical UV continuum slope of galaxies is ${\langle \beta_{\mathrm{obs}} \rangle \lesssim -2.5}$.
This implies that the galaxy population, across a relatively wide range of $M_{\mathrm{UV}}$, is consistent with the dust-free limit suggested by both theoretical models and observations of local galaxies.
Moreover, our data also suggest that the transition to these extremely blue slopes occurs over a narrow redshift range, with galaxies at $z\simeq9.5$ having on average blue, but not particularly extreme, UV continuum slopes of $\langle \beta \rangle \simeq -2.2$, consistent with a moderate amount of dust attenuation ($A_{\mathrm{UV}}\simeq0.5-1$, assuming $\beta_{\mathrm{int}}=-2.6$; see \citealp{mclure2018}).
Taken at face value, this implies a rapid buildup of dust in galaxies within the $\simeq 100$ Myr time frame between $z=11.5$ and $z=9.5$, consistent with models that predict an efficient ejection of dust in the earliest phases of galaxy formation \citep{ferrara2023, ziparo2023}, negligible amounts of dust being formed in the first phases of star formation \citep{jaacks2018} or efficient dust destruction at the highest redshifts \citep[e.g.][]{esmerian2023}.
    
\subsubsection{Selection effects and measurement bias}\label{subsec:recovery_sims}

It is worth considering the possibility that the extremely blue UV slopes at $z > 10$ are a result of selection effects and/or measurement biases.
Indeed, the measurement of $\beta$ from broadband photometry is known to be affected by subtle biases.
For example, the bias towards bluer values of $\beta$ at faint UV magnitudes has been extensively documented \citep[e.g.][]{bouwens2010, dunlop2012, rogers2014}.
In \citet{cullen2023}, we found that for our \emph{JWST} sample, the average value of $\beta$ was biased toward bluer values at $M_{\mathrm{UV, obs}} \geq -19.3$.
If we apply the relation derived in \citet{cullen2023} to the median $M_{\mathrm{UV}}$ of galaxies above and below the $z=10.55$ transition in Fig. \ref{fig:beta_z_redshift_transition} ($M_{\mathrm{UV, obs}}=-19.2$) we find that these $\langle \beta \rangle$ estimates could be biased blue by $\Delta \beta = -0.02$.
A bias at this level would clearly not affect our results. 
We have also confirmed that the same steep redshift-$\beta$ trend remains if we restrict our sample to the brightest galaxies in our wide-area \emph{JWST} sample with $M_{\mathrm{UV}} \leq -19.5$ (i.e. those which should not be affected by a measurement bias).
We therefore conclude that it is unlikely that our results are driven by a blue bias due to faint galaxies.

We have also explored potential redshift-dependent biases, which might arise from the fact that different combinations of filters are used and different portions of the UV spectrum are sampled depending on the redshift of the candidate.
For example, the UV continuum is sampled by the F150W, F200W, and F277W filters at $z < 11$, compared to the F200W, F277W, and F356W filters at $z > 11$.
To test for a redshift bias we ran a simple simulation in which we first constructed $20,000$ simple power-law SEDs with an intrinsic slope of $\beta_{\mathrm{int}}=-2.2$ spread uniformly across the redshift range $9 < z < 12$ and the $M_{\mathrm{UV}}$ range $-20.5 < M_{\mathrm{UV}} < -17.0$.
We then applied IGM attenuation using the prescription of \cite{inoue2014}.
Photometry was generated in each of the observed filters (see Table \ref{tab: depths}) and scattered according to the typical imaging depths (averaging the depths across multiple fields where appropriate).
The ‘observed’ UV continuum slopes ($\beta_{\mathrm{recovered}}$) were then recovered for the simulated galaxies that met our selection criteria.

    \begin{figure*}
        \centerline{\includegraphics[width=\textwidth]{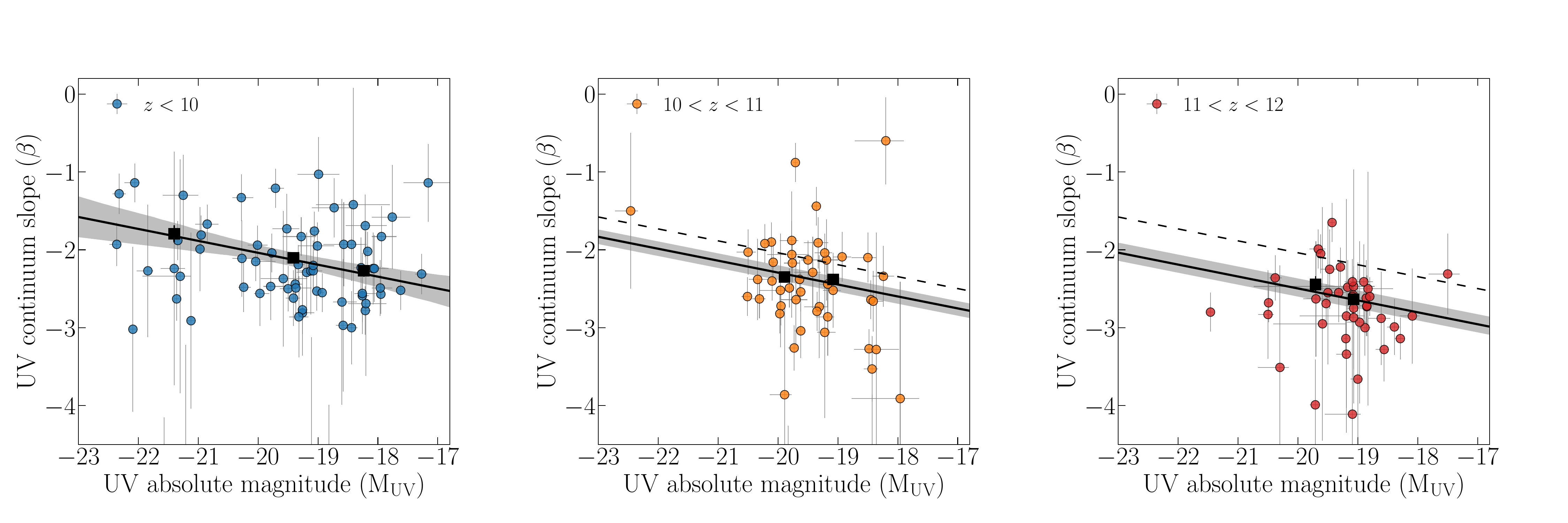}}
        \caption{Plots of $\beta$ versus $M_{\mathrm{UV}}$ for the combined sample in three bins of redshift.
        In each panel, the redshift interval is shown in the top left-hand corner, and coloured circular data points are the individual galaxy candidates in our sample. 
        The black square data points are the inverse-variance weighted values of $\langle \beta \rangle$ in the bins of $M_{\mathrm{UV}}$ given in Table \ref{tab:beta-muv-vs-redshift}.
        In the left-hand panel the black solid line is the best-fitting $\beta-M_{\mathrm{UV}}$ relation (Equation \ref{eq:beta_muv_lowz_bin}) with the light-grey shaded region showing the 95 per cent confidence interval around our best-fitting relation.
        We find evidence for a $\beta-M_{\mathrm{UV}}$ relation with best-fitting slope of ${\mathrm{d}\beta/\mathrm{d}M_{\mathrm{UV}}=-0.15 \pm 0.03}$ (i.e. $5\sigma$ significance).
        In the centre and right-hand panels, the black solid line shows the best-fitting $\beta-M_{\mathrm{UV}}$ relations where we have fixed the slope to be the same as the slope determined at $z<10$ (see text for details).
        In the centre and right-hand panels, the black dashed line shows the best-fitting $\beta-M_{\mathrm{UV}}$ relation at $z<10$ to highlight the redshift evolution of the normalisation.
        The best-fitting $\beta-M_{\mathrm{UV}}$ parameters are given in Table \ref{tab:beta_muv_fits}.}
        \label{fig:beta_muv_relations}
    \end{figure*}

The resulting bias as a function of redshift is shown in Fig. \ref{fig:uvslope_recovery}.
It can be seen that we do not observe a strong redshift-dependent effect.
The largest bias ($\Delta \beta = -0.11$) is seen in a narrow redshift interval around $z\simeq9.8$, which corresponds to the redshift at which the Lyman break passes between the F115W and F150W filters.
At this specific redshift, the photometric redshift is almost uniformly biased high, resulting in an underestimate of the true UV slope.
In general, the bias across all redshifts is negligible ($\Delta \beta \simeq -0.01$) and crucially, at $z>10.5$ there is no evidence for a strong blue bias.

    \begin{table}
        \centering
        \caption{Average $\beta$ values and standard errors for the combined sample as a function of $M_{\rm UV}$ in three redshift bins.
        The first column defines the $M_{\rm UV}$ range.
        In the second column we report the inverse-variance weighted mean and standard error of the individual $\beta$ values. 
        The third column gives the median $M_{\rm UV}$ and $\sigma_{\rm MAD}$.
        In the fourth column we report the median redshift ($z$) and $\sigma_{\rm MAD}$.
        The values in columns two and three correspond to the square black data points shown in Fig. \ref{fig:beta_muv_relations}.}
        \renewcommand{\arraystretch}{1.35}
        \begin{tabular}{r|c|r|r}
        \hline
        \multicolumn{1}{c}{$M_{\mathrm{UV}}$ range}&\multicolumn{1}{c}{$\langle \beta \rangle$}&\multicolumn{1}{c}{$\langle M_{\mathrm{UV}} \rangle$}&\multicolumn{1}{c}{$\langle z \rangle$} \\
        \hline
        \multicolumn{4}{|c|}{$7.5 < z < 10$} \\
        \hline
         $M_{\mathrm{UV}} < - 21$       & $-1.79 \pm 0.11$  & $-21.4 \pm 0.3$ &  $8.3 \pm 0.4$  \\
         $-21 < M_{\mathrm{UV}} < - 19$ & $-2.10 \pm 0.06$  & $-19.4 \pm 0.5$ &  $9.4 \pm 0.5$ \\
         $M_{\mathrm{UV}} > - 19$       & $-2.27 \pm 0.08$  & $-18.2 \pm 0.4$ &  $9.3 \pm 0.4$ \\
        \hline
        \multicolumn{4}{|c|}{$10 < z < 11$} \\
        \hline
         $-21 < M_{\mathrm{UV}} < - 19.5$ & $-2.35 \pm 0.06$  &  $-19.9 \pm 0.3$ &  $10.4 \pm 0.4$ \\
         $M_{\mathrm{UV}} > - 19.5$       & $-2.38 \pm 0.07$  &  $-19.1 \pm 0.5$ &  $10.5 \pm 0.3$ \\
        \hline
        \multicolumn{4}{|c|}{$11 < z < 12$} \\
        \hline
         $-21 < M_{\mathrm{UV}} < - 19.5$ & $-2.45 \pm 0.11$  &  $-19.7 \pm 0.2$ &  $11.4 \pm 0.3$  \\
         $M_{\mathrm{UV}} > - 19.5$       & $-2.63 \pm 0.07$  &  $-19.1 \pm 0.3$ &  $11.6 \pm 0.3$ \\
        \hline
        \end{tabular}
        \label{tab:beta-muv-vs-redshift}
    \end{table}
    
It is also clear from Fig. \ref{fig:uvslope_recovery} that the scatter in $\beta$ is driven primarily by the photometric redshift uncertainties.
When the photometric redshift is underestimated, $\beta$ is overestimated and vice versa.
We find that the typical scatter in the recovered $\beta$ at all $z$ and $M_{\mathrm{UV}}$ is $\sigma_{\beta} \simeq 0.25$, which motivates our decision to set this as a minimum error floor on individual $\beta$ estimates (see Section \ref{sec:sample_and_properties}).

Finally, we find that our results are unlikely to be driven by selection effects. 
This is partly highlighted by the similarity of the $M_{\mathrm{UV}}$ distributions above and below the $z=10.55$ transition in Fig. \ref{fig:beta_z_redshift_transition}, which demonstrates that our sample selection does not favour intrinsically fainter, and hence bluer, objects in the higher redshift bins.
Two further selection tests are described in detail in Appendix \ref{app:beta_recovery_tests}.
Across all tests, we find that the evolution of $\beta$ with redshift we observe is not caused by a more efficient selection of bluer objects at higher redshifts.

\subsection{The $\mathbf{\beta-M_{\mathrm{UV}}}$ relation at $\mathbf{z=8-12}$}\label{subsec:beta_muv_relations}

The $\beta-M_{\mathrm{UV}}$ relation, or colour-magnitude relation, traces the variation of the dust and stellar population properties of galaxies as a function of their UV luminosity.
Observations up to $z \simeq 10$ have shown that a correlation exists such that fainter galaxies are on average bluer \citep[e.g.][]{meurer1999, bouwens2009, rogers2014, bouwens2014, cullen2023, topping2023}.
These observations suggest that fainter galaxies are typically younger and less dust and metal enriched, in agreement with predictions of theoretical models \citep[e.g.][]{vijayan2021, kannan2022}.

In \citet{cullen2023} we provided a fit to $\beta-M_{\mathrm{UV}}$ for a sample of $61$ galaxies at $\langle z \rangle = 10$.
Due to the small sample size in \citet{cullen2023}, in this initial fit we included all galaxies across the full redshift range of the sample from $z=8$ to $z=16$\footnote{The $z=16$ candidate has subsequently been shown to be a $z\simeq5$ interloper \citep{arrabalharo2023b}, although the inclusion of this object does not affect the $\beta-M_{\mathrm{UV}}$ fit in \citet{cullen2023}.}.
We found evidence for a $\beta-M_{\mathrm{UV}}$ relation with $\mathrm{d}\beta/\mathrm{d}M_{\mathrm{UV}}=-0.17 \pm 0.05$.
This slope was consistent with the slope measured for large samples at lower redshift, for example, the $z=5$ relations of \citet{bouwens2014} and \citet{rogers2014} which have ${\mathrm{d}\beta/\mathrm{d}M_{\mathrm{UV}}=-0.12 \pm 0.02}$ and ${\mathrm{d}\beta/\mathrm{d}M_{\mathrm{UV}}=-0.14 \pm 0.02}$ respectively.
However, we found that the normalisation for the relation was lower at $z \simeq 10$, such that galaxies are on average bluer by $\Delta \beta = -0.4$  compared to $z \simeq 5$.

    \begin{table}
        \centering
        \caption{The best-fitting values for $\mathrm{d}\beta/\mathrm{d}M_{\mathrm{UV}}$ and  $\beta(M_{\mathrm{UV}}=-19)$ for the fits shown in Fig. \ref{fig:beta_muv_relations}.
        These fits assume the following functional form: ${\beta = \mathrm{d}\beta/\mathrm{d}M_{\mathrm{UV}} \times (M_{\mathrm{UV}}+19) + \beta(M_{\mathrm{UV}}=-19)}$.
        The first column gives the redshift range.
        In the second column we report the best-fitting value of $\mathrm{d}\beta/\mathrm{d}M_{\mathrm{UV}}$; note that this value is fixed for the two bins at $z>10$ (see text for details).
        In the third column we report the best-fitting value of $\beta(M_{\mathrm{UV}}=-19)$.}
        \renewcommand{\arraystretch}{1.35}
        \begin{tabular}{r|l|c}
        \hline
        Redshift range & $\mathrm{d}\beta/\mathrm{d}M_{\mathrm{UV}}$ & $\beta(M_{\mathrm{UV}}=-19)$ \\
        \hline
         $7.5 < z < 10$ & $-0.15\pm 0.03$  &  $-2.19 \pm 0.05$ \\
         $10 < z < 11$& $-0.15$ (fixed)  &  $-2.45 \pm 0.06$ \\
         $11 < z < 12$& $-0.15$ (fixed)  &  $-2.64 \pm 0.06$ \\
        \hline
        \end{tabular}
        \label{tab:beta_muv_fits}
    \end{table}

With our enlarged sample, we are now in a position to investigate the redshift evolution of the $\beta-M_{\mathrm{UV}}$ relation at $z\simeq8-12$.
For the analysis in this section, we make use of the combined sample, which includes the COSMOS/UltraVISTA galaxies at $7.5 < z < 10$.
These ground-based candidates trace the bright end of the UV luminosity distribution ($M_{\mathrm{UV}} < -21$) and are therefore crucial in providing a large dynamic range in $M_{\mathrm{UV}}$ (see Fig. \ref{fig:muv_z_sample}). 
To investigate the evolution of the $\beta-M_{\mathrm{UV}}$ relation with redsfhit, we split our sample into three bins of redshift; for each redshift bin, we then calculated the inverse-variance weighted mean $\beta$ in three bins of $M_{\mathrm{UV}}$.
The resulting values of $\langle \beta \rangle$, $\langle M_{\mathrm{UV}} \rangle$, and $\langle z \rangle$ are given in Table \ref{tab:beta-muv-vs-redshift}. 

In practice, only the redshift bin $7.5 < z < 10$ has a sufficient dynamic range in $M_{\mathrm{UV}}$ (from $M_{\mathrm{UV}}=-22.7$ to $M_{\mathrm{UV}}=-17.2$; Fig. \ref{fig:beta_muv_relations}) to allow an accurate estimate of the slope of the $\beta-M_{\mathrm{UV}}$ relation.
In this redshift range, we find evidence for a $M_{\mathrm{UV}}$ dependence, with $\langle \beta \rangle$ evolving from $\langle \beta \rangle=-1.79 \pm 0.11$ at the brightest magnitudes (median $M_{\mathrm{UV}}=-21.4$) to $\langle \beta \rangle=-2.27 \pm 0.08$ at the faintest magnitudes (median $M_{\mathrm{UV}}=-18.2$).
The best fitting colour-magnitude relation to the individual candidates (blue data points in Fig. \ref{fig:beta_muv_relations}) is
\begin{equation}\label{eq:beta_muv_lowz_bin}
    \beta = (-0.15 \pm 0.03) (M_{\mathrm{UV}} + 19) - (2.19\pm0.05).
\end{equation}
where we have accounted for the uncertainties in both $\beta$ and $M_{\mathrm{UV}}$.
In this formulation, the intercept, $\beta(M_{\mathrm{UV}}=-19)=-2.19\pm0.05$, represents the typical value of $\beta$ at $M_{\mathrm{UV}}=-19$.
Our new best-fit relation is slightly shallower than the fit in \citet{cullen2023} but is fully consistent within the uncertainties. 
We note that the median redshift of galaxies at the bright end ($M_{\mathrm{UV}}<-21$) is lower than the median redshift in the fainter bins by $\Delta z \simeq 1$, which may introduce a bias if there is a significant redshift evolution in $\langle \beta \rangle$ at fixed $M_{\mathrm{UV}}$ between $z=8$ and $z=9$.
The clear lack of galaxies at $z \geq 9$ and $M_{\mathrm{UV}}<-21$ with \emph{JWST} photometry is something that will hopefully be addressed with current and future wide-area surveys \citep[e.g.][]{casey2023,franco2023}.

In the two higher redshift bins we do not attempt to constrain $\beta-M_{\mathrm{UV}}$ due to the lack of galaxies with ${M_{\mathrm{UV}} < -21.0}$.
In both redshift bins, the data do not show any clear evidence for a $\beta-M_{\mathrm{UV}}$ relation between ${M_{\mathrm{UV}} = -20.0}$ and ${M_{\mathrm{UV}} = -18.0}$.
Due to the lack of sufficient dynamic range $M_{\mathrm{UV}}$ at these redshifts, we have assumed a fixed slope of ${\mathrm{d}\beta/\mathrm{d}M_{\mathrm{UV}}=-0.15}$ in these redshifts bins and fitted only the normalisation.
This approach yields ${\beta(M_{\mathrm{UV}}=-19)=-2.45 \pm 0.06}$ at $10 < z < 11$ and ${\beta(M_{\mathrm{UV}}=-19)=-2.64 \pm 0.06}$ at $11 < z < 12$ (where the uncertainties do not account for the slope uncertainty).
The two fits are shown in Fig. \ref{fig:beta_muv_relations}.
The resulting best-fitting $\beta-M_{\mathrm{UV}}$ parameters are given in Table \ref{tab:beta_muv_fits}.
These fits again highlight our main result: at fixed $M_{\mathrm{UV}}$, the typical value of $\langle \beta \rangle$ is evolving rapidly with redshift between $z=9$ and $z=12$, reaching $\langle \beta \rangle \simeq -2.6$ at the highest redshifts.

In summary, we find evidence ($5 \sigma$) for a ${\beta-M_{\mathrm{UV}}}$ relation in our sample at $\langle z \rangle \simeq 9$ with a slope of ${\mathrm{d}\beta/\mathrm{d}M_{\mathrm{UV}}=-0.15 \pm 0.03}$.
This slope is fully consistent with the slope derived at $z=5$ \citep[e.g.][]{bouwens2014, rogers2014}.
Our results currently suggest that the slope of the $\beta-M_{\mathrm{UV}}$ relation is already established within the first $\simeq 500$ Myr of cosmic time and does not evolve strongly thereafter.
At $z>10$ we lack the dynamic range in $M_{\mathrm{UV}}$ to constrain the slope, however, assuming a fixed ${\mathrm{d}\beta/\mathrm{d}M_{\mathrm{UV}}=-0.15}$, we find that the normalisation evolves rapidly from  $\beta(M_{\mathrm{UV}}=-19)=-2.19\pm0.05$ at $\langle z \rangle \simeq 9$ to $\beta(M_{\mathrm{UV}}=-19)=-2.64\pm0.06$ at $\langle z \rangle \simeq 11$.

    \begin{figure}
        \centerline{\includegraphics[width=\columnwidth]{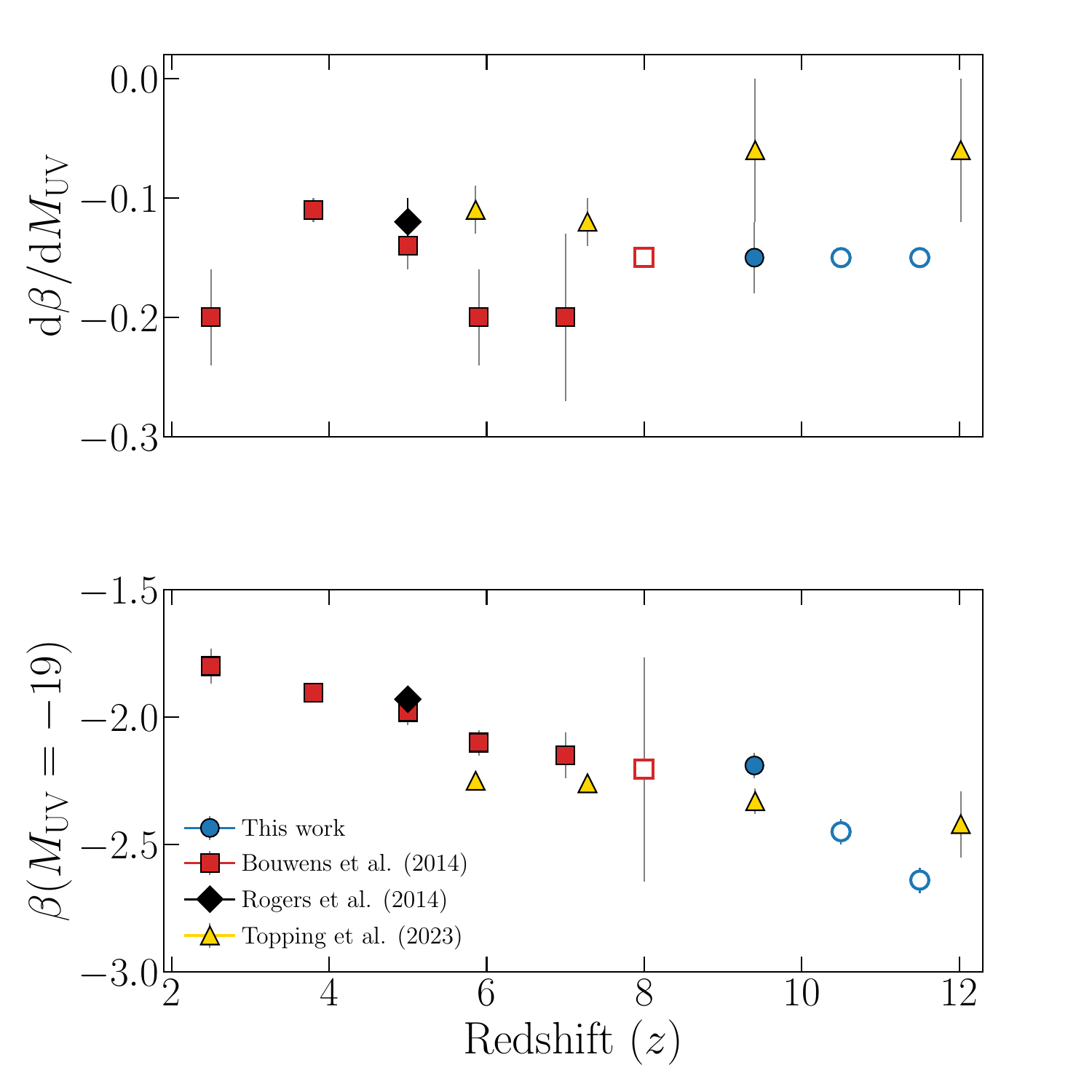}}
        \caption{Constraints on the $\beta-M_{\mathrm{UV}}$ relation as a function of redshift from this work (blue circles) and three literature studies: \citet{bouwens2014} (red squares), \citet{rogers2014} (black diamond) and \citet{topping2023} (yellow triangles).
        The upper panel shows the evolution of the slope of the relation (${\mathrm{d}\beta/\mathrm{d}M_{\mathrm{UV}}}$), which remains relatively constant with redshift, with constraints typically in the range ${-0.2 < \mathrm{d}\beta/\mathrm{d}M_{\mathrm{UV}} < -0.1}$.
        The open (i.e. unfilled) data points in the upper panel highlight cases in which the slope has been fixed when fitting (as is the case for the $z>10$ relations in this work; see text for discussion).
        The lower panel shows the evolution of normalisation, $\beta(M_{\mathrm{UV}}=-19)$,  which clearly evolves such that galaxies at higher redshifts have bluer UV slopes consistent with being typically younger and less dust and metal enriched.
        At $z\simeq12$, the current constraints (our work and \citealp{topping2023}) suggest that the galaxy population (at $20.5 \lesssim M_{\mathrm{UV}} \lesssim -18$) is on average extremely blue, approaching the dust-free limit of $\beta \simeq -2.6$.}
        \label{fig:beta_muv_param_evolution}
    \end{figure}

    \begin{figure*}
        \centerline{\includegraphics[width=\textwidth]{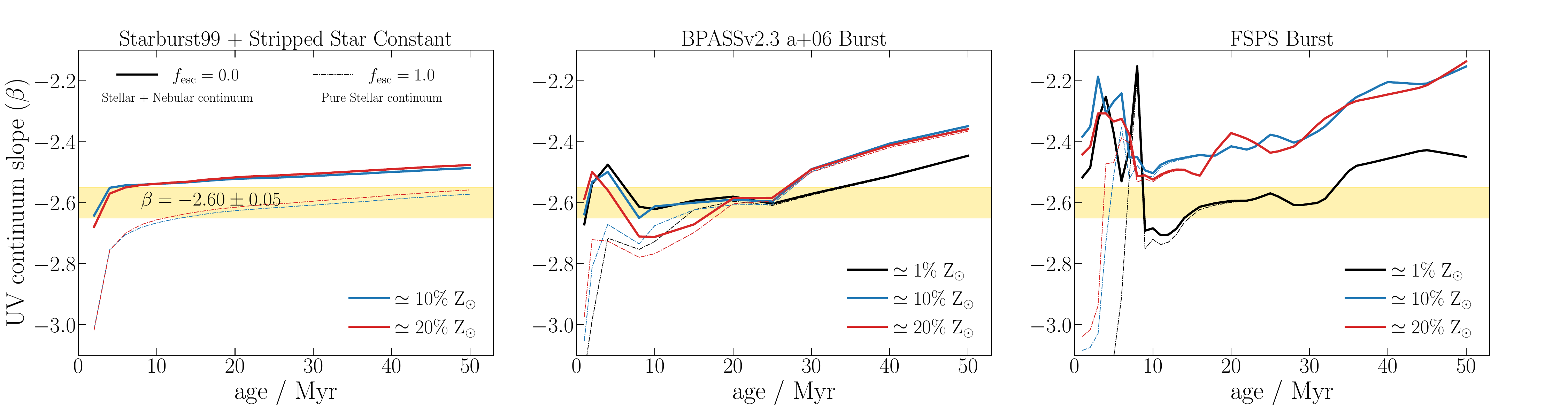}}
        \caption{Plots of $\beta$ versus stellar population age for four different SPS models.
        The stellar populations models considered are \textsc{starburst99} continuous star formation models including a contribution from stripped stars (\citealp{leitherer1999}; \citealp{gotberg2017}; left-hand panel), \textsc{bpassv2.3} single-burst models including $\alpha$-enhanced abundance ratios (\citealp{byrne2022}; centre panel) and Flexible Stellar Population Synthesis (\textsc{fsps}) single-burst models (\citealp{conroy2009, conroy2010, byler2017}; right-hand panel).
        In each panel, the dashed line shows the UV slopes for pure stellar continuum models.
        Each line is colour-coded by stellar metallicity, as indicated in the legend.
        The solid lines show the UV slopes when accounting for nebular continuum emission assuming an escape fraction of $0$ per cent (i.e. the maximum nebular continuum strength).
        The shaded yellow area shows our constraint of $\langle \beta \rangle = 2.60 \pm 0.05$ at $z > 10.5$ from the fit described in Section \ref{subsec:beta_vs_redshift}.
        All models suggest that the population average value we measure at $z > 10.5$ is consistent with pure stellar plus nebular emission in the absence of dust.}
        \label{fig:nebular_modeling}
    \end{figure*}

\subsubsection{Evidence for piecewise linear relation?}

We note that some studies have presented evidence for a change in the slope of the $\beta-M_{\mathrm{UV}}$ relation at faint magnitudes such that $\mathrm{d}\beta/\mathrm{d}M_{\mathrm{UV}}$ becomes shallower at the faint end.
For example, \citet{bouwens2014} presented tentative evidence for a piecewise-linear $\beta-M_{\mathrm{UV}}$ relation $z \simeq 4-6$, with the change in slope occurring at $M_{\mathrm{UV}} \gtrsim -19$.
On the other hand, \cite{rogers2014} find no significant evidence for a nonlinear $\beta-M_{\mathrm{UV}}$ relation $z = 5$.
From Fig. \ref{fig:beta_muv_relations} it can be seen that our data appear qualitatively consistent with a flattening at faint magnitudes.
However, we currently lack the dynamic range in $M_{\mathrm{UV}}$ to reliably confirm this trend.
Ultimately, larger sample sizes and an extension to fainter $M_{\mathrm{UV}}$ are needed to further explore this possibility at $z>9$.
In the following discussion, we compare our measurements with those in the literature that assume a simple linear form of the $\beta-M_{\mathrm{UV}}$ relation.

\subsubsection{Comparison to the literature}

In Fig. \ref{fig:beta_muv_param_evolution} we show a compilation of derived slopes and intercepts for the $\beta-M_{\mathrm{UV}}$ relation in star-forming galaxies from $z=2$ to $z=12$.
The compilation includes the results of this work and three studies from the literature: \citet{bouwens2014}, \citet{rogers2014} and \citet{topping2023}\footnote{This selection of literature sources shown in Fig. \ref{fig:beta_muv_param_evolution} have been restricted for clarity and are therefore not comprehensive. 
However, the general picture remains unchanged if other studies are included \citep[e.g.][]{finkelstein2012, dunlop2013, bahtawdeker2021}}.
At $2 < z < 8$, the majority of the constraints come from the large $\emph{HST}$ study of \citet{bouwens2014}\footnote{We note that the $z=2.5$ constraint reported in \citet{bouwens2014} and shown in Fig. \ref{fig:beta_muv_param_evolution} was first derived in \citet{bouwens2009}.}. 
Applying a consistent analysis across the full redshift range, \citet{bouwens2014} find a gradual decline in the normalisation of the $\beta-M_{\mathrm{UV}}$ relation from $\beta(M_{\mathrm{UV}}=-19) \simeq -1.80$ at $z=2.5$ to $\beta(M_{\mathrm{UV}}=-19) \simeq -2.15$ at $z=7$: an evolution of $\Delta \beta \simeq 0.4$ across $1.8$ Gyr of cosmic time.
Their derived slopes are scattered in the range ${-0.2 < \mathrm{d}\beta/\mathrm{d}M_{\mathrm{UV}} < -0.1}$, which is consistent with the general pattern followed by all data sets in Fig. \ref{fig:beta_muv_param_evolution}.
The inverse-variance weighted mean value of the slope across all redshifts in \citet{bouwens2014} is ${\langle \mathrm{d}\beta/\mathrm{d}M_{\mathrm{UV}} \rangle =-0.12}$.
We also show the wide-area ($\simeq 0.8 \, \mathrm{deg}^2$) analysis at $z=5$ of \citet{rogers2014} which, given the total sample size and area, is the benchmark constraint on $\beta-M_{\mathrm{UV}}$ at this redshift.
\citet{rogers2014} find ${\mathrm{d}\beta/\mathrm{d}M_{\mathrm{UV}}=-0.12\pm0.02}$ and $\beta(M_{\mathrm{UV}}=-19)=-1.93 \pm 0.03$, in excellent agreement with the $z=5$ constraints of \citet{bouwens2014}.
Therefore, we consider the $z=5$ constraints a robust anchor with which to compare the data at higher redshifts.

We also show recent determinations of the $\beta-M_{\mathrm{UV}}$ relation across the redshift range $6 < z <12$ from \citet{topping2023}.
Their analysis is based on deep \emph{JWST}/NIRCam observations (taken as part of the JADES survey) and is therefore directly comparable to ours.
In their two bins at $z > 9$ ($z=9.4$ and $z=12.0$), \citet{topping2023} report an essentially flat slope with ${\mathrm{d}\beta/\mathrm{d}M_{\mathrm{UV}}=-0.06 \pm 0.05}$ (at $z=12.0$ the slope is fixed to the $z=9.4$ value).
At $z=9.4$, this is shallower than our determination of ${\mathrm{d}\beta/\mathrm{d}M_{\mathrm{UV}}=-0.15 \pm 0.03}$.
The difference may be attributable in part to the lack of bright-end constraints in \citet{topping2023}: their brightest bin has an absolute UV magnitude of ${M_{\mathrm{UV}}=-20.0}$, whereas our sample probes to $M_{\mathrm{UV}}=-22.7$ thanks to the inclusion of the COSMOS/UltraVISTA sample (Fig. \ref{fig:muv_z_sample}).
However, in general, the tension is clearly not significant, and the two constraints are consistent within $2 \sigma$.
At $z \simeq 12$ our results agree fairly well within the uncertainties.
Crucially, \citet{topping2023} also finds that by $z=12$ the galaxy population is uniformly extremely blue.

Although there are tensions between the various studies, these differences are not highly significant (i.e. none of current constraints at fixed redshift are incompatible at the $\geq 3 \sigma$ level).
The general picture discernible from Fig. \ref{fig:beta_muv_param_evolution} is of a relatively shallow $\beta-M_{\mathrm{UV}}$ relation that is in place from $z \simeq 10$, with a fixed (or at least slowly evolving) slope, probably in the range ${-0.2 < \mathrm{d}\beta/\mathrm{d}M_{\mathrm{UV}} < -0.1}$.
At $z>10$ more data are needed to robustly constrain this slope, although current data are compatible with a slowly-/non-evolving $\mathrm{d}\beta/\mathrm{d}M_{\mathrm{UV}}$ scenario.
In contrast, the normalisation of the relation is clearly changing, with galaxies at fixed $M_{\mathrm{UV}}$ becoming bluer towards higher redshift.
Based on our results, the evolution is gradual below $z\simeq10$, with $\beta(M_{\mathrm{UV}}=-19)$ evolving from $-2.2$ at $z=10$ to $-1.9$ at $z=5$.
At $z>10$, the evolution becomes more rapid (especially when framed in terms of cosmic time), reaching $\beta(M_{\mathrm{UV}}=-19)\simeq-2.6$ by $z=12$.
The analysis of \citet{topping2023} suggests a more gradual evolution, but still suggests extremely blue UV colours at the earliest cosmic epochs.



\section{Discussion}\label{sec:discussion}

We have presented an investigation of the UV continuum slopes of a sample of $172$ galaxy candidates at $z \geq 7.5$, with the aim of understanding the dependence of $\beta$ on redshift and $M_{\mathrm{UV}}$.
The primary focus of our analysis has been a new wide-area sample of $121$ galaxies at $z \geq 9$ selected from $15$ public \emph{JWST}/NIRCam imaging datasets (covering an on-sky area of $\simeq 320$ arcmin$^2$).
The main result of this work is that we find a strong redshift evolution of $\beta$ in our wide-area sample, with a rapid transition from $\langle \beta \rangle \simeq -2.2$ at $z\simeq9$ to $\langle \beta \rangle \simeq -2.6$ at $z\simeq11$ (Figs. \ref{fig:beta-versus-redshift} and \ref{fig:beta_z_redshift_transition}).
The population average value of $\beta$ at $z>10.5$ is consistent with expectations for extremely dust-poor, even dust-free, stellar populations(see discussion below).
We then investigated the $\beta-M_{\mathrm{UV}}$ relation using our full sample of $172$ galaxies, which included the bright ($M_{\mathrm{UV}} < -21$) ground-based candidates at $\langle z \rangle \simeq 8.5$ from COSMOS/UltraVISTA.
We find evidence ($5 \sigma$) for a $\beta-M_{\mathrm{UV}}$ relation at $z\simeq9$ with a slope of $\mathrm{d}\beta/\mathrm{d}M_{\mathrm{UV}}=-0.15 \pm 0.03$, similar to the slope observed at lower redshifts (Figs. \ref{fig:beta_muv_relations} and \ref{fig:beta_muv_param_evolution}).
At $z>10$, our data are not sufficient to accurately constrain the $\beta-M_{\mathrm{UV}}$ slope due to a lack of bright galaxies; however, our analysis confirms a rapid evolution in the normalisation of the relation across the redshift range $9 < z < 12$.
In this section, we discuss the dust-free galaxy interpretation and consider the implications for dust formation in the early Universe and cosmic reionisation.

\subsection{A transition to dust-free star-formation at $\mathbf{z \gtrsim 10}$?}\label{subsec:dust_free_discussion}

The bluest/steepest possible value of $\beta$ is set by the intrinsic stellar spectrum of young metal-poor galaxies.
At the youngest ages ($\lesssim 1$ Myr) and the lowest metallicities ($\lesssim 1$ per cent solar), the UV slope can reach values as blue at $\beta \simeq -3$ \citep[e.g.][]{bouwens2010, robertson2010, stanway2016, topping2022}.
However, the presence of ionised gas in galaxies will result in continuum emission as a consequence of free-free, free-bound, and two-photon processes (i.e. the nebular continuum).
This nebular continuum acts to redden the slope of the UV continuum \citep[e.g.][]{byler2017}.
Models suggest that the bluest expected UV slopes when accounting for nebular continuum emission are $\beta \simeq -2.6$ \citep[e.g.][]{stanway2016,topping2022}.

We reproduce this result for a selection of stellar population synthesis models in Fig. \ref{fig:nebular_modeling} where we plot $\beta$ as a function of the stellar population age.
The stellar populations considered are \textsc{starburst99} continuous star formation models including a contribution from stripped stars \citep{leitherer1999,gotberg2017}, \textsc{bpassv2.3} single-burst models including $\alpha$-enhanced abundance ratios \citep{byrne2022} and Flexible Stellar Population Synthesis (\textsc{fsps}) single-burst models \citep{conroy2009, conroy2010, byler2017}.
Full details of the nebular continuum modelling are given in Appendix \ref{app:cloudy_nebcont}.
It can be seen that although variations exist between each model, the overall picture is consistent: when an escape fraction ($f_{\mathrm{esc}}$) of 0 per cent is assumed (i.e. all of the ionising photons emitted by the stellar population are converted into nebular emission), the bluest expected UV continuum slope is $\beta \simeq -2.6$.
Moreover, this occurs only for the youngest ages $\lesssim 30$ Myr.
If a nonzero $f_{\mathrm{esc}}$ is assumed, then bluer slopes are possible, with a minimum value of $\beta \simeq -3$, but this occurs at ages of ${\lesssim 3 \, \mathrm{Myr}}$ and for escape fractions close to 100 per cent (i.e. pure stellar emission).

    \begin{figure}
        \centerline{\includegraphics[width=\columnwidth]{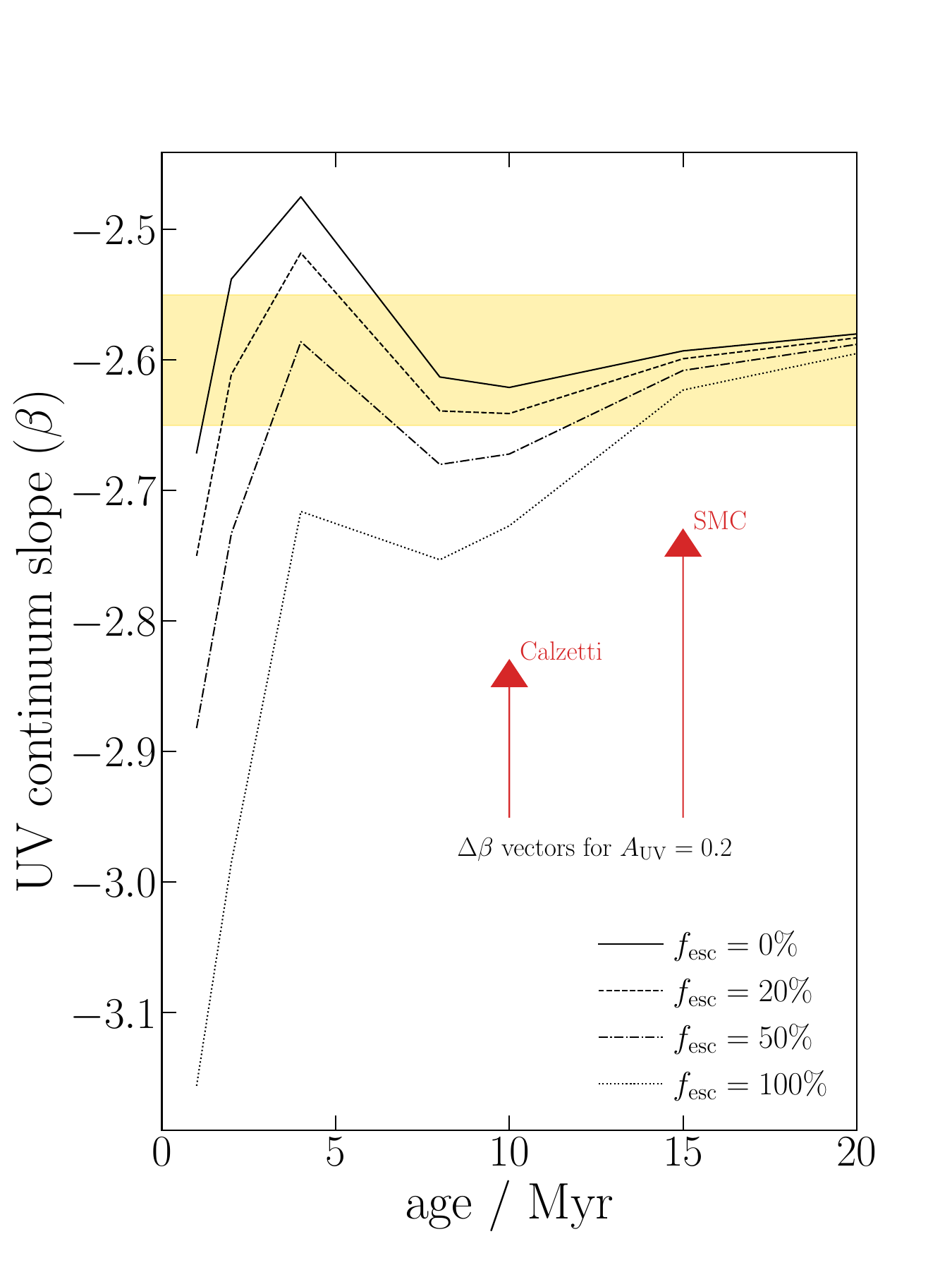}}
        \caption{The intrinsic UV continuum slope as a function of stellar population age for a \textsc{bpassv2.3} burst model \citep{byrne2022} with $1$ percent solar metallicity. 
        Each black line shows the relation for a different assumed value of $f_{\mathrm{esc}}$ as indicated in the legend.
        The shaded yellow area shows our constraint of $\langle \beta \rangle = 2.60 \pm 0.05$ at $z > 10.5$ from the model fit described in Section \ref{subsec:beta_vs_redshift}.
        The red arrows show the dust vectors (i.e. shift in $\beta$) assuming $0.2$ magnitudes of UV attenuation for the \citet{calzetti2000} attenuation curve and the SMC extinction curve \citep{gordon2003}.
        These models indicate that to achieve an observed UV continuum slope of $\beta_{\mathrm{obs}} = -2.6$ including dust would require a counterintuitive scenario in which a young burst, with a large ($\gtrsim 50$ per cent) escape fraction is at the same time experiencing $\gtrsim 0.2$ magnitudes of UV attenuation (see text for discussion). 
        }
        \label{fig:nebular_models_zoomin}
    \end{figure}
    
Another notable feature of Fig. \ref{fig:nebular_modeling} is that the youngest, most metal-poor stellar populations produce the strongest nebular continuum emission due to their harder ionising spectra. 
In some cases, the strong nebular emission can entirely compensate for their intrinsically bluer stellar UV slopes.
For example, considering the FSPS models (right-hand panel of Fig. \ref{fig:nebular_modeling}), the UV continuum slopes, including nebular continuum emission, are actually \emph{redder at the youngest ages}.
From this we can conclude that invoking more extreme stellar populations with harder ionising spectra (i.e. top-heavy IMFs, Pop-III stars) is unlikely to result in UV slopes intrinsically bluer than $\beta = -2.6$ when the effects of nebular continuum emission are included \citep[see e.g.][]{cameron2023}.

The model predictions in Fig. \ref{fig:nebular_modeling} imply that a stellar population with an observed UV continuum slope of $\beta_{\mathrm{obs}} \simeq -2.6$ is almost certainly experiencing negligible (essentially zero) dust attenuation, regardless of age.
For a stellar population to be dust-attenuated yet still have $\beta_{\mathrm{obs}} = -2.6$ would require a counter-intuitive scenario in which a significant fraction of ionising continuum photons are escaping into the IGM in the presence of dust.
We highlight this argument in Fig. \ref{fig:nebular_models_zoomin}, where we show $\beta$ versus stellar population age for a \textsc{bpassv2.3} burst model across a range of $f_{\mathrm{esc}}$ values. 
Even for extreme escape fractions of $f_{\mathrm{esc}} = 50$ per cent, the intrinsic UV continuum slope only falls below $\beta \simeq -2.6$ at the very youngest ages (i.e. $< 5$ Myr).
To achieve $\beta = -2.6$ including dust would therefore require a very young, dominant, burst with a large ($\gtrsim 50$ per cent) escape fraction that at the same time experiences $\gtrsim 0.2$ magnitudes of UV attenuation. 
Although we cannot completely rule such a scenario out (i.e. for very specific star/gas/dust geometries), the degree of fine-tuning required renders it an unlikely explanation for population-average values.

Indeed, it is arguably more likely that the stellar population ages at $z>10.5$ are $\gtrsim 30$ Myr, such that, on average, these galaxies will have intrinsic UV continuum slopes of $\beta_{\mathrm{int}} > -2.6$ (Fig. \ref{fig:nebular_modeling}).
For example, \citet{cullen2017} analysed the stellar populations of simulated galaxies at $z=5$ and found that the typical intrinsic UV slopes assuming $f_{\mathrm{esc}}=0$ were $\beta \simeq -2.4$ for galaxies with light-weighted stellar ages between $20$ Myr and $\simeq 100$ Myr.
Light-weighted ages of up to $100$ Myr are clearly plausible at ${z=11}$ as this would correspond to a formation redshift of ${z=13}$ and galaxies at this redshift have already been spectroscopically confirmed by \emph{JWST} \citep{curtis-lake2023}.
Therefore, it is possible and even likely that the typical intrinsic slopes of the $z>10.5$ galaxies in our sample are redder than $\beta = -2.6$ (for $f_{\mathrm{esc}}=0$).
In this case, an observed UV continuum slope of ${\beta_{\mathrm{obs}} = -2.6}$ indicates both negligible dust attenuation \emph{and} and a nonzero escape fraction of ionising photons.
We discuss implications for the escape fraction and reionisation in more detail below.

Finally, it is worth acknowledging that our $\langle \beta \rangle$ constraints are formally consistent with a value as red as $\langle \beta \rangle = -2.35$ within the uncertainties ($5\sigma$).
Assuming an intrinsic slope of $\beta_{\mathrm{int}}=-2.6$ (i.e. an extremely young burst), this value would correspond to an upper limit of ${A_{\mathrm{UV}} \lesssim 0.2-0.5}$ depending on the assumed attenuation curve \citep{mclure2018}.
However, for older and/or composite stellar populations (with $\beta_{\mathrm{int}}\simeq-2.4$), the implied upper limit on UV attenuation would be closer to ${A_{\mathrm{UV}} \lesssim 0.1 - 0.2}$.
Nevertheless, our formal best estimates are most consistent with negligible, essentially zero, dust attenuation.
Below we consider some physical interpretations of dust-free systems in the young Universe. 

\subsubsection{The physical motivation for dust-free galaxies}

The above comparison with theoretical UV spectra models suggests that the average $\beta$ value we observe at $z \simeq 11$ is consistent with the expectation for dust-free stellar populations.
It is important to ask whether such a scenario is physically plausible.
Dust is formed in the ejecta of core-collapse supernovae \citep[e.g.][]{todini2010}, and as a result galaxies might be expected to accumulate dust immediately after the onset of star formation.
In this case, a mechanism for either ejecting or destroying dust formed in supernovae is required to explain the absence of dust in star-forming galaxies.
\citet{ferrara2023} and \citet{ziparo2023} propose one potential scenario in which intense UV radiation pressure from ongoing star formation in these early galaxies drives dust out of the interstellar medium (see also \citealp{nath2023}; \citealp{tsuna2023}).
Most recently, \citet{ferrara2023b} have postulated that above $z\simeq10$ the specific star-formation rate of galaxies crosses a threshold above which powerful radiation-driven outflows are capable of clearing dust from the ISM.
Intriguingly, this is the same redshift threshold above which we begin to see uniformly extremely blue colours in our sample.

An alternative hypothesis, proposed by \citet{jaacks2018}, is that the first generation of star formation (i.e. Pop-III stars) produces negligible amounts of dust.
In this scenario the stars at $z>10.5$ are either forming from recently enriched but dust-free Pop-III gas, or alternatively we are seeing the effect of a significant fraction of current Pop-III star formation in our $z>10.5$ sample.
Indeed, \citet{jaacks2018} estimate ${\beta = -2.5 \pm 0.07}$ for Pop-III SEDs.
We note, however, that, other than their blue UV continuum slopes, the NIRSpec/PRISM spectra shown in Fig. \ref{fig:uvslope_fit_examples} do not show any obvious Pop-III signatures in their FUV spectra (e.g. strong broad \heii \ emission).
Finally, it is also possible that dust is more efficiently destroyed in the earliest star-forming systems.
Current estimates of dust grain destruction rates in the ISM vary by up to an order of magnitude due to different assumptions regarding dust mircrophysics (e.g. grain-grain collisions and shattering; \citealp[e.g.][]{kirchschlager2022}).
It is possible that specific physical conditions at $z > 10$ could result in enhanced dust destruction (e.g. enhanced surface densities of SFR), although direct evidence is currently lacking.
The dust destruction scenario is supported by the theoretical analysis of \citet{esmerian2023}, who used a cosmological fluid-dynamical simulation of galaxies within the first 1.2 Gyr ($z>5$) to investigate the effect of different dust models.
A comparison of our UV slope measurements up to $z\simeq12$ with their model favours enhanced dust destruction with destruction rates elevated by an order of magnitude or more relative to default assumptions.
There are therefore a number of possible physical processes which can explain dust-free galaxies.
However, our current data cannot distinguish between them.
Future observations, in particular deep NIRSpec/PRISM spectroscopy and improved constraints on the redshift evolution of $\langle \beta \rangle$, will be crucial in this regard.

\subsubsection{Direct constraints on dust emission at $z>10$}

    \begin{figure}
        \centerline{\includegraphics[width=\columnwidth]{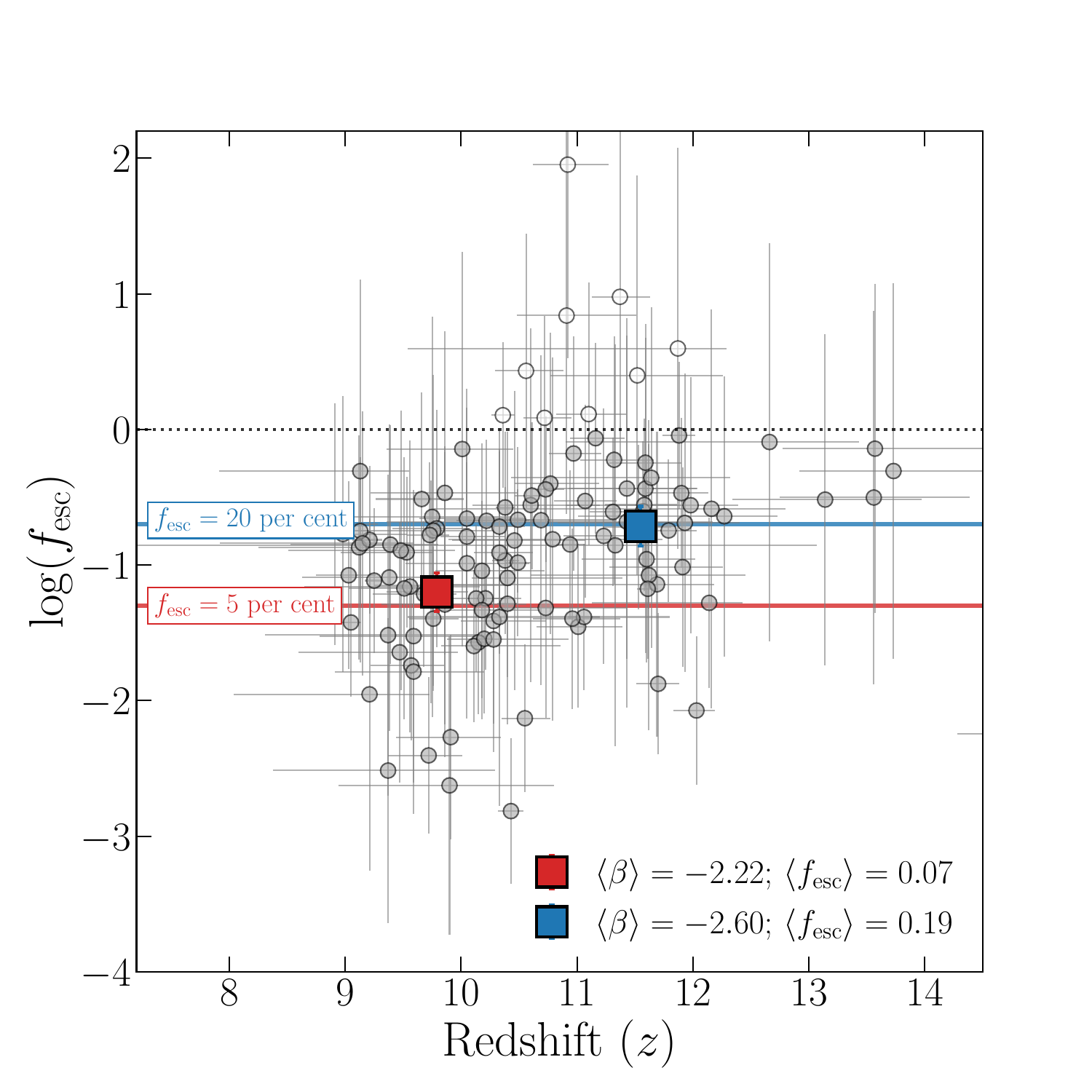}}
        \caption{The escape fraction of ionising photons ($f_{\mathrm{esc}}$) as a function of observed UV continuum slope ($\beta$) for our primary wide-area \emph{JWST} sample.
        The values of $f_{\mathrm{esc}}$ for individual galaxies (grey circles) were calculated using the $\beta-f_{\mathrm{esc}}$ relation at $z=0.3$ derived by \citet{chisholm2022} using (primarily) data from the LzLCS survey \citep{flury2022}.
        Open circular data points show galaxies with unphysical values of $f_{\mathrm{esc}} > 1$ based on this calibration (i.e. the galaxies in our sample that have been scattered to extremely blue $\beta$ values).
        The square data points in red and blue show the population average value of $f_{\mathrm{esc}}$ above and below $z=10.5$, respectively, using the best-fitting $\beta$ values from Fig. \ref{fig:beta_z_redshift_transition}.
        The red and blue horizontal lines show $f_{\mathrm{esc}}=0.05$ and $f_{\mathrm{esc}}=0.20$, respectively.
        Taking the predictions of the low-redshift calibration at face value implies that a substantial fraction of ionising photons ($\simeq 20$ per cent) are escaping into the IGM at $z > 10.5$.}
        \label{fig:fesc_vs_redshift}
    \end{figure}

Direct detection of dust emission in the rest-frame far-infrared at ${z>10}$ would represent a clear refutation of a dust-free star formation scenario.
However, it is worth noting that to date, all deep ALMA observations aimed at detecting dust continuum emission at these redshifts have been unsuccessful \citep[e.g.][]{fujimoto2022, bakx2023,popping2023, yoon2023, fudamoto23}.
Therefore, currently these observations are fully consistent with the results presented here.
Nevertheless, it is clear that further deep ALMA observations (enabling, for example, stacking of multiple targets) will be crucial for robustly testing the inferences that are made from rest-frame UV studies.
    
\subsection{The implications for cosmic reionisation}\label{subsec:fesc_reion}

    \begin{figure}
        \centerline{\includegraphics[width=\columnwidth]{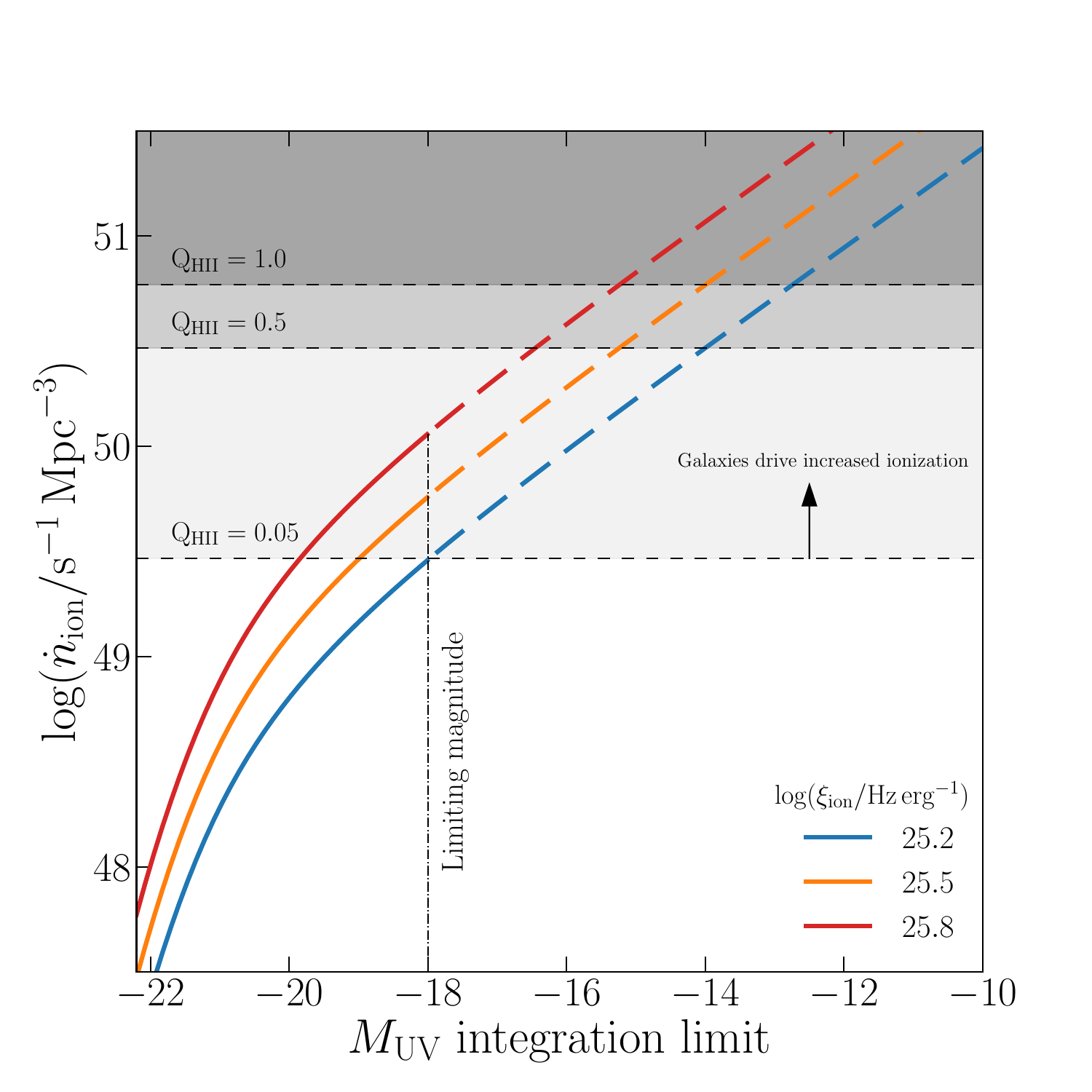}}
        \caption{The emission rate of ionising photons into the IGM per comoving Mpc$^3$ ($\dot{n}_{\mathrm{ion}} / s^{-1}$) for different assumed values of the ionising production efficiency ($\xi_{\mathrm{ion}}/\mathrm{Hz} \, \mathrm{erg}^{-1}$) and different integration limits of the UV LF.
        All calculations assume $f_{\mathrm{esc}}=0.2$, which we infer using the relation of \citet{chisholm2022} based on our measured value of $\langle \beta \rangle \simeq -2.6$ (see text and Fig. \ref{fig:fesc_vs_redshift}) and the $z=11$ UV LF of \citet{mcleod2024}.
        The dashed horizontal lines show the equilibrium value of $\dot{n}_{\mathrm{ion}}$ needed to maintain an ionised IGM fraction of $Q_{\hii}$ (as indicated by the labels).
        The red curve shows $\dot{n}_{\mathrm{ion}}$ as a function of $M_{\mathrm{UV, lim}}$ assuming ${\mathrm{log}(\xi_{\mathrm{ion}}/\mathrm{Hz}\,\mathrm{erg}^{-1}})=25.8$ (i.e. at the upper end of ${\xi_{\mathrm{ion}}}$ estimates for young metal-poor galaxies; \citealp[e.g.][]{tang2019}).
        The curve turns from solid to dashed below the limiting magnitude of current LF constraints ($M_{\mathrm{UV}} \simeq -18$).
        At this efficiency, the already observed population of galaxies (i.e. $M_{\mathrm{UV}} \lesssim -18$; vertical black line) would be supplying enough ionising photons to maintain ${Q_{\hii} > 0.1}$.
        The orange and blue curves represent ${\mathrm{log}(\xi_{\mathrm{ion}}/\mathrm{Hz}\,\mathrm{erg}^{-1}})=25.5$ and ${\mathrm{log}(\xi_{\mathrm{ion}}/\mathrm{Hz}\,\mathrm{erg}^{-1}})=25.2$, which may be more representative of the typical value during reionisation \citep[e.g.][]{bouwens2016, matthee2017}.
        At the lowest estimated ionising efficiencies, the observed population would be sufficient to maintain $Q_{\hii} \simeq 0.05$.
        Taking into account the faint, unseen, galaxy population at $M_{\mathrm{UV}} > -18$ implies relatively high (likely $\gtrsim 5$ per cent) ionised fractions at $z=11$.
        }
        \label{fig:ndot_ion_vs_muvlim}
    \end{figure}

As discussed above, while a UV continuum slope of $\beta \simeq -2.6$ clearly indicates negligible dust attenuation, it also implies that the ionising photon escape fraction ($f_{\mathrm{esc}}$) is likely to be high.
In fact, a direct connection between observed $\beta$ and $f_{\mathrm{esc}}$ has been established up to $z\simeq 3$ \citep[e.g.][]{chisholm2022, begley2022, keunho2023}.
It is likely that this connection extends to higher redshifts.
At $z \simeq 7.5$, \citet{topping2023} find that ultra-blue UV continuum slopes are associated with a lack of nebular emission line signatures in the rest-frame optical, again suggesting that bluer UV colours are associated with higher $f_{\mathrm{esc}}$.
Detailed cosmological radiative transfer simulations also predict that $\beta$ is a key indicator of $f_{\mathrm{esc}}$ at $z>4$ \citep{choustikov2023}.

Using a sample $89$ galaxies at $z\simeq0.3$ with direct LyC measurements from \emph{HST}/COS spectroscopy, \citet{chisholm2022} presented a strong ($5.7 \sigma$) correlation between $\beta$ and $f_{\mathrm{esc}}$.
The majority of the sample ($66$ galaxies) was drawn from the Low-redshift Lyman Continuum Survey \citep[LzLCS;][]{flury2022} which was designed to investigate LyC escape in analogues of young high redshift galaxies.
Interestingly, the bluest objects in this sample have $\beta \simeq -2.5$.
The result of applying the \citet{chisholm2022} $\beta-f_\mathrm{esc}$ relation (their equation 18) to our data is shown in Fig. \ref{fig:fesc_vs_redshift}.
While the scatter for individual objects is substantial, the population average values yield robust constraints.
Using the fitted values of $\langle \beta \rangle$ from Fig. \ref{fig:beta_z_redshift_transition}, the \citet{chisholm2022} relation yields average values of $\langle f_{\mathrm{esc}} \rangle = 0.07$ at $z<10.5$ to $\langle f_{\mathrm{esc}} \rangle = 0.19$ at $z>10.5$.

Interestingly, if $\langle f_{\mathrm{esc}} \rangle \simeq 0.2$, then the galaxy population that has already been observed with \emph{JWST} is likely able to provide enough ionising photons to drive reionisation at $z=11$.
In Fig. \ref{fig:ndot_ion_vs_muvlim} we demonstrate this by showing the emission rate of ionising photons into the IGM per comoving Mpc$^3$ ($\dot{n}_{\mathrm{ion}}$) as a function of the integration limit of the UV luminosity function.
The calculations assume $f_{\mathrm{esc}}=0.2$ and the $z=11$ UV LF of \citet{mcleod2024}.
We calculate $\dot{n}_{\mathrm{ion}}$ for three values of $\xi_{\mathrm{ion}}$ (i.e. the production efficiency of LyC photons in units of $\mathrm{Hz}/\mathrm{erg}$) that bracket the typical range of values inferred for young, low-metallicity galaxies at high redshifts \citep[i.e. ${\mathrm{log}(\xi_{\mathrm{ion}}/\mathrm{Hz}\,\mathrm{erg}^{-1}})=25.2-25.8$; e.g.][]{bouwens2016, atek2023, tang2019, tang2023}.
The ionising photon output needed to maintain an ionised IGM fraction of $Q_{\hii}$ is given by the prescription of \citet{madau1999}:
\begin{equation}\label{eq:ndot_ion_equation}
    \dot{n}_{\mathrm{ion}} \, (\mathrm{s}^{-1}\mathrm{Mpc}^{-3}) = 10^{47.4} Q_{\hii} \, C_{\mathrm{\hii}}(1+z)^3,
\end{equation}
where $C_{\mathrm{\hii}}$ is the IGM clumping factor which we set following the prescription of \citet{shull2012}.

Setting $Q_{\hii}=1$ in equation \ref{eq:ndot_ion_equation} represents the limiting case of maintaining a fully ionised Universe at redshift $z$.
However, at $z=11$ the Universe is expected to be only partially ionised.
Models assuming an early and slow reionisation typically have $Q_{\hii} \simeq 0.2$ \citep[e.g.][]{finkelstein2019} while late and rapid models have ${Q_{\hii} \simeq 0.01}$ \citep[e.g.][]{naidu2020}.
It can be seen from Fig. \ref{fig:ndot_ion_vs_muvlim} that, based on the above assumptions, the currently observed galaxy population (i.e. $M_{\mathrm{UV}} \lesssim -18$) is likely providing enough ionising photons to maintain ${Q_{\hii} \gtrsim 0.05}$.
Accounting for the population of fainter galaxies, examples of which have been uncovered down to $M_{\mathrm{UV}} \simeq -15$ \citep[e.g.][]{atek2023}, implies that the full galaxy population could in principle be supplying a sufficient number of photons to maintain an ionisation fraction of $\gtrsim 20$ per cent at $z=11$ (the exact value is strongly dependent on the uncertain faint-end slope of the UV LF).
Despite systematic uncertainties, it seems probable that the large abundance of blue galaxies at $z>10$ unveiled by early \emph{JWST} studies disfavours low ionised IGM fractions of $\lesssim 1$ per cent.

\section{Summary and Conclusions}\label{sec:conclusions}

We have measured the rest-frame ultraviolet (UV) continuum slopes ($\beta$) for a sample of $172$ galaxy candidates at $\langle z \rangle = 10.5$.
The majority of these candidates ($\simeq 70$ per cent) are drawn from our new wide-area \emph{JWST} sample with absolute UV magnitudes in the range $-20.5 < M_{\mathrm{UV}} < -17$ (Section \ref{subsec:widea-area-catalogue}).
This new sample has allowed us to improve on our previous work \citep{cullen2023} in terms of sample size, on-sky area, and median imaging depth.
Using a robust power-law fitting technique validated against available spectroscopy (Section \ref{subsec:measuring_uv_slope}), we report precise estimates of $\langle \beta \rangle$ at $z > 10$.
The main aim of this analysis is to use the \emph{average} values of $\beta$ for these galaxies to place constraints on the typical dust obscuration experienced at the earliest cosmic epochs.
Our main results can be summarised as follows:

\begin{enumerate}

    \item We measure an inverse-variance weighted mean value of ${\langle \beta \rangle = -2.37 \pm 0.03}$ for our primary wide-area sample at ${\langle z \rangle = 10.6}$.
    Incorporating candidates from \citet{cullen2023} $-$ that include UV-bright ground-based COSMOS/UltraVISTA sources at lower redshift (with $\langle z \rangle \simeq 8$; Fig. \ref{fig:muv_z_sample}) $-$ yields a slightly redder value of ${\langle \beta \rangle = -2.32 \pm 0.03}$ and a lower average redshift of $\langle z \rangle = 10.3$.
    Uniformly, therefore, our sample displays blue UV continuum slopes indicative of dust-poor galaxies at these redshifts.
    However, a value of $\langle \beta \rangle \simeq -2.37$ is not inconsistent with moderate amounts of dust attenuation, nor more extreme than blue objects observed in the local Universe (e.g. NGC 1705; $\beta = -2.46$).

    \vspace{0.1cm}

    \item Although the median value of the UV slopes of the wide-area sample is $\langle \beta \rangle \simeq -2.37$, we find evidence for a steep $\beta-z$ trend between $z=9$ and $z=12$ (Fig. \ref{fig:beta-versus-redshift}).
    The average value of $\beta$ evolves from ${\langle \beta \rangle = -2.17 \pm 0.06}$ at ${z \simeq 9.5}$ to ${\langle \beta \rangle = -2.59 \pm 0.06}$ at ${z \simeq 11.5}$.
    This represents a rapid evolution of $\Delta \langle \beta \rangle \simeq 0.4$ over $\simeq 100$ Myr of cosmic time.
    A slope of $\beta = -2.6$ represents the extreme end of the local distribution \citep[e.g.][]{chisholm2022} but appears to be typical for galaxies at the earliest cosmic epochs.
    Such steep UV slopes are consistent with pure stellar plus nebular continuum emission that is not attenuated by dust (Fig. \ref{fig:nebular_modeling}).
    
    \vspace{0.1cm}

    \item Fitting a step function model to the $\beta-z$ data we find that the galaxy population is becoming uniformly extremely blue at $z \gtrsim 10.5$ (Fig. \ref{fig:beta_z_redshift_transition}).
    Below this transition redshift, our sample displays ${\langle \beta \rangle = -2.22 \pm 0.05}$ while at $z \gtrsim 10.5$ the typical value is ${\langle \beta \rangle = -2.60 \pm 0.05}$.
    We have verified that this difference is not caused by measurement biases or selection effects.

    \vspace{0.1cm}
    
    \item In the redshift range $8 < z < 10$ we use the large dynamic range in $M_{\mathrm{UV}}$ enabled by the inclusion of COSMOS/UltraVISTA candidates to investigate the relationship between $\beta$ and absolute UV magnitude ($M_{\mathrm{UV}}$). 
    We find evidence for a $\beta-M_{\mathrm{UV}}$ relation such that brighter galaxies display redder UV slopes (Fig. \ref{fig:beta_muv_relations}).
    This is consistent with a picture in which the brighter (and more massive) galaxies at $z\simeq9$ are more obscured by dust.
    Fitting the $\beta-M_{\mathrm{UV}}$ at these redshifts, we find ${\mathrm{d}\beta/\mathrm{d}M_{\mathrm{UV}} = -0.15 \pm 0.03}$.
    This slope is consistent with values derived a lower redshifts (down to $z=2$; e.g. \citealp{bouwens2014}).
    Therefore, our results suggest that a $\beta-M_{\mathrm{UV}}$ relation has been in place since $z \simeq 10$ with a slope that does not strongly evolve with redshift.
    However, the normalisation of the relation is clearly evolving, with $\beta(M_{\mathrm{UV}}=-19) \simeq -1.8$ at $z \simeq 5$ compared to $\beta(M_{\mathrm{UV}}=-19) \simeq -2.2$ at $z \simeq 9$.
    At fixed $M_{\mathrm{UV}}$, galaxies are less dust-enriched (and metal-enriched) than at earlier cosmic epochs.

    \vspace{0.1cm}
    
    \item At $z > 10$ our data do not cover a sufficient dynamic range in $M_{\mathrm{UV}}$ to robustly constrain the ${\beta-M_{\mathrm{UV}}}$ relation (Fig. \ref{fig:beta_muv_relations}).
    However, the data at these redshifts remain consistent with a relatively shallow $\beta-M_{\mathrm{UV}}$ slope and a continued evolution to bluer values of $\beta(M_{\mathrm{UV}}=-19)$.
    Our results suggest that the normalisation evolves from $\beta(M_{\mathrm{UV}}=-19)=-2.2$ at $z \simeq 9$ to $\beta(M_{\mathrm{UV}}=-19)=-2.6$ at $z \simeq 11$.
    At $z>10$, wider-area surveys are required to probe the bright end of the UV LF and determine whether the most luminous galaxies ($M_{\mathrm{UV}} \lesssim -20$) show evidence for dust-attenuated UV SEDs.
    
\end{enumerate}

The primary new result of this work is the appearance of a uniformly extremely dust-poor, perhaps even dust-free, galaxy population at $z > 10.5$, with \emph{average} UV continuum slopes of ${\langle \beta \rangle \simeq -2.6}$.
Similarly blue UV slopes at these redshifts have also been reported in other independent analyses \citep{austin2023, topping2022, morales2023}.
Competing theoretical explanations for dust-free galaxies exist, but these different physical processes cannot be discriminated with our current data.
Nevertheless, these results place important constraints on the origin of dust and metals in the first galaxies.
Furthermore, our results imply that the ISM conditions in galaxies at $z>10$ favour a significant escape fraction of ionising photons, and that the already observed population of galaxies at these redshifts (i.e. with $M_{\mathrm{UV}} \lesssim -18$) is likely capable of maintaining ionised IGM fractions of $\gtrsim 5$ per cent. 

\section*{Acknowledgements}

F. Cullen, K. Z. Arellano-Cordova and T. M. Stanton acknowledge support from a UKRI Frontier Research Guarantee Grant (PI Cullen; grant reference EP/X021025/1).
R.\,J. McLure, D.\,J. McLeod, J.\,S.~Dunlop, C. Donnan, R. Begley and M.\,L. Hamadouche, acknowledge the support of the Science and Technology Facilities Council. 
A.\,C. Carnall thanks the Leverhulme Trust for their support via a Leverhulme Early Career Fellowship. 
R.\,A.\,A. Bowler acknowledges support from an STFC Ernest Rutherford Fellowship (grant number ST/T003596/1).
J.\,S.~Dunlop also acknowledges the support of the Royal Society via a Royal Society Research Professorship. S. R. Flury acknowledges support from NASA/FINESST (grant number 80NSSC23K1433).

Some of the data products presented herein were retrieved from the Dawn \emph{JWST} Archive (DJA). DJA is an initiative of the Cosmic Dawn Center, which is funded by the Danish National Research Foundation under grant No. 140.
This work is based on observations collected at the European Southern Observatory under ESO programme ID 179.A-2005 and 198.A-2003 and on data products produced by CALET and the Cambridge Astronomy Survey Unit on behalf of the UltraVISTA consortium.

This work is based in part on observations made with the NASA/ESA/CSA James Webb Space Telescope. 
The data were obtained from the Mikulski Archive for Space Telescopes at the Space Telescope Science Institute, which is operated by the Association of Universities for Research in Astronomy, Inc., under NASA contract NAS 5-03127 for JWST. 
These observations are associated with programs 1063, 1324, 1345, 1433, 1963, 2079, 2282, 2561, 2727, 2732, 2736, 2738, 2756 and 2767. The authors acknowledge the associated teams for developing their observing programs with a zero-exclusive-access period.
This work also utilises data from the JADES DR1 data release \citep[DOI: 10.17909/8tdj-8n28;][]{Rieke2023, eisenstein2023}.

This work is based in part on observations obtained with the NASA/ESA Hubble Space Telescope, retrieved from the Mikulski Archive for Space Telescopes at the Space Telescope Science Institute (STScI). STScI is operated by the Association of Universities for Research in Astronomy, Inc. under NASA contract NAS 5-26555.

This research used the facilities of the Canadian Astronomy Data Centre operated by the National Research Council of Canada with the support of the Canadian Space Agency.
 
This work is based on observations taken by the RELICS Treasury Program (GO 14096) with the NASA/ESA HST, which is operated by the Association of Universities for Research in Astronomy, Inc., under NASA contract NAS5-26555.

This research has made use of the SVO Filter Profile Service (http://svo2.cab.inta-csic.es/theory/fps/) supported from the Spanish MINECO through grant AYA2017-84089 \citep{rodrigo2012_svo, rodrigo2020_svo}

For the purpose of open access, the author has applied a Creative Commons Attribution (CC BY) licence to any Author Accepted Manuscript version arising from this submission.
We thank Andrea Ferrara, Gerg\"{o} Popping and Ken Duncan for useful suggestions and feedback.

\vspace*{-0.15in}

\section*{Data Availability}

All \emph{JWST} and HST data products are available via the Mikulski Archive for Space Telescopes (\url{https://mast.stsci.edu}). 
UltraVISTA DR5 data are available through the ESO portal (\url{http://archive.eso.org/scienceportal/home?data_collection=UltraVISTA&publ_date=2023-05-03}).
Additional data products are available from the authors upon reasonable request.



\bibliographystyle{mnras}
\bibliography{jwst_uvslopes} 



\appendix

\section{Overview of the wide-area \emph{JWST} sample}\label{appendix:datasets}

\setlength{\tabcolsep}{3.5pt}
    \begin{table*}
    \centering
    \caption{Overview of the different datasets utilised in this study. 
    The first three columns list the name of the field/survey, the \emph{JWST} proposal ID and the name of the principal investigator. 
    Columns four and five list references to the survey paper and ancillary \textit{HST} data (where applicable).
    Column six lists references for the lensing maps used to correct the photometry for the effect of gravitational lensing (where applicable), with the following key: $\mathrm{Z15}=\ $\citet{Zitrin2015}; $\mathrm{C19}=\ $\citet{Caminha2019}; $\mathrm{F23}=\ $\citet{furtak2023}; $\mathrm{P22}=\ $\citet{Pascale2022}; $\mathrm{O10}=\ $\citet{Oguri2010}; $\mathrm{K11}=\ $\citet{Kneib2011}.}
    \begin{tabular}{llllll}
    \hline
    Field & ID & PI & Reference & Ancillary Data & Lensing Map\\
    \hline
    CEERS & 1345 & S. Finkelstein & \citet{finkelstein2022} & \citet{Koekemoer2011},\citet{wang2020} & -\\
    Quintet & 2732 & K. M. Pontoppidan & \citet{Pontoppidan2022} & - & -\\
    Cartwheel & 2727 & K. M. Pontoppidan & \citet{Pontoppidan2022} & - & -\\
    SMACS 0723 & 2736 & K. M. Pontoppidan & \citet{Pontoppidan2022} & \citet{Coe2019} & P22\\
    J1235 & 1063 & B. Sunnquist & - & - & -\\
    GLASS & 1324 & T. Treu & \citet{treu2022} & - & F23\\
    A2744-DDT & 2756 & W. Chen & - & - & F23\\
    UNCOVER & 2561 & I. Labbe & \citet{Bezanson2022} & \citet{Lotz2017} & F23\\
    RXJ 2129 & 2767 & P. Kelly & - & \citet{Postman2012} & Z15, C19\\
    WHL 0137 & 2282 & D. Coe & - & \citet{Coe2019} & O10, K11, Z15\\
    MACS 0647 & 1433 & D. Coe & - & \citet{Postman2012} & Z15\\
    NEP TDF & 2738 & R. Windhorst & \citet{Windhorst2023} & - & -\\
    JEMS & 1963 & C. Williams & \citet{Williams2023udf} & \cite{Illingworth2016}, \cite{Whitaker2019} & -\\
    NGDEEP & 2079 & S. Finkelstein, C. Papovich, N. Pirzkal & \citet{Bagley2023} & \cite{Illingworth2016}, \cite{Whitaker2019} & -\\
    JADES & 1180, 1210 & D. Eisenstein, N. Luetzgendorf & \citet{eisenstein2023} & \cite{Illingworth2016}, \cite{Whitaker2019} & -\\
    \hline
    \end{tabular}
\label{tab: surveys}
\end{table*}

\begin{table*}
    \centering
        \caption{The effective area (accounting for cluster subtraction and de-lensing where applicable) and $5\sigma$ global limiting magnitudes for each of the fields analysed in this study.
        The depths have been measured in $0.35^{\prime\prime}-$diameter apertures on the PSF-homogenized imaging, and corrected to total assuming a point-source correction as described in \citet{mcleod2024}.}
        \begin{tabular}{llccccccccccccc}
        \hline
        Field & Area / acrmin$^2$ & F435W & F606W & F814W  & F090W & F115W & F150W & F200W & F277W & F300M & F356W & F410M & F444W \\
        \hline
        SMACS 0723 & 6.4 & 26.7 & 27.7 & 26.8 & 28.5 & - & 28.6 & 28.7 & 28.7 & - & 28.7 & - & 28.4 \\
        Quintet & 40.9 & - & - & - & 26.9 & - & 27.0 & 27.3 & 27.7 & - & 27.8 & - & 27.4 \\
        Cartwheel & 4.2 & - & - & - & 27.3 & - & 27.3 & 27.6 & 27.9 & - & 27.9 & - & 27.8 \\
        WHL 0137 & 5.9 & 26.6 & 27.4 & 26.8 & 27.7 & 27.9 & 28.0 & 28.1 & 28.3 & - & 28.2 & 27.9 & 27.9 \\
        MACS 0647 & 5.1 & 26.7 & 27.2 & 27.2 & - & 27.7  & 27.8 & 28.2 & 28.3 & - & 28.3 & - & 28.0 \\
        J1235 & 9.9 & - & - & - & 28.4 & 28.5 &  28.4 & 28.8 & 28.9 & 28.6 & 28.9 & - & 28.3 \\
        NEP TDF & 10.5 & - & - & - & 28.1 & 28.2  & 28.0 & 28.3 & 28.7 & - & 28.5 & 28.1 & 28.3 \\
        RXJ 2129 & 3.1 & 26.8 & 27.2 & 26.7 & - & 27.2  & 27.8 & 27.6 & 27.9 & - & 28.3 & - & 27.7 \\
        GLASS & 10.1 & - & - & - & 28.8 & 28.9  & 28.5  & 28.7 & 28.9 & - & 28.9 & - & 28.8 \\
        DDT 2756 & 6.6 & - & - & - & - & 28.5 & 28.6 & 28.8 & 28.7 & - & 28.9 & - & 28.3 \\
        UNCOVER & 12.2 & 28.2 & 28.6 & 28.4 & - & 28.6  & 28.6 & 28.8 & 28.8 & - & 28.9  & 28.5 & 28.5 \\
        UNCOVER-South & 20.1 & - & - & - & 28.5 & 28.4 & 28.6 & 28.7 & 28.7 & - & 28.8 & 28.1 & 28.5 & \\
        CEERS & 95.0 & 27.8 & 28.4 & 28.0 & - & 28.5 & 28.4 & 28.6 & 28.6 & - & 28.5  & 28.0 & 28.2 \\
        NGDEEP & 9.1 & 28.5 & 29.2 & 28.7 & - & 29.2 & 29.2 & 29.2 & 29.5 & -& 29.2 & - & 29.2 \\
        JADES/JEMS & 25.5 & 28.6 & 28.9 & 28.4 & 29.5 & 29.9 & 29.8 & 29.9 & 30.1 & - & 30.0 & 29.5 & 29.7 \\
        \hline
    \end{tabular}
    \label{tab: depths}
\end{table*}

\begin{table*}
\centering
\caption{Basic information about the additional candidates selected within the JADES field. We include a reference in the final column if this object has been reported previously in the literature: B11=\citet{bouwens2011}, E13=\citet{ellis2013}, M13=\citet{mclure2013}, O13=\citet{oesch2013}, A23=\citet{austin2023}, B23=\citet{bouwens2023}, D23=\citet{donnan2023b}, H23=\citet{hainline2023}, L23=\citet{leung2023}.}
\begin{tabular}{lccccc}
\hline
ID & RA & Dec & $z_{\mathrm{phot}}$ & $M_{\mathrm{UV}}$ & References\\
\hline
JADES-3213 & 03:32:39.83 & $-$27:50:03.28 & $10.73^{+0.13}_{-0.16}$ & $-19.50^{+0.07}_{-0.06}$ & H23 \\[1ex]
JADES-8695 & 03:32:34.87 & $-$27:49:24.94 & $10.18^{+0.42}_{-0.50}$ & $-18.49^{+0.14}_{-0.14}$ & H23 \\[1ex]
JADES-9079 & 03:32:39.73 & $-$27:49:22.94 & $10.56^{+0.27}_{-0.32}$ & $-18.22^{+0.14}_{-0.10}$ & H23 \\[1ex]
JADES-9320 & 03:32:30.03 & $-$27:49:21.49 & $11.64^{+1.21}_{-0.68}$ & $-18.15^{+0.23}_{-0.14}$ & - \\[1ex]
JADES-9585 & 03:32:44.03 & $-$27:49:19.96 & $9.03^{+0.28}_{-0.23}$ & $-17.27^{+0.11}_{-0.10}$ & - \\[1ex]
JADES-9996 & 03:32:39.92 & $-$27:49:17.62 & $12.14^{+1.01}_{-0.29}$ & $-18.54^{+0.08}_{-0.13}$ & H23 \\[1ex]
JADES-11367 & 03:32:32.87 & $-$27:49:10.86 & $13.14^{+0.80}_{-0.83}$ & $-17.99^{+0.13}_{-0.18}$ & - \\[1ex]
JADES-12993 & 03:32:44.65 & $-$27:49:02.94 & $9.79^{+0.38}_{-0.28}$ & $-17.71^{+0.07}_{-0.11}$ & H23 \\[1ex]
JADES-14905 & 03:32:38.07 & $-$27:48:53.05 & $11.32^{+0.29}_{-0.34}$ & $-17.86^{+0.11}_{-0.09}$ & H23 \\[1ex]
JADES-16929 & 03:32:41.63 & $-$27:48:43.82 & $10.18^{+0.40}_{-0.54}$ & $-17.97^{+0.18}_{-0.18}$ & H23 \\[1ex]
JADES-19283 & 03:32:36.26 & $-$27:48:32.85 & $9.59^{+0.81}_{-0.61}$ & $-17.39^{+0.20}_{-0.25}$ & - \\[1ex]
JADES-20624 & 03:32:40.17 & $-$27:48:27.04 & $9.53^{+0.57}_{-0.22}$ & $-19.06^{+0.06}_{-0.12}$ & H23 \\[1ex]
JADES-33477 & 03:32:40.47 & $-$27:47:33.94 & $11.62^{+0.25}_{-0.23}$ & $-18.56^{+0.08}_{-0.07}$ & B11, E13, M13, O13, B23, D23, H23 \\[1ex]
JADES-34414 & 03:32:32.42 & $-$27:47:29.77 & $11.16^{+0.22}_{-0.25}$ & $-17.87^{+0.09}_{-0.07}$ & H23 \\[1ex]
JADES-45865 & 03:32:48.07 & $-$27:46:43.13 & $9.47^{+0.87}_{-0.50}$ & $-17.50^{+0.16}_{-0.25}$ & - \\[1ex]
JADES-48153 & 03:32:30.96 & $-$27:46:34.48 & $9.39^{+0.86}_{-0.41}$ & $-17.75^{+0.15}_{-0.20}$ & - \\[1ex]
JADES-50455 & 03:32:38.12 & $-$27:46:24.56 & $9.51^{+0.48}_{-0.38}$ & $-18.07^{+0.12}_{-0.11}$ & B13, O13, B23, D23, H23 \\[1ex]
JADES-68549 & 03:32:45.72 & $-$27:44:59.33 & $11.88^{+0.14}_{-0.14}$ & $-18.95^{+0.04}_{-0.04}$ & H23 \\[1ex]
JADES-69507 & 03:32:37.53 & $-$27:44:54.37 & $10.05^{+0.41}_{-0.51}$ & $-18.15^{+0.11}_{-0.17}$ & H23 \\[1ex]
JADES-69979 & 03:32:44.25 & $-$27:44:51.27 & $9.13^{+1.22}_{-0.42}$ & $-17.60^{+0.12}_{-0.21}$ & - \\[1ex]
JADES-1015339 & 03:32:35.79 & $-$27:49:18.88 & $9.86^{+0.61}_{-0.57}$ & $-17.87^{+0.16}_{-0.18}$ & - \\[1ex]
JADES-1047091 & 03:32:30.58 & $-$27:47:56.45 & $11.62^{+1.02}_{-0.83}$ & $-17.28^{+0.32}_{-0.20}$ & - \\[1ex]
JADES-1058823 & 03:32:30.46 & $-$27:47:27.67 & $15.09^{+0.81}_{-0.75}$ & $-18.27^{+0.21}_{-0.17}$ & H23 \\[1ex]
JADES-1125442 & 03:32:39.27 & $-$27:44:28.46 & $9.76^{+0.67}_{-0.52}$ & $-17.86^{+0.13}_{-0.23}$ & - \\[1ex]
JADES-2016436 & 03:32:43.09 & $-$27:49:13.67 & $9.37^{+0.99}_{-0.92}$ & $-16.70^{+0.42}_{-0.38}$ & - \\[1ex]
JADES-2084090 & 03:32:45.59 & $-$27:46:17.37 & $11.59^{+0.31}_{-0.31}$ & $-18.20^{+0.12}_{-0.09}$ & H23 \\[1ex]
JADES-2103879 & 03:32:33.11 & $-$27:45:26.86 & $11.98^{+1.08}_{-0.41}$ & $-18.48^{+0.12}_{-0.12}$ & - \\[1ex]
NGDEEP-17469 & 03:32:57.75 & $-$27:51:43.68 & $11.10^{+0.28}_{-0.32}$ & $-18.03^{+0.14}_{-0.08}$ & L23 \\[1ex]
NGDEEP-23088 & 03:32:58.10 & $-$27:51:18.30 & $11.07^{+0.27}_{-0.27}$ & $-18.59^{+0.11}_{-0.08}$ & A23, L23 \\[1ex]
NGDEEP-26794 & 03:33:06.46 & $-$27:51:02.10 & $10.72^{+0.18}_{-0.23}$ & $-19.09^{+0.10}_{-0.07}$ & A23, L23 \\[1ex]
NGDEEP-51475 & 03:33:01.70 & $-$27:49:06.11 & $9.21^{+1.17}_{-0.52}$ & $-17.76^{+0.34}_{-0.30}$ & -\\[1ex]
NGDEEP-51295 & 03:33:01.30 & $-$27:49:03.53 & $11.23^{+0.35}_{-0.39}$ & $-18.21^{+0.19}_{-0.13}$ & -\\[1ex]
NGDEEP-54829 & 03:32:58.93 & $-$27:48:51.75 & $10.49^{+0.10}_{-0.11}$ & $-18.99^{+0.05}_{-0.04}$ & L23 \\[1ex]
NGDEEP-1003576 & 03:32:59.88 & $-$27:52:32.57 & $15.40^{+0.29}_{-0.29}$ & $-18.89^{+0.08}_{-0.07}$ & A23, L23 \\[1ex]
NGDEEP-1003750 & 03:33:06.53 & $-$27:52:31.05 & $11.69^{+1.96}_{-0.49}$ & $-18.83^{+0.15}_{-0.26}$ & -\\[1ex]
NGDEEP-1026946 & 03:32:57.85 & $-$27:49:43.77 & $9.13^{+0.08}_{-0.07}$ & $-17.79^{+0.02}_{-0.02}$ & L23 \\[1ex]
UNCOVER-SOUTH-7302 & 00:14:38.08 & $-$30:31:49.23 & $9.48^{+0.97}_{-0.47}$ & $-19.33^{+0.12}_{-0.26}$ & -\\[1ex]
UNCOVER-SOUTH-9195 & 00:14:29.80 & $-$30:31:38.53 & $9.15^{+1.23}_{-0.43}$ & $-19.31^{+0.14}_{-0.26}$ & -\\[1ex]
UNCOVER-SOUTH-22977 & 00:14:23.39 & $-$30:30:05.66 & $9.25^{+0.08}_{-0.07}$ & $-18.86^{+0.02}_{-0.02}$ & -\\[1ex]
UNCOVER-SOUTH-26383 & 00:14:23.94 & $-$30:29:50.55 & $9.73^{+0.20}_{-0.17}$ & $-18.79^{+0.02}_{-0.03}$ & -\\[1ex]
UNCOVER-SOUTH-31496 & 00:14:27.58 & $-$30:29:18.02 & $11.60^{+0.56}_{-0.42}$ & $-19.00^{+0.13}_{-0.12}$ & -\\[1ex]
UNCOVER-SOUTH-33059 & 00:14:36.78 & $-$30:29:09.30 & $9.76^{+0.21}_{-0.22}$ & $-19.20^{+0.04}_{-0.05}$ & -\\[1ex]
UNCOVER-SOUTH-43514 & 00:14:20.51 & $-$30:28:13.53 & $8.91^{+0.78}_{-0.37}$ & $-18.98^{+0.13}_{-0.11}$ & -\\[1ex]
\hline
\end{tabular}
\label{tab: extrafieldobjects}
\end{table*}

\begin{table*}
\ContinuedFloat
\centering
\caption{Continued.}
\begin{tabular}{lccccc}
\hline
ID & RA & Dec & $z_{\mathrm{phot}}$ & $M_{\mathrm{UV}}$ & References\\
\hline
UNCOVER-SOUTH-45015 & 00:14:34.92 & $-$30:28:06.74 & $9.74^{+0.38}_{-0.32}$ & $-19.00^{+0.08}_{-0.11}$ & -\\[1ex]
UNCOVER-SOUTH-51653 & 00:14:23.80 & $-$30:27:24.18 & $11.53^{+0.31}_{-0.28}$ & $-19.03^{+0.09}_{-0.07}$ & -\\[1ex]
UNCOVER-SOUTH-57007 & 00:14:22.33 & $-$30:26:59.23 & $10.61^{+0.15}_{-0.16}$ & $-19.22^{+0.07}_{-0.06}$ & -\\[1ex]
UNCOVER-SOUTH-59562 & 00:14:23.00 & $-$30:26:38.79 & $11.61^{+0.09}_{-0.09}$ & $-19.41^{+0.02}_{-0.02}$ & -\\[1ex]
UNCOVER-SOUTH-1017193 & 00:14:25.25 & $-$30:31:25.65 & $12.66^{+1.76}_{-0.77}$ & $-19.77^{+0.17}_{-0.26}$ & -\\[1ex]
UNCOVER-SOUTH-1089623 & 00:14:24.11 & $-$30:26:33.74 & $10.96^{+0.34}_{-0.41}$ & $-18.85^{+0.17}_{-0.12}$ & -\\[1ex]
UNCOVER-SOUTH-2017743 & 00:14:35.05 & $-$30:31:20.31 & $10.28^{+0.40}_{-0.65}$ & $-18.93^{+0.24}_{-0.16}$ & -\\[1ex]
\hline
\end{tabular}
\label{tab: extrafieldobjects}
\end{table*}

In Table \ref{tab: surveys} we give an overview of the datasets used in this work to construct our wide-area \emph{JWST} sample.
We include the field, proposal ID and PI name of each dataset as well as references to the relevant survey paper, ancillary data, and lensing maps where applicable.
In Table \ref{tab: depths} we list the $5\sigma$ global limiting magnitudes for each data set in each of the observed filters.
With the exception of JADES, NGDEEP and UNCOVER-South, which are described in Section \ref{sec:sample_and_properties}, the data reduction and catalogue creation for each dataset are described in \citet{mcleod2024}.
Across these various datasets, which cover a non-contiguous on-sky area of $\simeq 320$ arcmin$^2$, we select $121$ galaxy candidates at $z\geq9$.
The coordinates, $z_{\mathrm{phot}}$ and $M_{\mathrm{UV}}$ of the new candidates not reported in \citet{mcleod2024} are given in Table \ref{tab: extrafieldobjects}.
    
\section{Details of the nebular continuum models}\label{app:cloudy_nebcont}
In Section \ref{subsec:dust_free_discussion} we discuss the theoretical blue limit for the UV continuum slope of star-forming galaxies using theoretical stellar population models which include the effect of nebular continuum emission.
For the \textsc{starburst99} \citep{leitherer1999,gotberg2017} and \textsc{bpassv2.3}  \citep{byrne2022} stellar models, the nebular continuum predictions were generated using Cloudy v17.03 \citep{ferland2017}.
For a stellar population model of a given age and metallicity, the resulting nebular continuum emission was generated assuming a simple plane-parallel geometry. 
We assumed a constant \hii \ region electron density of $n_{e}=300$ cm$^{-3}$ and a dimensionless ionisation parameter of $\mathrm{log}(U)=-2.5$.
These values are typical of star-forming galaxies at $z\simeq2-3$ \citep[e.g.][]{steidel2016}, and early spectroscopic observations with \emph{JWST} have indicated that galaxies up to $z\simeq9$ are characterised by similar parameters \citep[][]{sanders2023}.
However, we note that although $n_{e}$ and $\mathrm{log}(U)$ do affect the resulting nebular continuum, the age and metallicity of the underlying stellar population remain the dominant factor.

For the \textsc{fsps} stellar models we generated nebular continuum emission using the \textsc{python-fsps} library \citep{pythonfsps}.
The photoionisation models underlying these nebular continuum are described in \citet{byler2017}.
In \textsc{python-fsps} the electron density is fixed at $n_{e}=100$ cm$^{-3}$ \citep{byler2017} and we assume $\mathrm{log}(U)=-2.5$ as before.
Despite subtle differences in the nebular modelling, the differences relative to \textsc{starburst99} and \textsc{bpassv2.3} visible in Fig. \ref{fig:nebular_modeling} are primarily due to differences in the underlying stellar models.

\section{Investigating selection effects}\label{app:beta_recovery_tests}

    \begin{figure}
        \centerline{\includegraphics[width=\columnwidth]{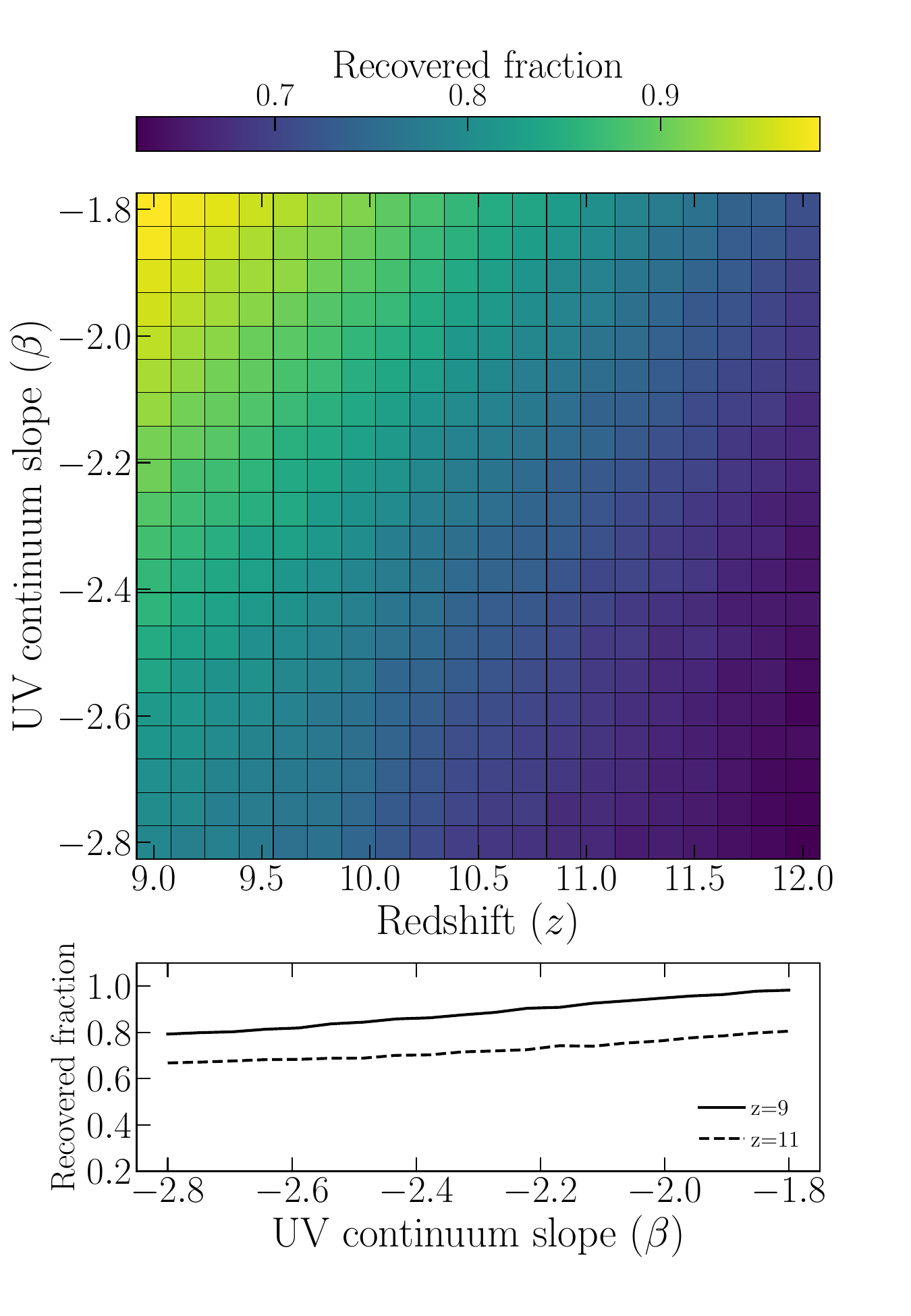}}
        \caption{Plots of the recovery fraction as a function of $\beta$ and redshift for a set of mock galaxy spectra.
        The construction of the mock sample is described in Appendix \ref{app:beta_recovery_tests}.
        The top panel shows a 2D heat map detailing the dependence of the recovered fraction on $\beta$ and $z$, highlighting the fact that the recovery fraction increases toward (i) lower $z$ at fixed $\beta$, and (ii) redder $\beta$ at fixed $z$.
        The lower panel shows two slices through the grid at $z=9$ (solid black line) and $z=11$ (dashed black line).
        The increase in selection efficiency towards redder $\beta$ is relatively small, increasing by $\simeq 20$ per cent between $\beta=-2.8$ and $\beta=-2.0$.
        Crucially, this test suggests that the relative selection efficiency as a function of $\beta$ is essentially the same at $z=9$ and $z=11$, indicating that the bluer $\beta$ values we observed in our data at $z=11$ should not be caused by a bias toward bluer/redder values of $\beta$ at different redshifts.}
        \label{fig:app_recov_fraction}
    \end{figure}

We performed a set of additional tests to investigate the extent to which the rapid evolution observed in $\langle \beta \rangle$ between $z=9$ and $z=11$ could be the result of sample selection effects.
We first estimated the recovered fraction of sources as a function of redshift ($z$) and intrinsic $\beta$ using a mock galaxy sample.
For each combination of $z$ and $\beta$, we generated mock spectra $5000$ with a uniform distribution of $M_{\mathrm{UV}}$ between $M_{\mathrm{UV}}=-20.5$ and $M_{\mathrm{UV}}=-18$ (i.e roughly the range of $M_{\mathrm{UV}}$ covered by our $z>9$ wide-area JWST sample).
We then generated NIRCam photometry for each mock galaxy and added noise consistent with the median image depths across the 15 independent datasets (see Table \ref{tab: depths}).
Finally, we applied the selection criteria described in Section \ref{sec:sample_and_properties} and determined the recovery fraction (averaged across all $M_{\mathrm{UV}}$ values).

    \begin{figure}
        \centerline{\includegraphics[width=\columnwidth]{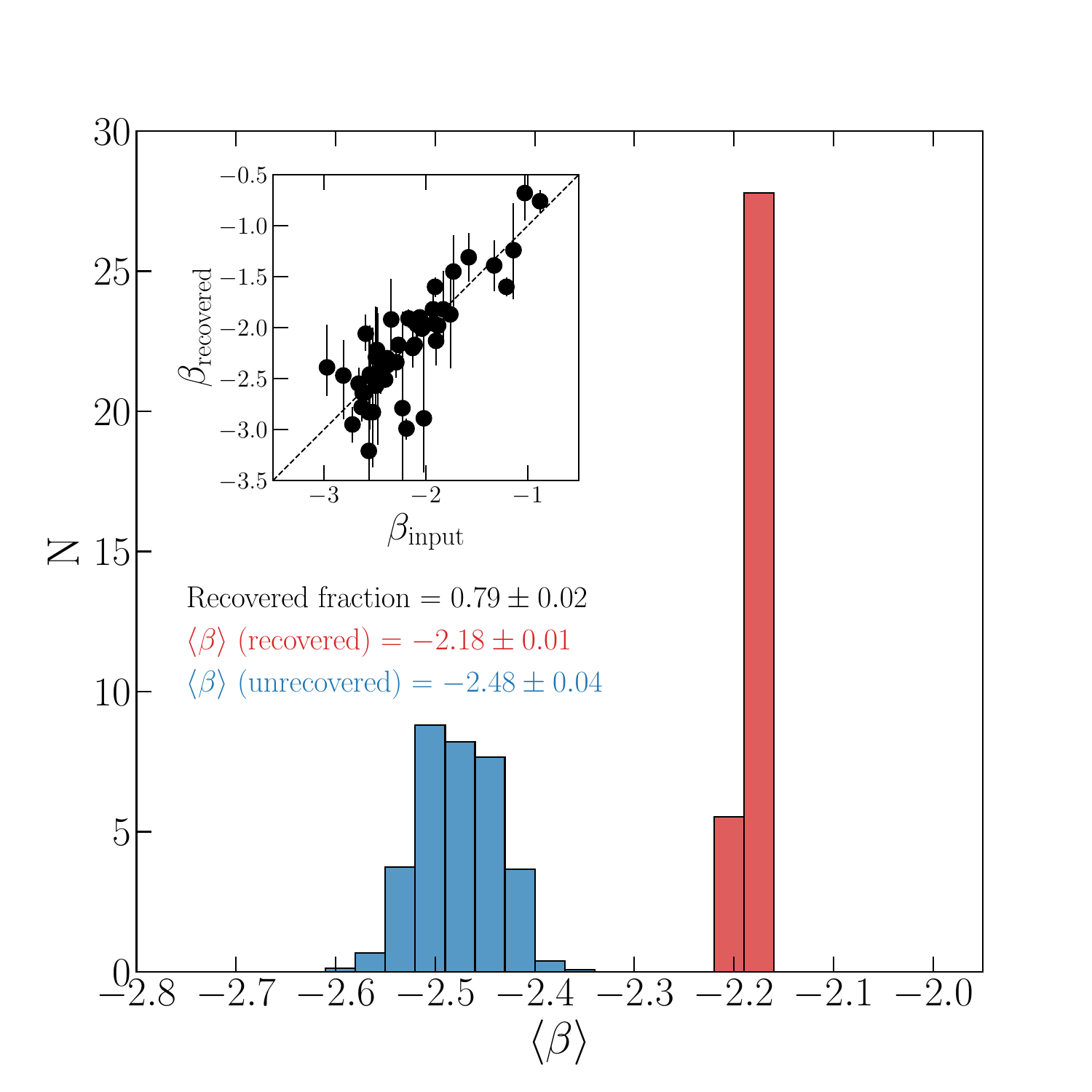}}
        \caption{Plot showing the recovered fraction and reestimated $\langle \beta \rangle$ after artificially redshifting the $9 < z < 10.5$ sources in our wide-area JWST sample (with $\langle z \rangle = 9.8$) to higher redshifts (such that $\langle z_{\mathrm{new}} \rangle = 11$).
        Details of the redshifting and source recovery process are given in Appendix \ref{app:beta_recovery_tests}.
        In the main panel, the red and blue histograms show the distribution of $\langle \beta \rangle$ for recovered and unrecovered sources, respectively, across 500 trials.
        On average, roughly $80$ per cent of the $z < 10.5$ sample would remain within our sample when shifted to higher redshift.
        The typical $\beta$ values of the recovered sources are redder than those of the unrecovered sources, in line with the simulation results shown in Fig. \ref{fig:app_recov_fraction}.
        The inset panel shows the remeasured $\beta$ values of the recovered sources versus their measured $\beta$ value at lower redshift.
        We find no evidence to suggest that the measurement of $\beta$ is biased depending on the redshift of the source (see also Fig. \ref{fig:uvslope_recovery}).}
        \label{fig:app_test2}
    \end{figure}
    
Fig. \ref{fig:app_recov_fraction} shows the recovered fraction as a function of $z$ and $\beta$ for redshifts in the range $9 < z < 11$ and UV continuum slopes in the range $-2.8 < \beta < -1.8$.
Focusing first on the upper panel of Fig. \ref{fig:app_recov_fraction}, two clear trends are visible.
First, at fixed $\beta$ galaxies at lower redshifts are more efficiently selected, which is as expected since galaxies at fixed $M_{\mathrm{UV}}$ will be brighter at lower redshift (i.e. the horizontal colour gradient).
Second, and crucially for our analysis, it can be seen that at fixed redshift the selection efficiency is relatively flat as a function of $\beta$ (i.e. the shallow vertical colour gradient).
We illustrate the shallow $\beta$ dependence explicitly in the bottom panel of Fig. \ref{fig:app_recov_fraction}, where we have taken two slices through the grid at $z=9$ and $z=11$.
At both redshifts, the selection efficiency increases slightly (by a factor $\simeq 1.2$) between $\beta=-2.8$ and $\beta=-2.0$.
This shallow increase in selection efficiency towards redder values of $\beta$ can be understood in terms of the fact that, for a fixed $M_{\mathrm{UV}}$, the median S/N across the full rest-frame UV spectrum will increase as $\beta$ increases\footnote{As $\beta$ increases from $\beta=-2.8$ the $f_{\nu}$ spectrum flattens/reddens (at $\beta=-2.0$ the spectrum is completely flat in $f_{\nu}$). For flatter/redder slopes the chance of at least one band redward of the break exceeding our $\mathrm{SNR}\geq8$ threshold is improved.}.
Because our wide-area JWST sample is selected by requiring an $8\sigma$ detection in \emph{any} band redward of the Lyman break, and because at $z>9$ the full rest-frame UV SED up to $\lambda_{\mathrm{rest}} \simeq 3000${\AA} is sampled by the JWST/NIRCam filters\footnote{Our selection filters trace the rest-frame UV SED up to $\lambda_{\mathrm{rest}} \simeq 2700${\AA} at $z=9$ and $\lambda_{\mathrm{rest}} \simeq 2300${\AA} at $z=11$.}, the moderate increase in selection efficiency towards redder $\beta$ values will apply at all redshifts. 
From Fig \ref{fig:app_recov_fraction} we can conclude that the evolution in $\langle \beta \rangle$ we observe cannot be a result of an increasing efficiency in selecting bluer galaxies at higher redshift as the relative selection efficiency as a function of $\beta$ is the same at $z=9$ and $z=11$.
These results highlight a more general point: galaxy selection up to $z\simeq15$ with JWST should not be biased towards the intrinsically bluer sources because deep JWST/NIRCam imaging allows us to select across their full rest-frame UV spectrum\footnote{We note that the increased efficiency in selecting redder objects discussed here is subtly different to the often-discussed $\beta$ `blue-bias' \citep[e.g.][]{bouwens2010, dunlop2012, rogers2013, cullen2023}.
The blue bias occurs for faint galaxies that are selected via the filter closest to the Lyman break (which will typically be the case for galaxies with $\beta \lesssim -2.0$, i.e. a falling $f_{\nu}$ SED).
For low S/N objects close to the detection threshold, this selection will favour objects whose photometry has been upscattered in the short-wavelength detection band, and these objects will appear bluer than they actually are.
However, because we have restricted our sample to high S/N detections ($8\sigma$), and allowed for selection in any band redward of the Lyman break, the sample average $\beta$ values for our current sample is not affected strongly by this bias (see discussion in Section \ref{subsec:recovery_sims}).}.

As a second test, we estimated the source recovery fraction and recovered $\beta$ when shifting the best-fitting SEDs of our $9 < z < 10.5$ sources to higher redshift and reapplying our selection criteria.
The inverse-variance weighted mean $\beta$ of these sources is $\langle \beta \rangle = -2.22$ with a mean redshift of $\langle z \rangle =9.8$.
The aim of the test was to discover whether the recovered galaxies were biased towards the bluer or redder objects, and whether there was any bias in the estimated $\beta$ for the recovered sources.
We redshifted each individual ${9 < z < 10.5}$ galaxy by $\Delta z = 1.2$ so that the median redshift of the artificially redshifted sample became $\langle z \rangle = 11$.
We then computed new model fluxes for each source and added noise consistent with the photometric uncertainties.
This process was repeated 500 times and for each iteration we calculated the fraction of artificially redshifted sources that were recovered using our selection criteria, as well as the inverse-variance weighted mean $\beta$ of both the recovered and unrecovered sources.

The results of the test are shown in Fig. \ref{fig:app_test2}.
Across the 500 iterations, the mean recovered fraction was $0.79 \pm 0.02$ and the recovered sources had $\langle \beta \rangle = -2.18 \pm 0.01$ compared to $\langle \beta \rangle = -2.48 \pm 0.04$ for the unrecovered sources.
The 79 per cent recovery fraction can be explained as a result of the redshifted sources becoming fainter such that $\simeq 20$ per cent drop below our selection thresholds.
It can be seen that, on average, the redder sources are more likely to be recovered, and as a result the average $\beta$ of the recovered sources is biased slightly red (by $\Delta \beta =0.04$).
This result is consistent with the simulations discussed above (i.e. at any given redshift, redder sources are slightly favoured by our selection; Fig. \ref{fig:app_recov_fraction}).
Crucially, for the sources that are recovered, there is no bias in the individual galaxy $\beta$ estimates at the new redshift (inset panel in Fig. \ref{fig:app_test2}).
This test conclusively rules out a scenario in which the blue $\langle \beta \rangle$ we observe at $z > 10.5$ is the result of a bias favouring the selection of bluer sources.
Indeed, this analysis suggests that if the $z\simeq9.5$ galaxy population were transplanted to $z\simeq11$, the redder sources would preferably be recovered.
The fact that we do not see these sources in our data is therefore indicative of a real evolution in the shape of the UV continuum SED with redshift.

We also ran this test in reverse, taking the $10.5 < z < 12.0$ sources (with $\langle \beta \rangle = -2.60$ and $\langle z \rangle = 11.4$) and blueshifting them to lower redshift by $\Delta z = -1.9$ so that the median redshift became ${\langle z \rangle=9.5}$.
In this case, we find a high average recovery fraction of $96$ per cent across the 500 iterations (as expected).
Typically, one or two sources per iteration failed to fulfil the selection criteria due to the random photometric perturbations, and these sources were always amongst the bluest in the sample (with $\beta < -2.7$).
However, because most of the sources remain within our selection criteria, the mean $\beta$ of the population was accurately recovered (the inverse variance-weighted mean value for the recovered sources was ${\langle \beta \rangle = -2.61 \pm 0.01}$).
From this test, we can further conclude that if the blue population we observe at ${z>10.5}$ also existed at ${z \simeq 9.5}$, these blue sources would be present in our sample.

\bsp	
\label{lastpage}
\end{document}